\documentclass[aps,prd,twocolumn,preprintnumbers,nofootinbib,superscriptaddress,amsmath]{revtex4-1}
\usepackage{graphicx}
\usepackage{amsmath}
\usepackage{amssymb}
\usepackage{amsfonts}
\usepackage{hyperref}
\usepackage{url}
\usepackage{color}
\usepackage{bm}
\usepackage{array}

\newcommand{\be}{\begin{equation}}
\newcommand{\ee}{\end{equation}}
\newcommand{\ba}{\begin{eqnarray}}
\newcommand{\ea}{\end{eqnarray}}
\newcommand{\nn}{\nonumber}

\begin{document}
\makeatother

\preprint{IFT-UAM/CSIC-21-097}

\title{Search for black hole hyperbolic encounters with gravitational wave detectors}

\author{Gonzalo Morr\'as}
\email{gonzalo.morras@estudiante.uam.es}
\affiliation{Instituto de F\'isica Te\'orica UAM-CSIC, Universidad Auton\'oma de Madrid,
Cantoblanco, 28049 Madrid, Spain}

\author{Juan Garc\'ia-Bellido}
\email{juan.garciabellido@uam.es}
\affiliation{Instituto de F\'isica Te\'orica UAM-CSIC, Universidad Auton\'oma de Madrid,
Cantoblanco, 28049 Madrid, Spain}

\author{Savvas Nesseris}
\email{savvas.nesseris@csic.es}
\affiliation{Instituto de F\'isica Te\'orica UAM-CSIC, Universidad Auton\'oma de Madrid,
Cantoblanco, 28049 Madrid, Spain}

\date{\today}

\begin{abstract}
\noindent In recent years, the proposal that there is a large population of primordial black holes living in dense clusters has been gaining popularity. One natural consequence of these dense clusters will be that the black holes inside will gravitationally scatter off each other in hyperbolic encounters, emitting gravitational waves that can be observed by current detectors. In this paper we will derive how to compute the gravitational waves emitted by black holes in hyperbolic orbits, taking into account up to leading order spin effects. We will then study the signal these waves leave in the network of gravitational wave detectors currently on Earth. Using the properties of the signal, we will detail the data processing techniques that can be used to make it stand above the detector noise. Finally, we will look for these signals from hyperbolic encounters in the publicly available LIGO-Virgo data. For this purpose we will develop a two step trigger. The first step of the trigger will be based on looking for correlations between detectors in the time-frequency domain. The second step of the trigger will make use of a residual convolutional neural network, trained with the theoretical predictions for the signal, to look for hyperbolic encounters. With this trigger we find 8 hyperbolic encounter candidates in the 15.3 days of public data analyzed. Some of these candidates are promising, but the total number of candidates found is consistent with the number of false alarms expected from our trigger.

\end{abstract}
\maketitle

\section{Introduction}
\label{sec:intro}

Ever since the first detection of gravitational waves in 2015 by the LIGO-Virgo collaboration \cite{GW150914} a new era of gravitational wave astronomy has begun. With the much improved sensitivity of the Advanced LIGO \cite{AdvLIGO_design} and Advanced Virgo \cite{AdvVirgo_design} detectors, the observation of gravitational wave events has become a common occurrence, with more than 50 black hole mergers confirmed \cite{GWTC-2}, as well as the detection of a binary neutron star inspiral \cite{GW170817} whose electromagnetic counterpart was also observed, opening the era of Multimessenger Astronomy \cite{Multimessenger}.

Astrophysical models have problems explaining the high rate of black hole mergers detected as well as some of their measured parameters. Observed black holes have a spin much smaller than what is expected if they were formed from the collapse of stars \cite{Paper_del_Nuno} and some of them have masses impossible to generate in this way. Stellar collapse models predict black hole masses to be between 5 and 50 solar masses \cite{StellarBHMass}, but there is experimental evidence in favor of black holes with masses above \cite{FatMerger} and below this range \cite{SkinnyMergers}.

These observations point at a new population of black holes that do not have a stellar origin. As a consequence, there has recently been renewed interest in primordial black holes (PBHs) \cite{PBHreview}, which are black holes formed shortly after the Big Bang by the gravitational collapse of the primordial density fluctuations \cite{PBHThermalHistory}. 

Even though there are some constraints on the primordial black hole population coming mainly from microlensing experiments as well as dwarf galaxy stellar dynamics and the cosmic microwave background \cite{PBHreview}, there is still enough freedom for a black hole population in the 0.1$M_{\odot}$ to 100$M_{\odot}$ mass range to comprise up to all the dark matter in the universe. This is the mass range which current gravitational wave detectors are sensitive to and could thus explain their observed anomalies.

A primordial black hole population in this $0.1-100 M_\odot$ mass range also has a strong theoretical motivation \cite{PBHThermalHistory}. A few fractions of a second after the Big bang, at a temperature of $\Lambda_{\mathrm{QCD}} \sim 200 \text{MeV}$, the $\mathrm{QCD}$ transition took place and quarks and gluons went from being free to being confined in hadrons. The sudden drop in the speed of sound due to the creation of non relativistic hadrons from relativistic quarks and gluons can lead to the collapse of high density regions, generating a large amount of primordial black holes. The natural value of the mass of the black holes formed in this way lies in the aforementioned $0.1-100 M_\odot$ mass range \cite{PBHThermalHistory}. Moreover, the spin of the black holes generated in this way is expected to be small, in accordance with the LIGO-Virgo observations. 

In addition to the experimental evidence for this population coming from gravitational waves, most relevant for this paper, one should also note other hints that point to the existence of these primordial black holes \cite{SevenHints} such as the microlensing of stars in M31 and distant quasars, the dynamics and star clusters of ultra-faint-dwarf-galaxies, the core galaxy profiles from primordial black hole scattering, the correlations between X-ray and infrared cosmic backgrounds and the Chandra Deep Field South.

The spatial distribution of the black holes will be highly dependent on the primordial power spectrum from inflation. In general, due to the nature of the primordial perturbations it will be natural for the black holes to form clusters \cite{ClusteringDiffusion}. This clustering will in addition relax the constraints on primordial black holes coming from microlensing \cite{UpdatedMicrolensingConstrains}, which usually assume a homogeneous distribution of black holes. 

Simulations of these clusters \cite{ClusterSimulation} show that not only are black hole binaries formed, but there will also be a large amount of ``hyperbolic encounters'' in which two black holes scatter off each other with emission of gravitational wave bremsstrahlung. If in the hyperbolic encounter the two black holes pass sufficiently near each other, the system will emit a large amount of gravitational waves that will be possible to detect in current and future gravitational wave detectors \cite{MiticoHiperbolicPaper, HiperbolicJetzer, HiperbolicMukherjee}. Finding these hyperbolic events would be a strong evidence for clustered primordial black holes and could be used to study their mass and spin~\cite{Jaraba:2021ces}. 
Moreover, dense stellar clusters may also accommodate collisions among neutron stars with impact parameters significantly larger than their physical sizes (to prevent tidal disruptions), as well as BH collisions, and could thus emit GW in the LIGO band. In order to see these events, those clusters should be sufficiently near us to be detectable, but there are plenty of these in the halo of our galaxy or nearby galaxies. 
The interpretation of the event GW190521 as a dynamical capture of two black holes \cite{GW190521_DynamicalCapture, waveform_DynamicalCapture} might already be a hint of dense clustering of black holes in our universe.

To our knowledge, no systematic search for close hyperbolic encounters has been carried out in the LIGO-Virgo data. The objective of this paper is to look for this type of events in the available open data \cite{OpenData_O1_O2}.

To this end, in Sec.~\ref{sec:waveforms} we will develop 1.5 post newtonian (PN) accurate templates for the gravitational waves emitted by spinning compact binaries in hyperbolic orbits and we will numerically implement them in an efficient way using \texttt{Python}.
In Sec.~\ref{sec:detec} we will study how these gravitational waves leave a signal in the network of gravitational wave detectors currently present on Earth. Since we are interested in studying the signal in real detectors, dominated by noise, we will see how to use signal processing techniques such as filtering, whitening and Q transforming to make the signal of these hyperbolic encounters stand out over the noise. 
Having determined the signal we are looking for, in Sec.~\ref{sec:Data} we will develop a two level trigger to extract the possible hyperbolic candidates from the data. Similarly as in standard burst searches \cite{cWB}, the first level of the trigger will be a loose trigger based on correlations between detectors and tuned to select possible close hyperbolic encounter event candidates while doing a large reduction of the data. The second level of the trigger will consist of a convolutional neural network (CNN) \cite{cWB_ML,MarioCNN} to classify the events of the first level trigger into either noise or close hyperbolic encounters (CHEs). This neural network will be trained using simulations of the close hyperbolic encounter signals as well as correlation triggers of noise.

\section{Development of templates for close hyperbolic encounters}
\label{sec:waveforms}

To compute the gravitational waves emitted by the scattering of two black holes in a close hyperbolic encounter, we will first need to determine the orbit that these black holes follow. Using this orbit we will then compute the gravitational wave strain emitted by the system. We will want to take into account up to the leading order effects of the spins of the black holes, since the spin is a critical quantity to distinguish between astrophysical and primordial black holes.

\subsection{1.5PN accurate hyperbolic orbit for spinning compact binaries}
\label{sec:waveforms:orbit}

The formulation of the problem we want to solve is very simple, we want to study the scattering of two gravitationally interacting masses $m_1$ and $m_2$ with spins $\vec{\mathcal{S}}_1$ and $\vec{\mathcal{S}}_2$ respectively. Nonetheless, this problem does not have a closed analytical solution in general relativity (GR), and numerically solving the full set of Einstein equations is in general not feasible, since the computational cost is prohibitive. Because of this, the problem will be studied using a perturbative expansion of general relativity in powers of $\frac{1}{c^2}$ known as the post newtonian (PN) approximation. 

The total mass of the binary is $m = m_1 + m_2$, while the reduced mass is $\mu = m_1 m_2 /m$ and the symmetric mass ratio is $\eta = \mu / m$. To find the dynamics of the system we will use the Hamiltonian formulation of general relativity. We are interested in studying the two body scattering problem taking into account up to leading order spin-orbit coupling, which is of order 1.5 PN. Because of this we will have to consider the following reduced hamiltonian ($\mathcal{H}/\mu$):

\begin{eqnarray}
H(\vec{r},\vec{p},\vec{S}_1,\vec{S}_2) &=& H_\mathrm{N} (\vec{r},\vec{p}) + H_\mathrm{1PN} (\vec{r},\vec{p}) \nn \\
&+& H_\mathrm{SO} (\vec{r},\vec{p},\vec{S}_1,\vec{S}_2) + O\left( \frac{1}{c^4} \right) ,
\label{eq:TotalHamiltonian}
\end{eqnarray}

\noindent where we have set $\vec{r} = \vec{\mathcal{R}}/(G m)$ and $\vec{p} = \vec{\mathcal{P}}/\mu$, being $\vec{\mathcal{R}}$ the relative separation vector between the black holes and $\vec{\mathcal{P}}$ its conjugate momentum. $\vec{S}_1$ and $\vec{S}_2$ are the reduced spin vectors, that is, $\vec{S}_1 = \vec{\mathcal{S}}_1/(\mu G m)$ and $\vec{S}_2 = \vec{\mathcal{S}}_2/(\mu G m)$. In the language of Poisson brackets $\{ \cdot, \cdot\}$, these variables satisfy:
\begin{subequations}
\label{eq:PoissonBrackets}
\begin{align}
 \{ r_i, p_j \} & = \delta_{ij} \, , \label{eq::PoissonBrackets:rp}\\
 \{ S_{1i}, S_{1j} \} & = \epsilon_{ijk} S_{1k} \, , \label{eq::PoissonBrackets:S1S1} \\
 \{ S_{2i}, S_{2j} \} & = \epsilon_{ijk} S_{2k} \, , \label{eq::PoissonBrackets:S2S2}
\end{align}
\end{subequations}

\noindent with all other brackets between the variables being zero. The Hamiltonians appearing in Eq.~\eqref{eq:TotalHamiltonian} can be derived directly from Einstein's theory of general relativity. The result can be found in the literature \cite{Hamiltonians} and has the following form:
\begin{subequations}
\label{eq:Hform}
\begin{align}
& H_\mathrm{N} (\vec{r},\vec{p})   = \frac{p^2}{2}-\frac{1}{r} \, , \label{eq:Hform:HN}\\
& H_\mathrm{1PN} (\vec{r},\vec{p}) = \frac{1}{c^2} \left(\frac{1}{8} (3\eta-1)(p^2)^2 \right. \nonumber \\
& \quad \quad \quad \quad \; \; \left. - \frac{1}{2}\left[ (3+\eta) p^2 + \eta (\hat{n} \cdot \vec{p})^2 \right] \frac{1}{r} + \frac{1}{2 r^2}\right) \, , \label{eq:Hform:H1PN}\\
& H_\mathrm{SO} (\vec{r},\vec{p},\vec{S}_1,\vec{S}_2)  = \frac{1}{c^2 r^3} (\vec{r} \times \vec{p}) \cdot \vec{S}_\mathrm{eff} \, , \label{eq:Hform:HSO}
\end{align}
\end{subequations}

\noindent where $\hat{n} = \vec{r}/r$ and the effective spin $\vec{S}_\mathrm{eff}$ is a quantity of order $O(1/c)$ defined as a combination of the spins of both objects given by:
\begin{equation}
   \vec{S}_\mathrm{eff} = \delta_1 \vec{S}_1 + \delta_2 \vec{S}_2 \, .
\label{eq:Seffdef}
\end{equation}

If we label the black holes such that $m_1 \geq m_2$, then $\delta_1$ and $\delta_2$ are given by:
\begin{subequations}
\label{eq:deltas}
\begin{align}
 \delta_1 & = 2\eta\left( 1 + \frac{3 m_2}{4 m_1} \right) = \frac{\eta}{2} + \frac{3}{4}\left(1 - \sqrt{1-4\eta} \right) \, , \label{eq:deltas:1}\\
 \delta_2 & = 2\eta\left( 1 + \frac{3 m_1}{4 m_2} \right) = \frac{\eta}{2} + \frac{3}{4}\left(1 + \sqrt{1-4\eta} \right) \, . \label{eq:deltas:2}
\end{align}
\end{subequations}

To find the equations of motion we will use the Poisson brackets. Instead of finding an equation of motion for the momentum $\vec{p}$, it will be more convenient to substitute this equation for an equation of motion for the angular momentum vector $\vec{L} = \vec{r} \times \vec{p}$. If we omit terms smaller than $O(1/c^3)$, the equations of motion are:
\begin{subequations}
\label{eq:eqmotion}
\begin{align}
 &\dot{\vec{L}} = \frac{d \vec{L}}{d t} = \{\vec{L},H\}  = \frac{1}{c^2 r^3} \vec{S}_\mathrm{eff} \times \vec{L} \, , \label{eq:eqmotion:L}\\
 &\dot{\vec{S}}_1 = \frac{d \vec{S}_1}{d t} = \{\vec{S}_1,H\}  = \frac{\delta_1}{c^2 r^3} \vec{L} \times \vec{S}_1 \, , \label{eq:eqmotion:S1}\\
 &\dot{\vec{S}}_2 = \frac{d \vec{S}_2}{d t} = \{\vec{S}_2,H\}  = \frac{\delta_2}{c^2 r^3} \vec{L} \times \vec{S}_2 \, , \label{eq:eqmotion:S2}\\
 &\dot{\vec{r}} = \frac{d \vec{r}}{d t} = \{\vec{r},H\}  = \vec{p} + \frac{\vec{S}_\mathrm{eff} \times \vec{r}}{c^2 r^3} \nonumber \\ 
 & + \frac{1}{c^2} \left(\frac{1}{2} (3\eta-1)(p^2)\vec{p}-\left[ (3+\eta) \vec{p} + \eta (\hat{n} \cdot \vec{p})\hat{n} \right] \frac{1}{r}\right) \, . \label{eq:eqmotion:r}
\end{align}
\end{subequations}

In addition to these equations of motion, we will have constants of motion that are conserved in time. The reduced energy $E = H$ is conserved because $\partial_t H = 0$. Multiplying Eq.~\eqref{eq:eqmotion:L} by $\vec{L}$ we immediately get that $L = |\vec{L}|$ is conserved. Doing the same multiplication by $\vec{L}$ to Eqs.~\eqref{eq:eqmotion:S1} and \eqref{eq:eqmotion:S2}, we get that $S_1 = |\vec{S}_1|$ and $S_2 = |\vec{S}_2|$ are also conserved. It can also be shown by direct computation that the magnitude $\vec{L}\cdot\vec{S}_\mathrm{eff}$ is conserved:
\begin{align}
    \frac{d}{dt} \left( \vec{L}\cdot\vec{S}_\mathrm{eff}\right) & =  \frac{d \vec{L}}{dt} \cdot \vec{S}_\mathrm{eff} +\vec{L} \cdot \frac{d \vec{S}_\mathrm{eff}}{dt} = \nonumber \\
    & = \frac{1}{c^2 r^3} \left( \vec{S}_\mathrm{eff} \times \vec{L} \right) \cdot \vec{S}_\mathrm{eff} \nn \\
    &+ \frac{1}{c^2 r^3} \vec{L} \cdot \left( \vec{L} \times \left(\delta_1^2 \vec{S}_1 + \delta_2^2 \vec{S}_2 \right) \right) = 0 \, .
    \label{eq:LSeffconst}
\end{align}

Finally, we observe that if we add Eqs.~\eqref{eq:eqmotion:L}, \eqref{eq:eqmotion:S1} and \eqref{eq:eqmotion:S2} we get that:
\begin{equation}
    \frac{d}{d t} \left( \vec{L} + \vec{S}_1 + \vec{S}_2 \right) = 0 \, .
    \label{eq:Jconst}
\end{equation}

And thus the total reduced angular momentum $\vec{J}$, defined as $\vec{J} = \vec{L} + \vec{S}_1 + \vec{S}_2 $, is conserved in both magnitude and direction. This allows us to introduce a set of Cartesian coordinates $(x,y,z)$ which generate the triad $(\hat{e}_x,\hat{e}_y,\hat{e}_z)$ such that at all times:
\begin{equation}
    \vec{J} = J \hat{e}_z \, .
    \label{eq:Jzconst}
\end{equation}

We will also define a set of spherical coordinates $(r,\theta,\phi)$ which generate the triad $(\hat{n},\hat{e}_\theta,\hat{e}_\phi)$. Using these coordinates we can write:
\begin{subequations}
\label{eq:rsph}
\begin{align}
 \vec{r} & = r \hat{n} \, , \label{eq:rsph:r}\\
 \dot{\vec{r}} & = \dot{r} \hat{n} + r \dot{\theta} \hat{e}_\theta + r \sin{\theta} \dot{\phi} \hat{e}_\phi \, , \label{eq:rsph:rdot} \\
 \vec{p} & = p_r \hat{n} + p_\theta \hat{e}_\theta + p_\phi \hat{e}_\phi \, . \label{eq:rsph:p}
\end{align}
\end{subequations}

\subsubsection{Radial motion}
\label{sec:waveforms:orbit:r}

Using Eqs.~\eqref{eq:rsph:r} and \eqref{eq:rsph:p} we have that $L^2 = |\vec{r} \times \vec{p}|^2 = r^2 (p_\theta^2+p_\phi^2)$ and thus:
\begin{equation}
    p^2 = p_r^2+\frac{L^2}{r^2} \, .
    \label{eq:pofprL}
\end{equation}

Using the Newtonian Hamiltonian of Eq.~\eqref{eq:Hform:HN} we have:
\begin{equation}
    E = H = \frac{p^2}{2}-\frac{1}{r} + O\!\left( \frac{1}{c^2} \right)  \quad \rightarrow \quad p^2 = 2E +\frac{2}{r} +  O\!\left( \frac{1}{c^2} \right) \, .
    \label{eq:pofEN}
\end{equation}

We can substitute Eqs.~\eqref{eq:pofprL} and \eqref{eq:pofEN} in the 1.5PN Hamiltonian of Eq.~\eqref{eq:TotalHamiltonian} and get a 1.5PN accurate expression for the energy:
\begin{eqnarray}
    E &=& H = \frac{p^2}{2}-\frac{1}{r} \nn \\
    &+& \frac{1}{c^2} \left( \frac{3\eta-1}{2} E^2 + \frac{1}{r} (\eta-4)E - \frac{1}{r^2} \frac{\eta+6}{2}+\frac{1}{r^3} \frac{\eta L^2}{2} \right)\nn \\
    &+& \frac{\vec{L} \cdot \vec{S}_\mathrm{eff}}{c^2 r^3} +  O\!\left( \frac{1}{c^4} \right) \, .
    \label{eq:E1PNofp}
\end{eqnarray}

Projecting the equation of motion for $\vec{r}$ of Eq.~\eqref{eq:eqmotion:r} in the radial direction, we obtain:
\begin{equation}
    \dot{r} = \{\vec{r},H\} \cdot \hat{n} = p_r +\frac{1}{c^2} \left( \frac{3\eta -1}{2} p^2 p_r - (3+2\eta) \frac{p_r}{r} \right).
    \label{eq:rdotppr}
\end{equation}

And substituting Eq.~\eqref{eq:pofprL} into this equation of motion we have:
\begin{eqnarray}
    \frac{p^2}{2} &=& \frac{p_r^2}{2}+\frac{L^2}{2 r^2} = \frac{\dot{r}^2}{2}+\frac{L^2}{2 r^2}\nn \\
    &-&\frac{1}{c^2} \left( \frac{3\eta -1}{2} p^2 - (3+2\eta) \frac{1}{r} \right) p_r^2 + O\!\left( \frac{1}{c^4} \right).
    \label{eq:p2ofppr}
\end{eqnarray}

Finally, substituting Eqs.~\eqref{eq:pofprL} and \eqref{eq:pofEN} in Eq.~\eqref{eq:p2ofppr}, then using the result in Eq.~\eqref{eq:E1PNofp}, solving for $\dot{r}^2$ and omitting terms $O(1/c^4)$ or smaller, we get:
\begin{equation}
    \dot{r}^2 = A + \frac{2B}{r}+\frac{C}{r^2}+\frac{D}{r^3} \, ,
    \label{eq:rdotfinal}
\end{equation}

\noindent where A, B, C and D are defined only in terms of the constants of motion and are given by the following expressions:
\begin{subequations}
\label{eq:coefrdot}
\begin{align}
 A & = 2E \left(1 + \frac{3}{2} (3\eta -1) \frac{E}{c^2}  \right) \, , \label{eq:coefrdot:A}\\
 B & = 1 + (7\eta - 6) \frac{E}{c^2} \, ,  \label{eq:coefrdot:B}\\
 C & = -L^2 \left( 1 + 2(3\eta-1) \frac{E}{c^2}  \right) + 5 (\eta-2) \frac{E}{c^2} \, , \label{eq:coefrdot:C}\\
 D & = (-3\eta + 8) \frac{L^2}{c^2} - 2 \frac{\vec{L} \cdot \vec{S}_\mathrm{eff}}{c^2} \, . \label{eq:coefrdot:D}
\end{align}
\end{subequations}

Since D is of order $O(1/c^2)$, we can define a new variable $\overline{r}$ with the following equation:
\begin{equation}
    r = \overline{r}-\frac{D}{2 L^2} \, .
    \label{eq:conctransf}
\end{equation}

And then, omitting terms of order $O(1/c^4)$, we can write Eq.~\eqref{eq:rdotfinal} as:
\begin{equation}
    \dot{\overline{r}}^2 = A + \frac{2B}{\overline{r}}+\frac{\overline{C}}{\overline{r}^2} \, ,
    \label{eq:rdotconc}
\end{equation}

with:
\begin{equation}
    \overline{C} = C + \frac{DB}{L^2} \, .
    \label{eq:Cbardef}
\end{equation}

Eq.~\eqref{eq:rdotconc} is completely analogous to the expression one gets for Kepler's problem in Newtonian dynamics, and thus it accepts hyperbolic solutions given in the following parametric way:
\begin{subequations}
\label{eq:NHiperbdef}
\begin{align}
 \overline{n}(t-t_0) & = e \sinh{v} - v \, , \label{eq:NHiperbdef:t}\\
 \overline{r} & = \overline{a}(e \cosh{v} - 1) \, . \label{eq:NHiperbdef:r}
\end{align}
\end{subequations}

Substituting this in Eq.~\eqref{eq:rdotconc} and imposing the equality of both sides, one gets the following orbital parameters:
\begin{subequations}
\label{eq:NHiperbparam}
\begin{align}
 \overline{a} & = \frac{B}{A} \, , \label{eq:NHiperbparam:a}\\
 \overline{n} & = \frac{A^{3/2}}{B} \, , \label{eq:NHiperbparam:n}\\
 e & = \sqrt{1-\frac{A\overline{C}}{B^2}} \, . \label{eq:NHiperbparam:e}
\end{align}
\end{subequations}

We can now find a parametric expression for $r$ undoing the transformation of Eq.~\eqref{eq:conctransf}:
\begin{subequations}
\label{eq:GRHiperbdef}
\begin{align}
 \overline{n}(t-t_0)  = e \sinh{v} - v   \; \longrightarrow \; \overline{n}(t-t_0)  = e_t \sinh{v} - v \, ,  \label{eq:GRHiperbdef:t}\\
 r  = \overline{a}(e \cosh{v} - 1) - \frac{D}{2L^2}  \; \longrightarrow  \; r   = \overline{a}_r(e_r \cosh{v} - 1) \, , \label{eq:GRHiperbdef:r}
\end{align}
\end{subequations}

\noindent where the new parameters are given by:

\begin{subequations}
\label{eq:GRHiperbparam}
\begin{align}
 \overline{n} & = \frac{A^{3/2}}{B} = (2E)^{3/2}\left( 1- \frac{1}{4}(\eta-15)\frac{E}{c^2} \right) \, , \label{eq:GRHiperbparam:n}\\
 e_t^2 & = e^2 = 1-\frac{A\overline{C}}{B^2} = 1 + 2 E L^2 - \nonumber
 \\ & -\left(4(\eta-1) + (7\eta-17) E L^2 - \frac{4 \vec{L} \cdot \vec{S}_\mathrm{eff}}{L^2} \right) \frac{E}{c^2} \, , \label{eq:GRHiperbparam:et2}\\
 e_r^2 & = \left(1 - \frac{AD}{2BL^2} \right)^2 e_t^2 \nonumber \\ 
 &= \left(1 + \left( (3\eta - 8) +\frac{2 \vec{L} \cdot \vec{S}_\mathrm{eff}}{L^2}\right) \frac{E}{c^2} \right)^2 e_t^2  \nonumber \\
 & = 1 + 2 E L^2 + \Bigg(2(\eta-6) + 5(\eta-3) E L^2  \nonumber \\ 
 &  + 8 \vec{L} \cdot \vec{S}_\mathrm{eff} \left(E + \frac{1}{L^2} \right) \Bigg) \frac{E}{c^2} \, , \label{eq:GRHiperbparam:er2}\\
 \overline{a}_r & = \frac{B}{A} +\frac{D}{2 L^2} = \frac{1}{2E} \left( 1 - \left( \frac{1}{2} (\eta - 7) + \frac{2 \vec{L} \cdot \vec{S}_\mathrm{eff}}{L^2}  \right) \frac{E}{c^2} \right) \, . \label{eq:GRHiperbparam:a}
\end{align}
\end{subequations}

And thus we have been able to write the radial motion in a parametric manner only in terms of the constants of motion. Nonetheless, the constants of motion E and L will not be the most convenient to express our results. It will be better to express them in terms of orbital parameters, namely the eccentricity $e_t$ and the mean motion $\overline{\xi}$ and defining a new constant $\Sigma$ to quantify the effect of the spin. That is:
\begin{subequations}
\label{eq:orbitparamindep}
\begin{align}
 \overline{\xi} & = \frac{\overline{n}}{c^3}  = \left(\frac{2E}{c^2}\right)^{3/2}\left( 1- \frac{1}{4}(\eta-15)\frac{E}{c^2} \right) \nonumber 
 \\ & = \left(\frac{2E}{c^2}\right)^{3/2} + O\!\left( \frac{1}{c^2} \right) \, ,  \label{eq:orbitparamindep:xi}\\
 e_t^2-1 &  =  2 E L^2 - \left(4(\eta-1) + (7\eta-17) E L^2 - \frac{4 }{c } \frac{\Sigma}{L} \right) \frac{E}{c^2} \nonumber \\ 
 & = 2 E L^2 + O\!\left( \frac{1}{c^2} \right) \, , \label{eq:orbitparamindep:et}\\
 \Sigma & = c \frac{\vec{L} \cdot \vec{S}_\mathrm{eff}}{L} \, . \label{eq:orbitparamindep:sigma}
\end{align}
\end{subequations}

The remaining orbital parameters can be expressed in terms of these new variables in the following way:
\begin{subequations}
\label{eq:orbitparamdep}
\begin{align}
 e_r & = \left(1 + \overline{\xi}^{2/3} \frac{3 \eta - 8}{2}  + \overline{\xi} \frac{\Sigma}{\sqrt{e_t^2-1}}  \right) e_t \, , \label{eq:orbitparamdep:er}\\
 a_r & = \frac{1}{c^2} \frac{1}{\overline{\xi}^{2/3}} \left(1 - \overline{\xi}^{2/3} \frac{\eta - 9}{3}  - \overline{\xi} \frac{\Sigma}{\sqrt{e_t^2-1}}  \right) \, . \label{eq:orbitparamdep:ar}
\end{align}
\end{subequations}

Substituting this in the expressions of Eq.~\eqref{eq:GRHiperbdef} for the orbit, we finally get:
\begin{subequations}
\label{eq:rrdotxi}
\begin{align}
 r & = \frac{1}{c^2 \overline{\xi}^{2/3}} \Bigg[ e_t \cosh{v} -1 \nonumber \\ 
 & + \overline{\xi}^{2/3} \frac{(7 \eta -6) e_t \cosh{v} + 2 (\eta - 9) }{6} + \overline{\xi} \frac{\Sigma}{\sqrt{e_t^2-1}} \Bigg] \, , \label{eq:rrdotxi:r}\\
 \dot{r} & = c \overline{\xi}^{1/3} \frac{e_t \sinh{v}}{e_t \cosh{v} -1} \left[ 1 + \overline{\xi}^{2/3} \frac{7 \eta -6}{6}\right] \, , \label{eq:rrdotxi:rdot}
\end{align}
\end{subequations}

\noindent where $\dot{r}$ is computed using the fact that $\frac{dr}{dt} = \frac{dr}{dv}/\frac{dt}{dv}$.

\subsubsection{Angular motion}
\label{sec:waveforms:orbit:ang}

If we try to naively solve the equations of motion for the angular part of the two body system using Eq.~\eqref{eq:eqmotion:r}, we will have that:
\begin{subequations}
\label{eq:angeqmotionnaive}
\begin{align}
  r \dot{\theta} & = \{\vec{r},H\} \cdot \hat{e}_\theta = p_\theta + \frac{\hat{e}_\phi \cdot \vec{S}_\mathrm{eff}}{c^2 r^2} \nonumber \\ 
  & + \frac{1}{c^2} \left( \frac{1}{2} (3\eta-1) p^2 -(3+\eta) \frac{1}{r} \right) p_\theta \, , \label{eq:angeqmotionnaive:theta}\\
  r \sin{\theta} \dot{\phi} & = \{\vec{r},H\} \cdot \hat{e}_\phi = p_\phi - \frac{\hat{e}_\theta \cdot \vec{S}_\mathrm{eff}}{c^2 r^2}  \nonumber \\ 
  & + \frac{1}{c^2} \left( \frac{1}{2} (3\eta-1) p^2 -(3+\eta) \frac{1}{r} \right) p_\phi \, . \label{eq:angeqmotionnaive:phi}
\end{align}
\end{subequations}

These equations are not in an optimal form since they explicitly depend on $ p_\theta$ and $p_\phi$ and they are coupled to each other and to the equation of motion for $\vec{S}_\mathrm{eff}$. Part of the difficulty lies in the fact that in our reference frame, the orbital angular momentum $\vec{L} = \vec{r} \times \vec{p}$ precesses, and thus, the plane of the orbit is not constant. This can be solved going to a new non-inertial coordinate system in which $\vec{L} = L \hat{k}$ at all times. One of the unit vectors of the new coordinate system will then be $\hat{k}$. Since $\vec{r}$ is perpendicular to $\hat{k} = \frac{\vec{r} \times \vec{p}}{L}$, then the radial direction $\hat{n} = \vec{r}/r$ can be used as another unit vector of the new basis, which will allow us to use the radial solution found in Sec.~\ref{sec:waveforms:orbit:r}. With these two unit vectors chosen, the new triad is fully specified: 
\begin{equation}
    (\hat{n},\hat{\xi},\hat{k}) \equiv \left(\frac{\vec{r}}{r}, \hat{k} \times \hat{n}, \frac{\vec{L}}{L}\right) \, .
    \label{Lztriad}
\end{equation}

The transformation between our previous triad $(\hat{e}_x,\hat{e}_y,\hat{e}_z)$ and this new one can be written in terms of three Euler angles $\Phi$, $\alpha$ and $\iota$ in the following manner \cite{EulerAngles}:
\begin{subequations}
\label{eq:EulerJtoL}
\begin{align}
 \hat{n} & = (\cos{\alpha}\cos{\Phi}-\cos{\iota}\sin{\alpha}\sin{\Phi}) \hat{e}_x \nonumber \\ 
  & + (\sin{\alpha}\cos{\Phi}+\cos{\iota}\cos{\alpha}\sin{\Phi}) \hat{e}_y \nonumber \\ 
  & + (\sin{\iota}\sin{\Phi}) \hat{e}_z \, , \label{eq:EulerJtoL:n}\\
 \hat{\xi} & = (-\cos{\alpha}\sin{\Phi}-\cos{\iota}\sin{\alpha}\cos{\Phi}) \hat{e}_x \nonumber \\ 
  & + (-\sin{\alpha}\sin{\Phi}+\cos{\iota}\cos{\alpha}\cos{\Phi}) \hat{e}_y \nonumber \\ 
  & + (\sin{\iota}\cos{\Phi}) \hat{e}_z \, , \label{eq:EulerJtoL:xi} \\
 \hat{k} & = \sin{\alpha}\sin{\iota} \hat{e}_x - \cos{\alpha}\sin{\iota} \hat{e}_y + \cos{\iota} \hat{e}_z \, . \label{eq:EulerJtoL:k}
\end{align}
\end{subequations}

In these coordinates we have that the relative position is given by:
\begin{subequations}
\label{eq:rEuler}
\begin{align}
 \vec{r} & = r \hat{n} \, , \label{eq:rEuler:r}\\
 \dot{\vec{r}} & = \dot{r} \hat{n} + r ( \dot{\Phi} +  \dot{\alpha} \cos{\iota} ) \hat{\xi} + r ( \dot{\iota} \sin{\Phi} - \dot{\alpha} \sin{\iota} \cos{\Phi} ) \hat{k} \, , \label{eq:rEuler:rdot} 
\end{align}
\end{subequations}

\noindent where we can see that the radial part is left unchanged and thus the results of Sec.~\ref{sec:waveforms:orbit:r} for $r$ and $\dot{r}$ are still valid. In addition, we can use the equation of motion Eq.~\eqref{eq:eqmotion:r} for $\vec{r}$ to compute:
\begin{align}
    r ( \dot{\iota} \sin{\Phi} - \dot{\alpha} \sin{\iota} \cos{\Phi} ) & = \dot{\vec{r}} \cdot \hat{k} = \{\vec{r},H\} \cdot \frac{\vec{L}}{L} \nonumber \\ 
    & = \frac{\vec{L} \cdot \left(\vec{S}_\mathrm{eff} \times \vec{r} \right)}{c^2 r^3 L} \, ,
    \label{eq:eqmotionk}
\end{align}

\noindent where we have used that $\vec{p} \cdot \vec{L} =\vec{p} \cdot ( \vec{r} \times \vec{p}) = 0$. From Eq.~\eqref{eq:eqmotionk} we can deduce that $\dot{\alpha}$ and $\dot{\iota}$ are both of order $O(1/c^3)$.

The equation of motion in the $\hat{\xi}$ direction is:
\begin{align}
    r ( \dot{\Phi} +  \dot{\alpha} \cos{\iota} ) & = \dot{\vec{r}} \cdot \hat{\xi} = \{\vec{r},H\} \cdot \frac{\vec{L} \times \vec{r}}{L r} \nonumber \\
    & = \vec{p} \cdot \frac{\vec{L} \times \vec{r}}{L r} + O\!\left( \frac{1}{c^2} \right) \, .
    \label{eq:eqmotionxi}
\end{align}

Using that $\vec{p} \cdot (\vec{L} \times \vec{r}) = \vec{L} \cdot (\vec{r} \times \vec{p}) = L^2$ and $\dot{\alpha}$ is $O(1/c^3)$, we can solve Eq.~\eqref{eq:eqmotionxi} for $\dot{\Phi}$:
\begin{equation}
    \dot{\Phi} = \frac{L}{r^2} + O\!\left( \frac{1}{c^2} \right) \, .
    \label{eq:phiO2}
\end{equation}

With all of these results we can compute the modulus squared of $\dot{\vec{r}}$ given in Eq.~\eqref{eq:rEuler:rdot}, ignoring $O(1/c^4)$ terms we have that:
\begin{equation}
    |\dot{\vec{r}}|^2 = \dot{r}^2 + r^2 \dot{\Phi}^2 + 2 L^2 \dot{\alpha} \cos{\iota} \, .
    \label{eq:rdot2Euler}
\end{equation}

The modulus squared of $\dot{\vec{r}}$ can also be computed using the equations of motion (ignoring terms $O(1/c^4)$):
\begin{align}
    |\dot{\vec{r}}|^2 & = \{\vec{r},H\} \cdot \{\vec{r},H\} =  p^2 + 2  \frac{\vec{L} \cdot \vec{S}_\mathrm{eff}}{c^2 r^3} \nonumber \\ 
  & + \frac{2}{c^2} \left(\frac{1}{2} (3\eta-1)(p^2)^2-\left[ (3+\eta) p^2 + \eta (\hat{n} \cdot \vec{p})^2 \right] \frac{1}{r}\right) \, ,
    \label{eq:rdot2motionraw}
\end{align}

\noindent where we have used that $\vec{p} \cdot (\vec{S}_\mathrm{eff} \times \vec{r}) = \vec{S}_\mathrm{eff} \cdot (\vec{r} \times \vec{p}) = \vec{L} \cdot \vec{S}_\mathrm{eff}$.

Substituting the expression for $p^2$ found in Eq.~\eqref{eq:p2ofppr} into the first term of Eq.~\eqref{eq:rdot2motionraw} we have that:
\begin{align}
    |\dot{\vec{r}}|^2  & =  \dot{r}^2 + \frac{L^2}{r^2} + 2  \frac{\vec{L} \cdot \vec{S}_\mathrm{eff}}{c^2 r^3} \nonumber \\ 
  &  + \frac{2}{c^2} \left[\frac{1}{2} (3\eta-1)(p^2)-(3+\eta) \frac{1}{r}\right] (p^2 - p_r^2) \, .
    \label{eq:rdot2motionppr}
\end{align}

And finally using expressions of Eqs.~\eqref{eq:pofprL} and \eqref{eq:pofEN} to express $p$ in the $O(1/c^2)$ term of Eq.~\eqref{eq:rdot2motionppr} in terms of constants of motion we get:
\begin{equation}
    |\dot{\vec{r}}|^2  =  \dot{r}^2 + \frac{L^2}{r^2} + 2  \frac{\vec{L} \cdot \vec{S}_\mathrm{eff}}{c^2 r^3} + \frac{2}{c^2} \left[ (3\eta-1)E+(2\eta-4) \frac{1}{r}\right] \frac{L^2}{r^2} \, .
    \label{eq:rdot2motion}
\end{equation}

Since we have now two expressions for $|\dot{\vec{r}}|^2$ in Eqs.~\eqref{eq:rdot2Euler} and \eqref{eq:rdot2motion}, we can equate them and solve for $\dot{\Phi}$:
\begin{align}
    \dot{\Phi}^2 & = \frac{L^2}{r^4} + 2  \frac{\vec{L} \cdot \vec{S}_\mathrm{eff}}{c^2 r^5} + \frac{2}{c^2} \left[ (3\eta-1)E+(2\eta-4) \frac{1}{r}\right] \frac{L^2}{r^4} \nonumber \\ 
  &  - 2 \frac{L}{r^2} \dot{\alpha} \cos{\iota} \, .
    \label{eq:phidot2}
\end{align}

And taking the square root of this expression ignoring $O(1/c^4)$ terms, we finally get:
\begin{equation}
    \dot{\Phi} = \frac{L}{r^2} + \frac{\hat{k} \cdot \vec{S}_\mathrm{eff}}{c^2 r^3 } + \frac{1}{c^2} \left[ (3\eta-1)E+(2\eta-4) \frac{1}{r}\right] \frac{L}{r^2} - \dot{\alpha} \cos{\iota} \, .
    \label{eq:phidotraw}
\end{equation}

If we rewrite the constants of motion in terms of the parameters of the orbit defined in Eq.~\eqref{eq:orbitparamindep} we get:
\begin{align}
    \dot{\Phi} & = \frac{1}{r^2} \frac{\sqrt{e_t^2-1}}{c \overline{\xi}^{1/3}} \Bigg[ 1 + \overline{\xi}^{2/3} \left( \frac{\eta-1}{e_t^2-1} + \frac{7 \eta -6}{3} + \frac{2(\eta - 2)}{c^2 \overline{\xi}^{2/3} r} \right) \nonumber \\ 
    & - \overline{\xi} \frac{\Sigma}{\sqrt{e_t^2-1}} \left(\frac{1}{e_t^2-1} - \frac{1}{c^2 \overline{\xi}^{2/3} r} \right) \Bigg] - \dot{\alpha} \cos{\iota} \, , 
    \label{eq:phidotofr}
\end{align}

\noindent where we can now substitute our expression of Eq.~\eqref{eq:rrdotxi:r} for $r$ obtained when solving the radial motion:
\begin{align}
    \dot{\Phi} & =  \frac{c^3 \overline{\xi} \sqrt{e_t^2-1}}{(e_t \cosh{v}-1)^2} \Bigg[ 1 - \overline{\xi}^{2/3} \left( \frac{\eta-4}{e_t \cosh{v}-1} - \frac{ \eta -1}{e_t^2-1} \right) \nonumber\\
    & - \overline{\xi} \frac{\Sigma}{\sqrt{e_t^2-1}} \left(\frac{1}{e_t \cosh{v}-1} + \frac{1}{e_t^2-1} \right) \Bigg] - \dot{\alpha} \cos{\iota} \, .
    \label{eq:phidotofv}
\end{align}

An important parameter when talking about hyperbolic encounters is the impact parameter, which is defined as usual:
\begin{equation}
    b = \lim_{r \to \infty} \frac{|\vec{r} \times \dot{\vec{r}}|}{|\dot{\vec{r}}|} \, .
    \label{eq:bdefraw}
\end{equation}

Using the expressions of Eq.~\eqref{eq:rEuler} for $\vec{r}$ and $\dot{\vec{r}}$, the impact parameter takes the following 1.5 PN accurate form:
\begin{equation}
    b = \lim_{r \to \infty} \frac{r^2 (\dot{\Phi} + \dot{\alpha} \cos{\iota})}{\dot{r}} \, .
    \label{eq:bphir}
\end{equation}

Finally, substituting Eq.~\eqref{eq:phidotofr} for $\dot{\Phi}$ and Eq.~\eqref{eq:rrdotxi:rdot} for $\dot{r}$, and taking the limit $r \to \infty$ and consequently $v \to \infty$ we have:
\begin{equation}
    b = \frac{\sqrt{e_t^2-1}}{c^2 \overline{\xi}^{2/3}} \left[  1 + \overline{\xi}^{2/3} \left( \frac{\eta-1}{e_t^2-1} + \frac{7 \eta -6}{6}\right) - \overline{\xi} \frac{\Sigma}{(e_t^2-1)^{3/2}} \right] \, .
    \label{eq:bfinal}
\end{equation}

In addition, to get an idea of the validity of the approximation, it will also be convenient to compute the maximum velocity, which is reached at periastron. Computing this to leading order we obtain:
\begin{align}
    \text{v}_{\text{max}} & = |\dot{\vec{r}}|_{v=0} = \left( \sqrt{r^2 \dot{\Phi}} \right|_{v = 0} + O\!\left( \frac{1}{c^3} \right) = \nonumber \\ 
  &  = \left( \frac{L}{r} \right|_{v = 0} + O\!\left( \frac{1}{c^2} \right) = c \overline{\xi}^{1/3} \sqrt{\frac{e_t + 1}{e_t - 1} } + O\!\left( \frac{1}{c^2} \right) \, .
    \label{eq:vmax}
\end{align}

However, the problem is not complete, to be able to fully characterize the orbit and to solve Eq.~\eqref{eq:phidotofv} for $\dot{\Phi}$, we need to know the evolution of the Euler angles $\alpha$ and $\iota$. To do this we must find how the unit vector $\hat{k} = \vec{L}/L$ given in Eq.~\eqref{eq:EulerJtoL:k} precesses. This can be done using the equation of motion found in Eq.~\eqref{eq:eqmotion:L}:
\begin{equation}
        \dot{\hat{k}} = \frac{1}{L} \{\vec{L},H\} = \frac{1}{c^2 r^3} \vec{S}_\mathrm{eff} \times \hat{k} = \frac{1}{c^2 r^3} (\delta_1 \vec{S}_1 + \delta_2 \vec{S}_2) \times \hat{k} \, .
    \label{eq:keqmotionraw}
\end{equation}

And thus we see that we will need also the evolution of $\vec{S}_1$ and $\vec{S}_2$ given in Eqs.~\eqref{eq:eqmotion:S1} and \eqref{eq:eqmotion:S2} respectively. We can simplify these equations somewhat using that the magnitude of the individual spins $S_1$ and $S_2$ is conserved, and thus we can write:
\begin{subequations}
\label{eq:Sofs}
\begin{align}
 \vec{S}_1 & = S_1 \hat{s}_1 = S_1 (\sin{\theta_1} \cos{\phi_1} \hat{e}_x + \sin{\theta_1} \sin{\phi_1} \hat{e}_y + \cos{\theta_1} \hat{e}_z) \, ,  \label{eq:Sofs:s1}\\
 \vec{S}_2 & = S_2 \hat{s}_2 = S_2 (\sin{\theta_2} \cos{\phi_2} \hat{e}_x + \sin{\theta_2} \sin{\phi_2} \hat{e}_y + \cos{\theta_2} \hat{e}_z) \, . \label{eq:Sofs:s2} 
\end{align}
\end{subequations}
 
And then the equations of motion that we need to solve will be:
\begin{subequations}
\label{eq:angeqmotion}
\begin{align}
  \dot{\hat{k}} & = \frac{1}{L} \{\vec{L},H\} = \frac{1}{c^2 r^3} (\delta_1 S_1 \hat{s}_1 + \delta_2 S_2 \hat{s}_2) \times \hat{k} \, , \label{eq:angeqmotion:k} \\
  \dot{\hat{s}}_1 &  = \frac{1}{S_1} \{\vec{S}_1,H\} = \frac{\delta_1 L}{c^2 r^3} \hat{k} \times \hat{s}_1 \, , \label{eq:angeqmotion:s1} \\
  \dot{\hat{s}}_2 &  = \frac{1}{S_2} \{\vec{S}_2,H\} = \frac{\delta_1 L}{c^2 r^3} \hat{k} \times \hat{s}_2 \, . \label{eq:angeqmotion:s2}
\end{align}
\end{subequations}

Rewriting $L$ in terms of the parameters defined in Eq.~\eqref{eq:orbitparamindep} and $r$ in terms of Eq.~\eqref{eq:rrdotxi:r} we have the following equations we have to solve:
\begin{subequations}
\label{eq:angeqmotionxi}
\begin{align}
  \dot{\hat{k}} &  = \frac{c^4 \overline{\xi}^2}{(e_t \cosh{v}-1)^3} (\delta_1 S_1 \hat{s}_1 + \delta_2 S_2 \hat{s}_2) \times \hat{k}  \, , \label{eq:angeqmotionxi:k} \\
  \dot{\hat{s}}_1 &  = \delta_1 \frac{ c^3 \overline{\xi}^{5/3} \sqrt{e_t^2-1}}{(e_t \cosh{v}-1)^3} \hat{k} \times \hat{s}_1 \, , \label{eq:angeqmotionxi:s1} \\
  \dot{\hat{s}}_2 &   = \delta_2 \frac{ c^3 \overline{\xi}^{5/3} \sqrt{e_t^2-1}}{(e_t \cosh{v}-1)^3} \hat{k} \times \hat{s}_2 \, . \label{eq:angeqmotionxi:s2}
\end{align}
\end{subequations}

These equations will have to be solved numerically and to integrate them it will be necessary to find the initial conditions. Assuming that the initial orientations of the spins are given by $(\theta_1^i, \phi_1^i)$ and $(\theta_2^i, \phi_2^i)$ for $\hat{s}_1$ and $\hat{s}_2$ respectively, we can use the conservation of the total angular momentum $\vec{J} = \vec{L} + \vec{S}_1 + \vec{S}_2 = J \hat{e}_z$ to find the initial conditions for $\hat{k}$:
\begin{subequations}
\label{eq:initialconk}
\begin{align}
  k_x |_{t=t_0} &  = \frac{-S_{1x}|_{t=t_0} - S_{2x} |_{t=t_0} }{L} \, , \label{eq:initialconk:x} \\
  k_y |_{t=t_0} &  = \frac{-S_{1y}|_{t=t_0} - S_{2y} |_{t=t_0} }{L} \, .  \label{eq:initialconk:y} 
\end{align}
\end{subequations}

Using the expressions of Eq.~\eqref{eq:Sofs} for $\vec{S}_1$ and $\vec{S}_2$ and expressing $L$ in terms of the parameters of Eq.~\eqref{eq:orbitparamindep}, we can rewrite Eq.~\eqref{eq:initialconk} in the following way:
\begin{subequations}
\label{eq:initialconkxi}
\begin{align}
  k_x |_{t=t_0} &  = - c \overline{\xi}^{1/3} \frac{ S_1 \sin{\theta_1^i} \cos{\phi_1^i} + S_2 \sin{\theta_2^i} \cos{\phi_2^i} }{\sqrt{e_t^2-1}} \, , \label{eq:initialconkxi:x} \\
  k_y |_{t=t_0} &  = - c \overline{\xi}^{1/3} \frac{ S_1 \sin{\theta_1^i} \sin{\phi_1^i} + S_2 \sin{\theta_2^i} \sin{\phi_2^i} }{\sqrt{e_t^2-1}} \, , \label{eq:initialconkxi:y} 
\end{align}
\end{subequations}

\noindent where $k_z$ is fully specified by the fact that $\hat{k}$ is a unit vector ($|\hat{k}| = 1$).

Using these initial conditions to integrate the equations of motion we can extract the values of $\alpha$, $\dot{\alpha}$ and $\iota$ from Eq.~\eqref{eq:EulerJtoL:k}, obtaining:
\begin{subequations}
\label{eq:anglesofk}
\begin{align}
    \alpha & = -\arctan \! \left( \frac{k_x}{k_y} \right) \, , \label{eq:anglesofk:a} \\
    \iota & = \arccos{k_z} \, , \label{eq:anglesofk:i} \\
    \dot{\alpha} & = \frac{k_x \dot{k}_y - \dot{k}_x k_y}{k_x^2+k_y^2} \, . \label{eq:anglesofk:da}
\end{align}
\end{subequations}

The values of these angles can be plugged into Eq.~\eqref{eq:phidotofv} for $\dot{\Phi}$ and solve this equation numerically to get an expression for $\Phi$ using the initial condition of $\Phi |_{t = t_0} = \Phi_0$. This together with the expressions in Eq.~\eqref{eq:rrdotxi} for the radial part fully specifies the orbit.

\subsection{Gravitational wave emission}
\label{sec:waveforms:GW}

Now that we have obtained a method to compute the orbit followed by the black holes, we will want to determine the gravitational waves emitted by the system. In this section we will first derive the leading order gravitational wave emission to get a better grasp of the physical process as well as to introduce the variables and degrees of freedom involved. This leading order computation will also allow us to understand how to derive the more cumbersome expressions for the gravitational wave emission that take into account up to spin-orbit corrections. The full step by step derivation of these higher order expressions is out of the scope of this project and they will be taken from the literature, since they are the same formulas used in the compact binary coalescence templates \cite{hphc_15PN}, which are very well documented.

To compute the leading order gravitational wave emission we start from the quadrupolar formula derived by Einstein in 1918:

\begin{equation}
    \overline{h}_{ij} = \frac{2 G}{c^4 R} \Ddot{Q}_{ij} (t_r) \, ,
    \label{eq:QuadrupEinst}
\end{equation}

\noindent where $R$ is the distance from the encounter to our detector, $t_r$ is the retarded time $(t_r = t - R/c)$ and $Q_{ij}$ are the components of the quadrupole moment defined as:

\begin{equation}
    Q_{ij} (t_r) = \int d^3 \vec{x} \, x_i x_j T^{00}(t_r,\vec{x}) \, .
    \label{eq:QuadrupMomentdef}
\end{equation}

The 0-component of the stress energy tensor $T^{00}(t_r,\vec{x})$ is given in our system by the following Newtonian accurate expression:

\begin{equation}
    T^{00}(t_r,\vec{x}) = \mu \, \delta^{(3)} (\vec{x}-\vec{r}(t)) \, ,
    \label{eq:T00}
\end{equation}

\noindent where $\vec{r}(t)$ is the solution for the orbit found in the previous section. Eq.~\eqref{eq:T00} can be substituted in Eq.~\eqref{eq:QuadrupMomentdef} obtaining:

\begin{equation}
    Q_{ij} (t_r) = \mu \, r_i r_j \, .
    \label{eq:Qmurirj}
\end{equation}

To compute the gravitational waves using Eq.~\eqref{eq:QuadrupEinst} we have to differentiate $Q_{ij}$ twice with respect to time:

\begin{equation}
    \Ddot{Q}_{ij} (t_r) = \mu (2 \dot{r}_i \dot{r}_j + \Ddot{r}_i r_j + r_i \Ddot{r}_j ) \, ,
    \label{eq:ddotQmurirj}
\end{equation}

\noindent where we can introduce the Newtonian accurate expression for $\Ddot{r}_i$:

\begin{equation}
    \Ddot{r}_i = -\frac{r_i}{r^3} \, .
    \label{eq:ddotriN}
\end{equation}

Substituting this in Eq.~\eqref{eq:ddotQmurirj} we obtain that the second derivative of the quadrupole moment of the system is:

\begin{equation}
    \Ddot{Q}_{ij} (t_r) = 2 \mu \left(\dot{r}_i \dot{r}_j - \frac{r_i r_j}{r^3}\right) \, .
    \label{eq:ddotQfin}
\end{equation}

Which when substituted in Einstein's quadrupolar formula of Eq.~\eqref{eq:QuadrupEinst} gives:

\begin{equation}
    \overline{h}_{ij} = \frac{4 G \mu}{c^4 R} \left(\dot{r}_i \dot{r}_j - \frac{r_i r_j}{r^3}\right) \, .
    \label{eq:QuadrupSolved}
\end{equation}

For the gravitational waves we will be interested in the transverse traceless part of this strain tensor. To compute it we will use the transverse traceless projection operator ${P}_{ijkm}(\hat{N})$ which projects vectors into the plane orthogonal to $\hat{N}$, where $\hat{N}$ is the unit vector pointing from the encounter to the detector. This projector is defined as:

\begin{equation}
    \mathcal{P}_{ijkm}(\hat{N}) =  \mathcal{P}_{ik} \mathcal{P}_{jm} - \frac{1}{2} \mathcal{P}_{ij} \mathcal{P}_{km} \, ,
    \label{eq:TTPdef}
\end{equation}

\noindent where $\mathcal{P}_{ij} = \delta_{ij} - N_i N_j$. And thus, using this projector we have that:

\begin{equation}
    h_{ij}^{TT} = \frac{4 G \mu}{c^4 R} \mathcal{P}_{ijkm}(\hat{N}) \left( \dot{r}_k \dot{r}_m - \frac{r_k r_m}{r^3} \right) \, ,
    \label{eq:hijTT}
\end{equation}

\noindent where we have used that $h_{ij}^{TT} = \overline{h}_{ij}^{TT}$. To write down the two polarizations of the gravitational waves $h_+$ and $h_{\times}$, it will be convenient to define a new triad:

\begin{equation}
    (\hat{p},\hat{q},\hat{N}) = \left(\frac{\hat{N} \times \hat{e}_z}{|\hat{N} \times \hat{e}_z|}, \frac{\hat{N} \times \hat{p}}{|\hat{N} \times \hat{p}|}, \hat{N}\right) \, ,
    \label{eq:detectortriad}
\end{equation}

\noindent where $\hat{e}_z$ is the unit vector pointing in the direction of the conserved total angular momentum $\vec{J}$ introduced in Eq.~\eqref{eq:Jzconst}. In this new triad, the two polarizations will be given by:

\begin{subequations}
\label{eq:hphxofhij}
\begin{align}
    h_+ & = \frac{1}{2}(p_i p_j - q_i q_j) h_{ij}^{TT} \, , \label{eq:hphxofhij:hp} \\ 
    h_{\times} & = \frac{1}{2}(p_i q_j + p_j q_i) h_{ij}^{TT} \, . \label{eq:hphxofhij:hx}
\end{align}
\end{subequations}

Substituting here Eq.~\eqref{eq:hijTT} for $h_{ij}^{TT}$ and expressing it in a vector form, we obtain the following expressions for the leading order gravitational waves emitted:

\begin{subequations}
\label{eq:hphxvec}
\begin{align}
    h_+ & =  \frac{2 G \mu}{c^4 R} \left[(\hat{p} \cdot \dot{\vec{r}})^2 - (\hat{q} \cdot \dot{\vec{r}})^2 - \frac{1}{r} \left[ (\hat{p} \cdot \hat{n})^2 - (\hat{q} \cdot \hat{n})^2 \right] \right] \, , \label{eq:hphxvec:hp} \\ 
    h_{\times} & = \frac{4 G \mu}{c^4 R} \left[(\hat{p} \cdot \dot{\vec{r}})(\hat{q} \cdot \dot{\vec{r}}) - \frac{1}{r} (\hat{p} \cdot \hat{n})(\hat{q} \cdot \hat{n})  \right] \, . \label{eq:hphxvec:hx}
\end{align}
\end{subequations}

To find expressions for $\hat{p}$ and $\hat{q}$ in the frame in which the orbits were solved, we note that if we define $\Theta$ as the angle formed between $\hat{N}$ and $\hat{e}_z$, then invoking the arbitrariness in the orientation of the triad $(e_x,e_y,e_z)$, we can always write:

\begin{equation}
    \hat{N} = \sin{\Theta} \hat{e}_x + \cos{\Theta} \hat{e}_z \, .
    \label{eq:Nexez}
\end{equation}

From here, using Eq.~\eqref{eq:detectortriad}, $\hat{p}$ and $\hat{q}$ can be obtained by multiplication:
\begin{subequations}
\label{eq:pqexeyez}
\begin{align}
    \hat{p} & = -e_y \, , \label{eq:pqexeyez:p}\\
    \hat{q} & = \cos{\Theta} \hat{e}_x - \sin{\Theta} \hat{e}_z \, . \label{eq:pqexeyez:q}
\end{align}
\end{subequations}

And using Eq.~\eqref{eq:EulerJtoL}, the new frame $(\hat{p},\hat{q},\hat{N})$ can be written in terms of the non inertial frame $(\hat{n},\hat{\xi},\hat{k})$ in which the equations of motion were solved:
\begin{subequations}
\label{eq:pqNofnxik}
\begin{align}
    \hat{p} = & (-\sin{\alpha} \cos{\Phi}- \cos{\iota} \cos{\alpha} \sin{\Phi}) \hat{n} \nonumber\\ 
    & + (\sin{\alpha} \sin{\Phi}- \cos{\iota} \cos{\alpha} \cos{\Phi}) \hat{\xi} \nonumber\\ 
    & + \cos{\alpha} \sin{\iota} \hat{k} \, , \label{eq:pqNofnxik:p}\\
    \hat{q} = & (\cos{\alpha} \cos{\Phi} \cos{\Theta}  - \cos{\iota} \sin{\alpha} \sin{\Phi} \cos{\Theta} - \nonumber \\
    &\sin{\iota} \sin{\Phi} \sin{\Theta}) \hat{n} - (\cos{\alpha} \sin{\Phi} \cos{\Theta} 
    \nonumber \\
    & + \cos{\iota} \sin{\alpha} \cos{\Phi} \cos{\Theta} + \sin{\iota} \cos{\Phi} \sin{\Theta}) \hat{\xi}  \nonumber \\
    &+ (\sin{\iota} \sin{\alpha}\cos{\Theta} - \cos{\iota} \sin{\Theta}) \hat{k} \, , \label{eq:pqNofnxik:q}\\
    \hat{N} = & (\cos{\alpha} \cos{\Phi} \sin{\Theta} - \cos{\iota} \sin{\alpha} \sin{\Phi} \sin{\Theta} \nonumber \\
    &+ \sin{\iota} \sin{\Phi} \cos{\Theta}) \hat{n}  - (\cos{\alpha} \sin{\Phi} \sin{\Theta} \nonumber \\
    &+ \cos{\iota} \sin{\alpha} \cos{\Phi} \sin{\Theta} - \sin{\iota} \cos{\Phi} \cos{\Theta}) \hat{\xi} \nonumber \\
    & + (\sin{\iota} \sin{\alpha}\sin{\Theta} + \cos{\iota} \cos{\Theta}) \hat{k} \, . \label{eq:pqNofnxik:N}
\end{align}
\end{subequations}
The gravitational waves of Eq.~\eqref{eq:hphxvec} emitted by the system will have a 2.5 PN order dissipating effect on the orbit that can be modeled using the following coupled differential equations \cite{Dissipation, De_Vittori_2014}:

\begin{subequations}
\label{eq:dndtdedt}
\begin{align}
    \frac{d \overline{\xi}}{dt} & = - c^3 \frac{\overline{\xi}^{11/3} 8 \eta}{5 \beta^7} \big[-49 \beta^2-32 \beta^3 + 35 (e_t^2-1) \beta \nonumber\\ 
    & - 6 \beta^4 + 9 e_t^2 \beta^2 \big] \, , \label{eq:dndtdedt:dndt}\\
     \frac{d e_t}{dt} & = - c^3 \frac{\overline{\xi}^{8/3} 8 \eta (e_t^2-1)}{15 \beta^7 e_t} \big[-49 \beta^2-17 \beta^3 \nonumber\\ 
    & + 35 (e_t^2-1) \beta - 3 \beta^4 + 9 e_t^2 \beta^2 \big] \, , \label{eq:dndtdedt:dedt}
\end{align}
\end{subequations}

\noindent where $\beta = e_t \cosh{v} - 1$.

The formulas for the gravitational wave strain $h_+$, $h_\times$ of Eq.~\eqref{eq:hphxvec} only take into account the leading order post-newtonian contribution, but since we have computed the orbit to order 1.5 PN, we can use this orbit to compute gravitational waves with higher order contributions. Taking into account up to leading order spin-orbit contributions to the gravitational wave emission follows a similar procedure to the one used to arrive at the leading order expression of Eq.~\eqref{eq:hphxvec}, but taking into account higher order corrections of the quadrupole moment. These formulas are also used in the computation of gravitational waves from compact binary coalescence and the derivation of the contributions up to spin-orbit corrections can be seen in the literature \cite{derivation_hphc_15PN}. The result one obtains is:

\begin{widetext}
\begin{subequations}
\label{eq:hphc_15PN}
\begin{align}
h_{+} &=  2\, \frac {G\, \mu  } {c^4\, R } \,  \bigg\{  \bigg[ \left( {(\hat{q} \cdot \hat{n} )}^{2}-{(\hat{p} \cdot \hat{n} )}^{2} \right) \,z +
	{(\hat{p} \cdot \dot{\vec{r}} )}^{2}- {(\hat{q} \cdot \dot{\vec{r}} )}^{2}  \bigg] -
	\frac{\delta}{2\,c }\,  \bigg[  (  (\hat{N} \cdot \hat{n} )\, {\dot r} -(\hat{N} \cdot \dot{\vec{r}} )\,  )\, z\, {(\hat{p} \cdot \hat{n} )}^{2} \nonumber \\
&-6\,z\,(\hat{N} \cdot \hat{n} )\, (\hat{p} \cdot \hat{n} )\,(\hat{p} \cdot \dot{\vec{r}} )+ \left( -3\,(\hat{N} \cdot \hat{n} )\, {\dot r} +(\hat{N} \cdot \dot{\vec{r}} )\, \right) \,z\, {(\hat{q} \cdot \hat{n} )}^{ 2}
	+6\,z\,(\hat{N} \cdot \hat{n} )\, (\hat{q} \cdot \hat{n} )\,(\hat{q} \cdot \dot{\vec{r}} ) \nonumber \\
&+ 2 \left( {(\hat{p} \cdot \dot{\vec{r}} )}^{ 2}-\,{(\hat{q} \cdot \dot{\vec{r}} )}^{2} \right) (\hat{N} \cdot \dot{\vec{r}} )  \bigg]
	+ \frac{1}{6\,c^2}\,  \bigg[ 6\,{(\hat{N} \cdot \dot{\vec{r}} )}^{2}\,  (  (\hat{p} \cdot \dot{\vec{r}} )^2 - (\hat{q} \cdot \dot{\vec{r}} )^2   )
	\left( 1-3\,\eta \right) +  (  \left[ 6\,\eta-2 \right] {(\hat{N} \cdot \dot{\vec{r}} )}^{2}{(\hat{p} \cdot \hat{n} )}^{2} \nonumber \\
&+ \left( 96\,\eta-32 \right) \, (\hat{N} \cdot \dot{\vec{r}} ) \,(\hat{N} \cdot \hat{n} )\,(\hat{p} \cdot \dot{\vec{r}} )\,(\hat{p} \cdot \hat{n} )+ \left( -6\,\eta+2 \right) 
	\, (\hat{N} \cdot \dot{\vec{r}} )^2\, {(\hat{q} \cdot \hat{n} )}^{2} + \left( -96\,\eta+32 \right) (\hat{N} \cdot \dot{\vec{r}} )\, (\hat{N} \cdot \hat{n} ) \nonumber \\
&\times (\hat{q} \cdot \dot{\vec{r}} )\,(\hat{q} \cdot \hat{n} ) +  [   \left( -14+42\,\eta \right) (\hat{N} \cdot \hat{n} )^2 - 4+6\,\eta  ]  {(\hat{p} \cdot \dot{\vec{r}} )}^{2}+
	 [   \left( -42\, \eta+14 \right) {(\hat{N} \cdot \hat{n} )}^{2}+4-6\,\eta  ]  {(\hat{q} \cdot \dot{\vec{r}} )}^{2}  ) \, z \nonumber \\
&+  \left(  \left( -9\,\eta+3 \right) {(\hat{p} \cdot \dot{\vec{r}} )}^{2} + \left( - 3+9\,\eta \right) {(\hat{q} \cdot \dot{\vec{r}} )}^{2}  \right) 
	\,v^2 +  (   [  29+ \left( 7-21\,\eta \right) {(\hat{N} \cdot \hat{n} )}^{2}  ]  {(\hat{p} \cdot \hat{n} )}^{2} +  [ -29+ \left( 21\,\eta-7 \right) \nonumber \\
&\times {(\hat{N} \cdot \hat{n} )}^{2}  ] \, {(\hat{q} \cdot \hat{n} )}^{2}  ) {z}^{2} +  (  (  \left( -9\,\eta+3 \right) {(\hat{N} \cdot \hat{n} )}^{2}-10-3\,\eta )
	{(\hat{p} \cdot \hat{n} )}^{2}+ \left(  \left( -3+9\,\eta \right) {(\hat{N} \cdot \hat{n} )}^{2}+10+3\,\eta \right) \nonumber \\
&\times {(\hat{q} \cdot \hat{n} )}^{2} )\, z\, v^2 +  (  \left( -36\,\eta+12 \right) (\hat{N} \cdot \dot{\vec{r}} )\,(\hat{N} \cdot \hat{n} )\,{(\hat{p} \cdot \hat{n} )}^{2}+
	\left(  \left( -90\,\eta+30 \right) {(\hat{N} \cdot \hat{n} )}^{2}+20+12\,\eta \right) (\hat{p} \cdot \dot{\vec{r}} )\,(\hat{p} \cdot \hat{n} ) \nonumber \\
&+ \left( -12+36\,\eta \right) (\hat{N} \cdot \dot{\vec{r}} )\, (\hat{N} \cdot \hat{n} ) \,{(\hat{q} \cdot \hat{n} )}^{2}+ \left(  \left( 90\,\eta-30 \right) {(\hat{N} \cdot \hat{n} )}^{2}
	-12\,\eta-20 \right) (\hat{q} \cdot \dot{\vec{r}} )\,(\hat{q} \cdot \hat{n} )  ) \, z\, {\dot r} \nonumber \\
&+ \left(   [  \left( 45\,\eta-15 \right) {(\hat{N} \cdot \hat{n} )}^{2}-9\,\eta+3  ] \, {(\hat{p} \cdot \hat{n} )}^{2}+ \left(  \left( 15-45\,\eta \right) {(\hat{N} \cdot \hat{n} )}^{2}-3+9\,
	\eta \right) {(\hat{q} \cdot \hat{n} )}^{2}  \right)\,  \bigg] + \frac{ z^2  }{ c^2} \,  \bigg[ (\hat{p} \cdot \hat{n} )\nonumber \\
& \times  (X_2 \chi_2 \left [ \hat{p} \cdot \left ( {\hat{s}_2} \times  \hat{N} \right ) \right ] - X_1\chi_1\left [ 
	\hat{p} \cdot \left ( {\hat{s}_1} \times  \hat{N} \right ) \right ]  ) + (\hat{q} \cdot \hat{n} )  ( X_1 \chi_1 \left [
	\hat{q} \cdot \left ( {\hat{s}_1} \times  \hat{N} \right ) \right ] - X_2 \chi_2 \left [ \hat{q} \cdot \left ( {\hat{s}_2} \times  \hat{N} \right ) \right ]  )  \bigg] \bigg\}~, \label{eq:hphc_15PN:hp} \\
h_{\times} &=4\, { \frac {G\, \mu  }{ {c^4\, R } } }\,  \bigg\{  \bigg[ -(\hat{p} \cdot \hat{n} )\,(\hat{q} \cdot \hat{n} )\,z +(\hat{p} \cdot \dot{\vec{r}} )\,(\hat{q} \cdot \dot{\vec{r}} )  \bigg] 
	- \frac{\delta}{c} \, \bigg[  \bigg( \big\{  [  3\,(\hat{N} \cdot \hat{n} )\,{\dot r} -(\hat{N} \cdot \dot{\vec{r}}) ]  (\hat{q} \cdot \hat{n} )
	-3\,(\hat{N} \cdot \hat{n} ) \,(\hat{q} \cdot \dot{\vec{r}} ) \big\} \nonumber \\
&\times (\hat{p} \cdot \hat{n} ) -3\,(\hat{N} \cdot \hat{n} )\,(\hat{q} \cdot \hat{n} )\,(\hat{p} \cdot \dot{\vec{r}} ) \bigg)\, z +2\,(\hat{p} \cdot \dot{\vec{r}} )\,(\hat{q} \cdot \dot{\vec{r}} )\,(\hat{N} \cdot \dot{\vec{r}} )
	\bigg] + \frac{1}{6\, c^2}\,  \bigg[ 6\,  \left( 1-3\,\eta \right) {(\hat{N} \cdot \dot{\vec{r}} )}^{2}\, (\hat{p} \cdot \dot{\vec{r}})\,(\hat{q} \cdot \dot{\vec{r}} ) \nonumber \\
&+  (    [   \left( 6\,\eta-2 \right) (\hat{N} \cdot \dot{\vec{r}} )^{2}\, (\hat{q} \cdot \hat{n} )+ \left( 48\,\eta-16 \right)
	(\hat{N} \cdot \dot{\vec{r}} )\,(\hat{N} \cdot \hat{n} )\,(\hat{q} \cdot \dot{\vec{r}} )  ]  (\hat{p} \cdot \hat{n} ) + \left( 48\,\eta-16 \right) (\hat{N} \cdot \dot{\vec{r}} )\,(\hat{N} \cdot \hat{n} )\,
	(\hat{p} \cdot \dot{\vec{r}} )\,(\hat{q} \cdot \hat{n} ) \nonumber \\
&+ \left(  \left( -14+42\,\eta \right) {(\hat{N} \cdot \hat{n} )}^{2} -4+6\,\eta \right) (\hat{q} \cdot \dot{\vec{r}} )\,(\hat{p} \cdot \dot{\vec{r}} )  ) z
	+  ( -9\,\eta+3  ) (\hat{q} \cdot \dot{\vec{r}} )\,(\hat{p} \cdot \dot{\vec{r}} )\,v^2+  ( 29+ \left( 7-21\,\eta \right) {(\hat{N} \cdot \hat{n} )}^{2}  ) \nonumber \\
&\times (\hat{q} \cdot \hat{n} )\,(\hat{p} \cdot \hat{n} )\,{z}^{2}+  (  \left( -9\,\eta+3 \right) {(\hat{N} \cdot \hat{n} )}^{2}-10-3\,\eta  ) (\hat{q} \cdot \hat{n} )\,(\hat{p} \cdot \hat{n} )\,z\, v^2+
	 (  [   \left( -36\,\eta+12 \right) (\hat{N} \cdot \dot{\vec{r}} )\,(\hat{N} \cdot \hat{n} ) \,(\hat{q} \cdot \hat{n} ) \nonumber \\
&+ \left(  \left( 15-45\,\eta \right) {(\hat{N} \cdot \hat{n} )}^{2}+10+6\,\eta  \right) (\hat{q} \cdot \dot{\vec{r}} )  ] (\hat{p} \cdot \hat{n} ) +  [  \left( 15-45\,\eta
	\right) {(\hat{N} \cdot \hat{n} )}^{2}+10+6\,\eta  ] (\hat{p} \cdot \dot{\vec{r}} )\,(\hat{q} \cdot \hat{n} )  ) \dot r\,z +  (  \left( 45\,\eta-15 \right) \nonumber \\
&\times {(\hat{N} \cdot \hat{n} )}^{2}-9\,\eta+3  ) (\hat{q} \cdot \hat{n} )\,(\hat{p} \cdot \hat{n} )\,{\dot r}^{2}\,z  \bigg]  
	+ \frac{ z^2 }{ c^2} \, (\hat{q} \cdot \hat{n} )  \bigg[ X_2 \chi_2  ( \hat{p} \cdot \left ( {\hat{s}_2} \times  \hat{N} \right )  )
      - X_1 \chi_1 ( \hat{p} \cdot \left ( {\hat{s}_1} \times  \hat{N} \right )  )  \bigg]  \bigg\}~. \label{eq:hphc_15PN:hc}
\end{align}
\end{subequations}
\end{widetext}

\noindent where $\delta = |m_1 - m_2|/m$, $z = 1/r$, $X_1 = m_1/m$ and $X_2 = m_1/m$. It can be seen that the first terms of $h_+$ and $h_\times$ in Eq.~\eqref{eq:hphc_15PN} coincide with the leading order expressions derived in Eq.~\eqref{eq:hphxvec}. Subsequent terms correspond to the higher order contributions and will in general be smaller.

\subsection{Numerical implementation}
\label{sec:waveforms:num}

In this section we will gather all the formulas that we have been discussing, to show how the templates are actually computed. For the numerical implementation we create a \texttt{Python} code that takes the following input:

\begin{equation}
    \text{Input: } m_1, m_2, \overline{\xi}_0, e_{t0}, \Phi_0, \Theta, \chi_1, \theta_1^i, \phi_1^i, \chi_2,  \theta_2^i, \phi_2^i, t_0, t_f \, ,
    \label{eq:Imput}
\end{equation}

\noindent where all parameters have been discussed in previous sections except $t_0$ and $t_f$ which are the initial and final times of the simulation, and $\chi_1$ and $\chi_2$ which are the Kerr parameters of the two black holes that take values between 0 and 1 and measure the magnitude of the component spins. They are defined in the following way:

\begin{equation}
    \vec{\mathcal{S}}_i = G m \mu \vec{S}_i = \frac{G m_i^2 \chi_i}{c} \hat{s}_i \;\; \rightarrow \;\; S_1 = \frac{1}{c} \frac{m_1}{m_2}\chi_1 \, , \;\; S_2 = \frac{1}{c} \frac{m_2}{m_1}\chi_2 \, .
    \label{eq:KerrParam}
\end{equation}

Using this input, the problem is completely specified and the linear system of differential equations that has to be solved is:

\begin{subequations}
\label{eq:dy}
\begin{align}
    \frac{d \overline{\xi}}{dt}  = & - c^3 \frac{\overline{\xi}^{11/3} 8 \eta}{5 \beta^7} \big[-49 \beta^2-32 \beta^3 + 35 (e_t^2-1) \beta \nonumber\\
    & - 6 \beta^4 + 9 e_t^2 \beta^2 \big] \, , \label{eq:dy:0}\\
    \frac{d e_t}{dt}  = & - c^3 \frac{\overline{\xi}^{8/3} 8 \eta (e_t^2-1)}{15 \beta^7 e_t} \big[-49 \beta^2-17 \beta^3 \nonumber\\
    & + 35 (e_t^2-1) \beta - 3 \beta^4 + 9 e_t^2 \beta^2 \big] \, , \label{eq:dy:1} \\
    \frac{d \Phi}{dt} = &  \frac{c^3 \overline{\xi} \sqrt{e_t^2-1}}{(e_t \cosh{v}-1)^2} \Bigg[  1 - \overline{\xi}^{2/3} \left( \frac{\eta-4}{e_t \cosh{v}-1} - \frac{ \eta -1}{e_t^2-1} \right) \nonumber\\
    & - \overline{\xi} \frac{\Sigma}{\sqrt{e_t^2-1}} \left(\frac{1}{e_t \cosh{v}-1} + \frac{1}{e_t^2-1} \right) \Bigg] - \dot{\alpha} \cos{\iota} \, ,
    \label{eq:dy:2}\\
    \frac{d\hat{s}_1}{dt} = & \delta_1 \frac{ c^3 \overline{\xi}^{5/3} \sqrt{e_t^2-1}}{(e_t \cosh{v}-1)^3} \hat{k} \times \hat{s}_1 \, , \label{eq:dy:678} \\
    \frac{d \hat{s}_2}{dt} = & \delta_2 \frac{ c^3 \overline{\xi}^{5/3} \sqrt{e_t^2-1}}{(e_t \cosh{v}-1)^3} \hat{k} \times \hat{s}_2 \, , \label{eq:dy:91011} \\
    \frac{d \hat{k}}{dt}  = & \frac{c^4 \overline{\xi}^2}{(e_t \cosh{v}-1)^3} (\delta_1 S_1 \hat{s}_1 + \delta_2 S_2 \hat{s}_2) \times \hat{k} \, . \label{eq:dy:345}
\end{align}
\end{subequations}

To be able to evaluate the derivatives in Eq.~\eqref{eq:dy}, we need to compute the values of the parameter $v$ at each time $t$. This is done solving the following transcendental equation:

\begin{equation}
    c^3 \overline{\xi} t = e_t \sinh{v} - v \, .
    \label{eq:Eulereq}
\end{equation}

To obtain an accurate solution to this transcendental equation in an efficient way we use Mikkola's method \cite{Mikkola}, which was specifically designed to solve this common type of equation in orbital dynamics. Note that writing the relation between $v$ and $t$ in this manner fixes the periastron time at $t=0$. The values of $\alpha$, $\dot{\alpha}$, $\iota$ and $\dot{\iota}$ appearing in Eq.~\eqref{eq:dy} and that are needed to fully characterize the orbit can be derived using the following formulas:

\begin{subequations}
\label{eq:adaidi}
\begin{align}
    \alpha & = -\arctan \! \left( \frac{k_x}{k_y} \right) \, , \label{eq:adaidi:a} \\ 
    \dot{\alpha} & = \frac{k_x \dot{k}_y - \dot{k}_x k_y}{k_x^2+k_y^2} \, , \label{eq:adaidi:da} \\
    \iota & = \arccos{k_z} \, , \label{eq:adaidi:i} \\
    \dot{\iota} & = -\frac{\dot{k}_z}{\sqrt{1 - k_z^2}} \, . \label{eq:adaidi:di}
\end{align}
\end{subequations}

To integrate the system of differential equations of Eq.~\eqref{eq:dy} we will also need the initial conditions, which will be given by:

\begin{subequations}
\label{eq:y0}
\begin{align}
    \overline{\xi}(t = t_0) &  = \overline{\xi}_0 \, , \label{eq:y0:0}\\
    e_t (t = t_0) & = e_{t0} \, , \label{eq:y0:1} \\
    \Phi (t = t_0) & = \Phi_0 \, , \label{eq:y0:2}\\
    \hat{s}_1 (t = t_0) &  = (\sin{\theta_1^i} \cos{\phi_1^i}, \sin{\theta_1^i} \sin{\phi_1^i},\cos{\theta_1^i}) \, , \label{eq:y0:678} \\
    \hat{s}_2 (t = t_0) & = (\sin{\theta_2^i} \cos{\phi_2^i}, \sin{\theta_2^i} \sin{\phi_2^i},\cos{\theta_2^i}) \, , \label{eq:y0:91011}\\
    \hat{k} (t = t_0) & = \Bigg(- c \overline{\xi}^{1/3}_0 \frac{ S_1 s_{1x}|_{t_0} + S_2 s_{2x}|_{t_0} }{\sqrt{e_{t0}^2-1}},  \nonumber\\
    & - c \overline{\xi}^{1/3}_0 \frac{ S_1 s_{1y}|_{t_0} + S_2 s_{2y}|_{t_0}  }{\sqrt{e_{t0}^2-1}}, \sqrt{1-k_x^2|_{t_0}-k_y^2|_{t_0}} \Bigg) \, . \label{eq:y0:345} 
\end{align}
\end{subequations}

From these initial conditions the value of the constant of motion $\Sigma$ is also deduced:

\begin{equation}
    \Sigma = c \; \hat{k}|_{t_0} \cdot (\delta_1 S_1 \hat{s}_1|_{t_0} + \delta_2 S_2 \hat{s}_2|_{t_0}) \, .
    \label{eq:sigmainit}
\end{equation}

With the initial conditions and all the variables appearing in the differential equations of Eq.~\eqref{eq:dy} now specified, we can numerically integrate them. To do this in an efficient and fast way, we use an explicit Runge-Kutta method of order 5(4) \cite{RungeKutta45}, where the error is controlled assuming accuracy of the fourth-order method, but steps are taken using the fifth-order accurate formula.

Having the solution for the differential equations of Eq.~\eqref{eq:dy}, we can compute $\vec{r}$, $\dot{\vec{r}}$, $\hat{p}$, $\hat{q}$ and $\hat{N}$ given by:

\begin{subequations}
\label{eq:rdrpq}
\begin{align}
    \vec{r} = & \left(r,0,0\right) \, , \label{eq:rdrpq:r}\\
    \dot{\vec{r}} = & \left(\dot{r}, r ( \dot{\Phi} +  \dot{\alpha} \cos{\iota} ), r ( \dot{\iota} \sin{\Phi} - \dot{\alpha} \sin{\iota} \cos{\Phi} )\right) \, , \label{eq:rdrpq:dr} \\
    \hat{p} = & \big(-\sin{\alpha} \cos{\Phi}- \cos{\iota} \cos{\alpha} \sin{\Phi},  \nonumber\\
    & \sin{\alpha} \sin{\Phi}- \cos{\iota} \cos{\alpha} \cos{\Phi}, \cos{\alpha} \sin{\iota} \big) \, , \label{eq:rdrpq:p}\\
    \hat{q} = & \big(\cos{\alpha} \cos{\Phi} \cos{\Theta} - \cos{\iota} \sin{\alpha} \sin{\Phi} \cos{\Theta} \nonumber \\
    &- \sin{\iota} \sin{\Phi} \sin{\Theta},  - \cos{\alpha} \sin{\Phi} \cos{\Theta} \nonumber \\
    &- \cos{\iota} \sin{\alpha} \cos{\Phi} \cos{\Theta} - \sin{\iota} \cos{\Phi} \sin{\Theta}, \nonumber \\
    & \sin{\iota} \sin{\alpha}\cos{\Theta} - \cos{\iota} \sin{\Theta}\big) \, , \label{eq:rdrpq:q} \\
    \hat{N} = & \big(\cos{\alpha} \cos{\Phi} \sin{\Theta} - \cos{\iota} \sin{\alpha} \sin{\Phi} \sin{\Theta} \nonumber \\
    &+ \sin{\iota} \sin{\Phi} \cos{\Theta}, - (\cos{\alpha} \sin{\Phi} \sin{\Theta} \nonumber \\
    &+ \cos{\iota} \sin{\alpha} \cos{\Phi} \sin{\Theta} - \sin{\iota} \cos{\Phi} \cos{\Theta}) , \nonumber \\
    &\sin{\iota} \sin{\alpha}\sin{\Theta} + \cos{\iota} \cos{\Theta}\big) \, . \label{eq:rdrpq:N}
\end{align}
\end{subequations}

And these vectors, which are the solution of the orbit, can be substituted in Eq.~\eqref{eq:hphc_15PN} to obtain the gravitational waves emitted by the system taking into account up to spin-orbit corrections.

\begin{figure}[t!]
\begin{center}
\includegraphics[width = 0.47\textwidth]{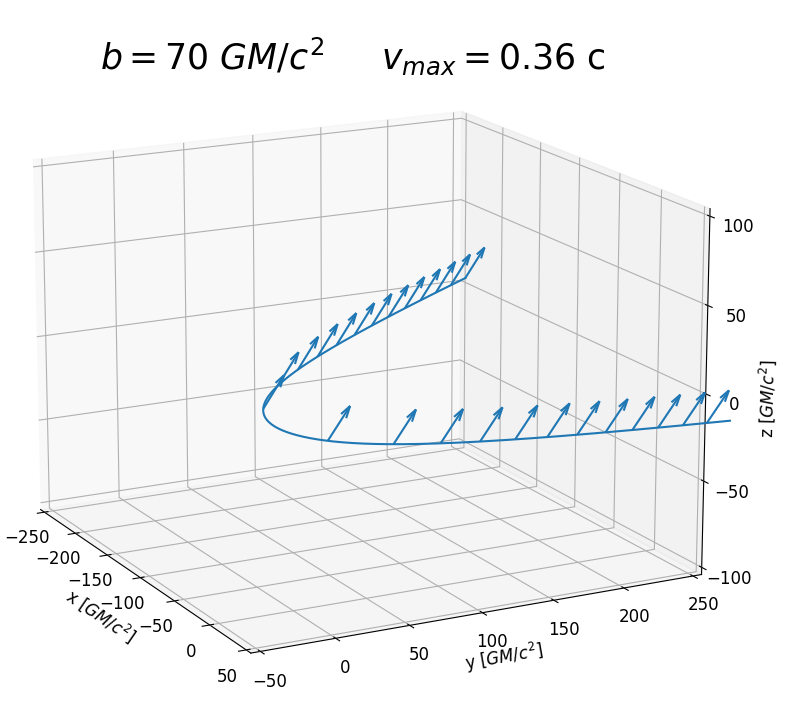}
\end{center} 
\caption{Example of an orbit obtained when solving the 1.5 PN equations of motion for a system of black holes in a close hyperbolic encounter with $m_1 = 20 M_{\odot}$, $ m_2 = 15{M_\odot}$, $\chi_1 = 1$, $\chi_2 = 1$, $\overline{\xi}_0 = 0.0005$, $ e_{t0} = 1.1$,  $ \Phi_0 = 0$, $ \theta_1^i = 0.5 \text{ rad}$, $ \phi_1^i = 0.35 \text{ rad}$, $  \theta_2^i = 0.8 \text{ rad}$, $ \phi_2^i = 1 \text{ rad}$. We have superimposed on the orbit the effective spin of the system $\vec{S}_\mathrm{eff}$, showing its direction and magnitude with an arrow.}
\label{fig:Orbit}
\end{figure}

In Fig.~\ref{fig:Orbit} we can observe a representative example of the type of orbit $\vec{r}(t)$ one obtains when solving the equations of motion of Eq.~\eqref{eq:dy}. We have drawn arrows on the orbit representing the effective spin of the system $\vec{S}_\mathrm{eff}$ as defined in Eq.~\eqref{eq:Seffdef}. Even though the maximum velocity reached is quite high, at around $0.36$c (the expansion parameter of the PN approximation will be $\frac{\text{v}^2}{c^2} \sim 0.13$), the orbit is still hyperbolic and the effective spin is almost constant, meaning the higher order post newtonian corrections are ``small'', as should happen for the expansion to be valid. 

\begin{figure}[t!]
\begin{center}
\includegraphics[width = 0.47\textwidth]{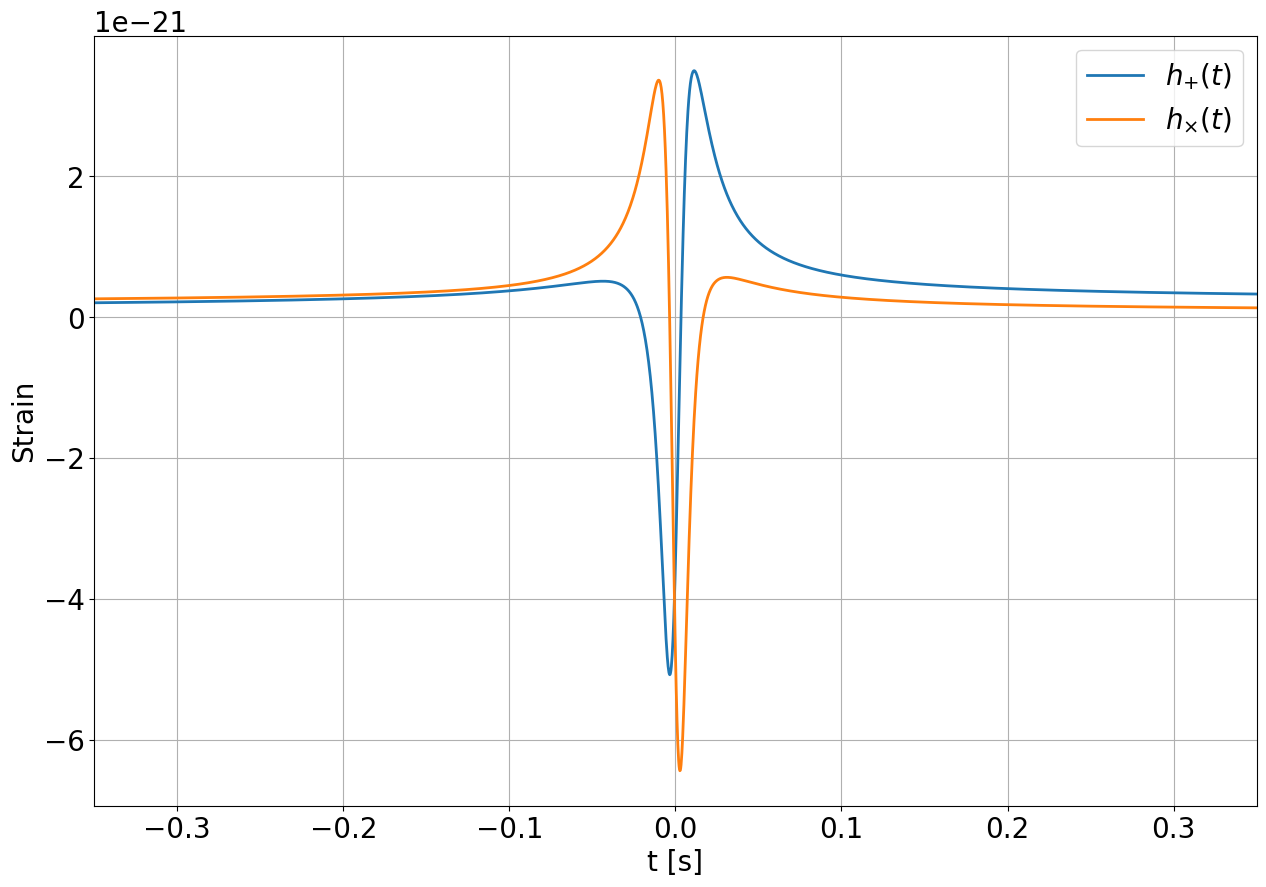}
\end{center} 
\caption{Gravitational waves emitted by the system shown in Fig.~\ref{fig:Orbit} assuming it happens at a distance $R = 20$Mpc and with an inclination of the orbit $\Theta = 45 ^{\circ}$.}
\label{fig:PolarizationsTime}
\end{figure}

The hyperbolic event shown in Fig.~\ref{fig:Orbit} will emit gravitational waves, in accordance with Eq.~\eqref{eq:hphc_15PN}. We show in Fig.~\ref{fig:PolarizationsTime} the strain from these gravitational waves for the two polarizations. This image is very representative of the shape of the gravitational wave strain we expect from close hyperbollic encounters. We can see that as the black holes get closer to each other, the absolute value of the strain increases for the two polarizations and it reaches a maximum before they get to the point of closest approach (at $t = 0$). Due to the quadrupole nature of the gravitational waves (the frequency of the waves is twice the orbital frequency), the absolute value of the strain falls down before the closest approach and changes sign. The absolute value reaches again a maximum after closest approach and then as the black holes drift further appart, the absolute value diminishes again until it reaches a constant value.  All in all, the gravitational waves emitted during hyperbolic encounters perform only one oscillation.

Note that the asymptotic value of the strain after the encounter is different from the one before the encounter. This is known as the gravitational wave memory effect \cite{MemoryEffect}, and has been studied in the literature for hyperbolic encounters \cite{De_Vittori_2014}.

\section{Signal in a gravitational wave detector}
\label{sec:detec}

In this paper we are not only interested in the purely theoretical aspects of the gravitational wave emission of black holes in hyperbolic orbits, we also want to look for this type of events in current gravitational wave detectors. To this end, we will need to determine and characterize the signal that these gravitational waves leave in a detector.

Right now there are many observatories looking for gravitational waves via different experimental setups, using resonant mass antennas \cite{ResonantMass}, pulsar timing arrays \cite{PulsarTimingArrays} and laser interferometers.

Laser interferometry is to date the only method that has been able to confidently observe gravitational waves, and it offers by far the best sensitivity to detect transient gravitational wave events. Because of this, we will focus our search to the data from laser interferometer gravitational wave observatories, in particular from Advanced LIGO and Advanced Virgo.

\subsection{Laser interferometers for gravitational wave detection}
\label{sec:detec:laser}

\begin{figure*}[t!]
\begin{center}
\includegraphics[width = 0.85\textwidth]{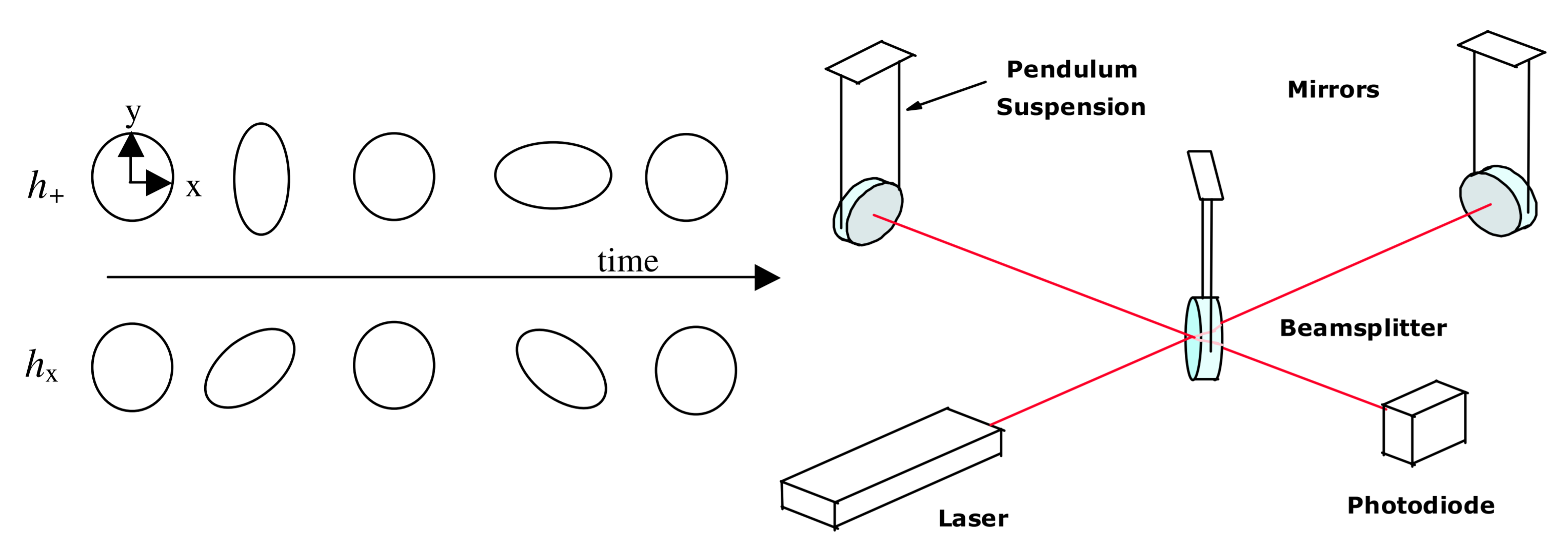}
\end{center} 
\caption{Sketch of the design of a Michelson interferometer for gravitational wave detection. On the left we can also see a schematic diagram of how gravitational waves interact with a ring of matter, showing the quadrupolar nature of the interaction. Image from Ref.~\cite{MichelsonSketch}.}
\label{fig:InterferometerSketch}
\end{figure*}

The basic design of a laser interferometer for gravitational wave detection is to have a Michelson interferometer decoupled from outside forces. This is done using pendulum suspension for the mirrors and for the optical benches. The interferometer is tuned to be in destructive interference in the absence of gravitational waves. Whenever a gravitational wave passes through the detector, the strain will change the effective length of the arms, the interference will not be completely destructive anymore and a signal will be observed at the photodiode. This basic design is schematically shown in Fig.~\ref{fig:InterferometerSketch}.

Of course to improve sensitivity the simple design of Fig.~\ref{fig:InterferometerSketch} gets more complicated \cite{MichelsonSketch}. All Earth-based laser interferometers looking for gravitational waves add a Fabry-Perot resonant cavity in each arm to build up the phase shift produced by the arm length change. Another standard feature added is power recycling, a technique to increase the effective power of the laser by adding power recycling mirrors to form a resonant cavity between the laser source and the Michelson.

\subsubsection{Antenna patterns}
\label{sec:detec:laser:antenna}

Laser interferometers are only sensitive to the differential change in the length of their arms. Because of this, we will have to determine how the gravitational wave strain $h_+$ and $h_\times$ will affect the length of the arms, and in this way go from the theoretical prediction obtained in Sec.~\ref{sec:waveforms} to a physical observable.

We will consider the case in which the wavelength of the gravitational wave is much larger than the size of the arms of the detector ($L/\lambda_{gw} \ll 1$). For detectors like LIGO and Virgo with an arm length of 3-4 km, this is satisfied for frequencies smaller than ${\sim 10 \text{kHz}}$. In this large wavelength approximation, we can ignore spatial variations of the metric inside the detector. Using the proper detector frame, the motion of the mirrors at the end of the interferometer arms will be governed by the geodesic deviation equation \cite{Maggiore_Vol1}:

\begin{equation}
    \ddot{\xi}^i = \frac{1}{2} \ddot{h}_{ij} \xi^j \, .
    \label{eq:GeodesicMirrors}
\end{equation}

\begin{figure}[t!]
\begin{center}
\includegraphics[width = 0.4\textwidth]{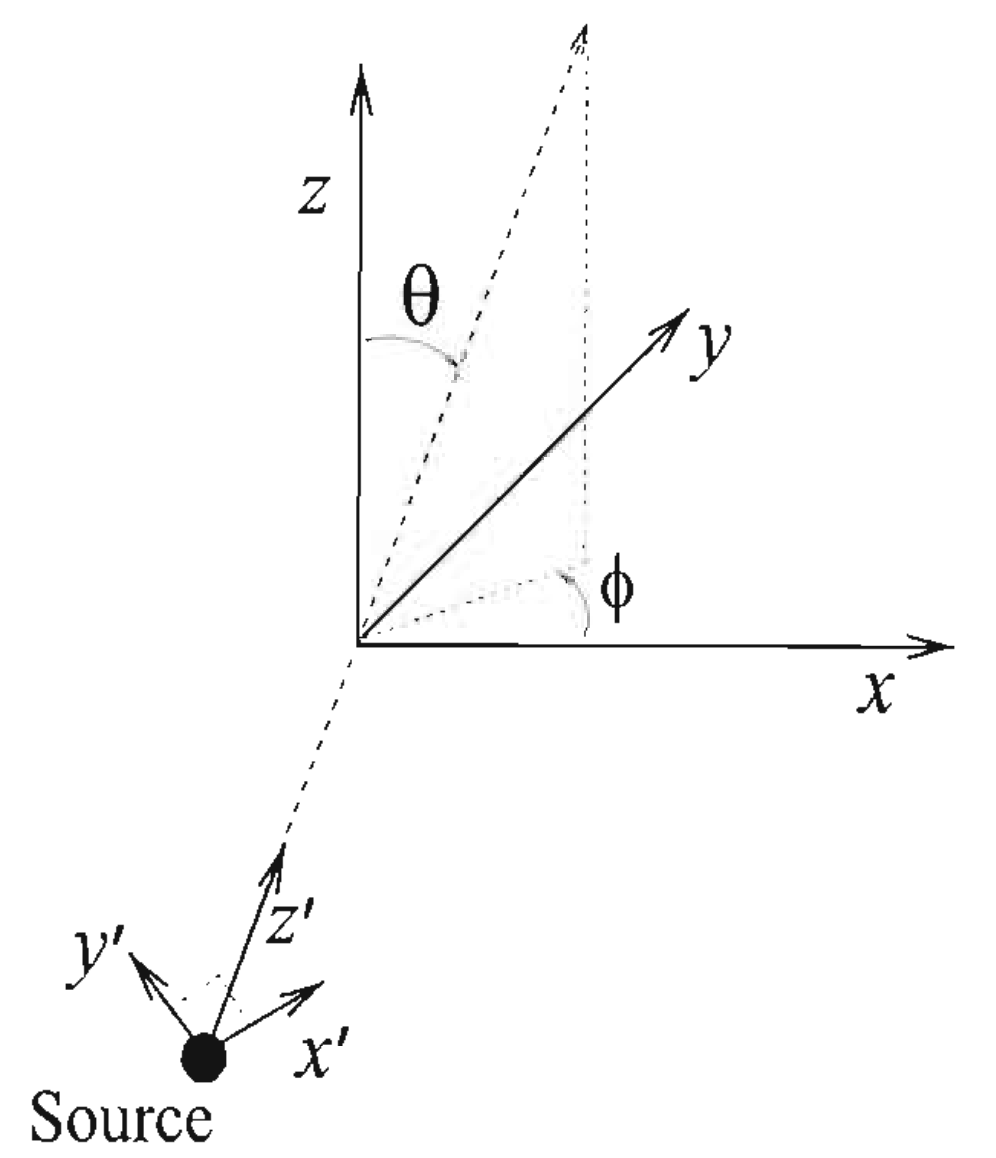}
\end{center} 
\caption{Geometry used in the computation of the antenna patterns. The arms of the interferometer are along the $x$ and $y$ axes. As shown in the figure, $\theta$ and $\phi$ are the polar angles of $z'$ in the detector frame. $\psi$ measures the polarization angle of the wave and it represents a rotation about $z'$. Image from Ref.~\cite{Maggiore_Vol1}.}
\label{fig:AntennaCoord}
\end{figure}

We will now need to define a coordinate system with which to carry our computation. We choose a coordinate system as shown in Fig.~\ref{fig:AntennaCoord}, where we have a detector reference frame $(x,y,z)$ such that the arms of the interferometer are along the $x$ and $y$ axis, and we have a source reference frame $(x',y',z')$ such that the propagation direction of the gravitational wave coincides with the $z'$ axis. The transformation between these two coordinate systems can be given in terms of the usual Eulerian angles $\theta$, $\phi$ and $\psi$. That is, the transformation from $(x',y',z')$ to $(x,y,z)$ will be given by \cite{AntennaPatternsPaper}:

\begin{widetext}
\begin{equation}
    A^{i}_{j'} = 
    \begin{pmatrix}
        \cos{\psi}\cos{\phi}-\cos{\theta}\sin{\phi}\sin{\psi} & -(\sin{\psi}\cos{\phi}+\cos{\theta}\sin{\phi}\cos{\psi}) & \sin{\theta}\sin{\phi}\\
        \cos{\psi}\sin{\phi}+\cos{\theta}\cos{\phi}\sin{\psi} & -\sin{\psi}\sin{\phi}+\cos{\theta}\cos{\phi}\cos{\psi} & -\sin{\theta}\cos{\phi}\\
        \sin{\theta}\sin{\psi}& \sin{\theta}\cos{\psi}  & \cos{\theta}
    \end{pmatrix} \, .
    \label{eq:EulerSourceToDetec}
\end{equation}
\end{widetext}

Since the value of the gravitational wave strain $h_{i j}$ is assumed to be very small (usually $h \sim 10^{-21}$), then in the detector frame, the test mass at the end of the arm along the x-axis will be displaced from its equilibrium position by a small amount, that is $\xi^j = (L,0,0)+O(hL)$. Looking at how this test mass moves along the length of the arm, we have:

\begin{equation}
    \ddot{\xi}_x = \frac{1}{2} \ddot{h}_{x x} L + O(h^2 L) \, .
    \label{eq:GeodesicMirrors_x}
\end{equation}

We can get an equivalent expression for how the test mass in the y-axis moves in the y-direction:

\begin{equation}
    \ddot{\xi}_y = \frac{1}{2} \ddot{h}_{y y} L + O(h^2 L) \, .
    \label{eq:GeodesicMirrors_y}
\end{equation}

And the differential change in the length of the arms can be gotten subtracting Eq.~\eqref{eq:GeodesicMirrors_y} from Eq.~\eqref{eq:GeodesicMirrors_x}:

\begin{equation}
    \delta \ddot{L} = \frac{1}{2} (\ddot{h}_{x x} - \ddot{h}_{y y}) L + O(h^2 L) \, .
    \label{eq:GeodesicMirrors_x-y_dif}
\end{equation}

We can integrate this expression using that in the absence of gravitational waves the interferometer is tuned to have $\delta L = 0$. Keeping only linear terms in $h$ we obtain:

\begin{equation}
    \delta L = \frac{1}{2} (h_{x x} - h_{y y})L \, .
    \label{eq:GeodesicMirrors_x-y}
\end{equation}

To connect $h_{x x}$ and $h_{y y}$ in the detector frame with $h_+$ and $h_\times$ in the source frame in which the wave propagates in the $z'$ direction, we use that in this source frame the strain is given by:

\begin{equation}
    h_{i' j'} = 
    \begin{pmatrix}
        h_+ & h_\times & 0 \\
        h_\times & -h_+ & 0 \\
        0 & 0 & 0
    \end{pmatrix} \, .
    \label{eq:h_source}
\end{equation}

And since the gravitational waves are tensorial, they transform in the following way:

\begin{equation}
    h_{i j} = A^{i}_{i'} A^{j}_{j'} h_{i' j'} \, .
    \label{eq:h_source_to_detec_ez}
\end{equation}

Substituting this in the expression for the differential change in arm length of Eq.~\eqref{eq:GeodesicMirrors_x-y} we get:

\begin{equation}
    \frac{\delta L}{L} = F_+(\theta,\phi,\psi)h_+ + F_\times(\theta,\phi,\psi)h_\times \equiv h \, ,
    \label{eq:def_h}
\end{equation}

\noindent where $h \equiv \delta L/L$ is called the strain amplitude and it is the actual quantity that can be measured by laser interferometers. $F_+(\theta,\phi,\psi)$ and $F_\times(\theta,\phi,\psi)$ are the antenna patterns and they contain all the dependence with the angles that appears in the transformation from the source to the detector frames. They are given by:

\begin{subequations}
\label{eq:AntennaPatterns}
\begin{align}
    F_+(\theta,\phi,\psi) & = \frac{1}{2} (1+\cos^2 \theta) \cos{2\phi} \cos{2\psi} \nonumber\\
    &  - \cos{\theta}\sin{2\phi} \sin{2\psi} \, , \label{eq:AntennaPatterns:F_+}\\
    F_\times(\theta,\phi,\psi) & = \frac{1}{2} (1+\cos^2 \theta) \cos{2\phi} \sin{2\psi} \nonumber\\
    &  + \cos{\theta}\sin{2\phi} \cos{2\psi} \, . \label{eq:AntennaPatterns:F_x}
\end{align}
\end{subequations}

\begin{figure}[t!]
\begin{center}
\includegraphics[width = 0.47\textwidth]{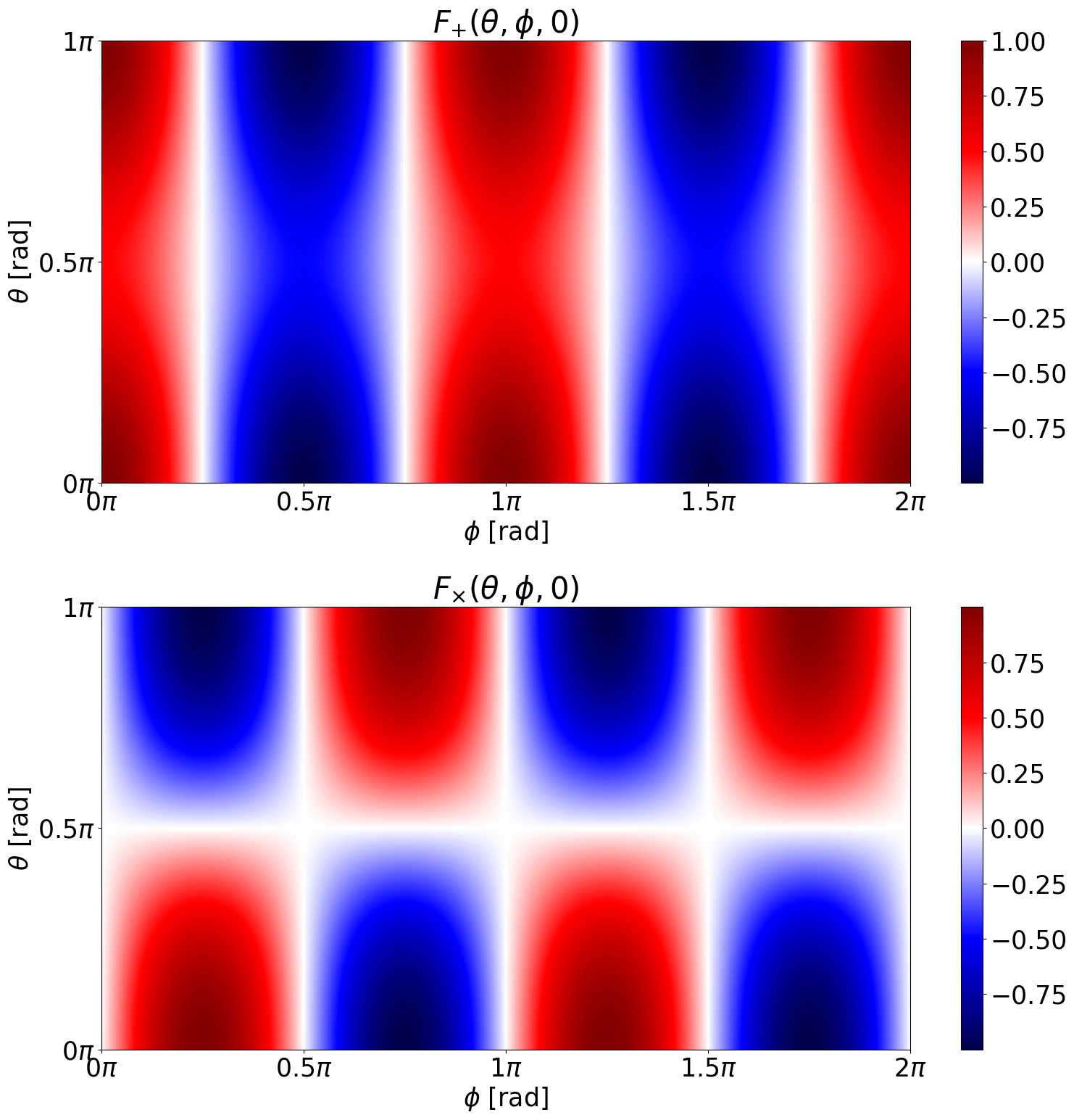}
\end{center} 
\caption{Antenna patterns $F_+(\theta,\phi, 0)$ and $F_\times(\theta,\phi, 0)$ of a two arm interferometric detector of gravitational waves. The wave is assumed to have $0$ polarization angle ($\psi = 0$) with $\theta$ and $\phi$ being the polar angles of its direction, shown in Fig.~\ref{fig:AntennaCoord}.}
\label{fig:AntennaPatterns}
\end{figure}

We show the antenna patterns for $0$ polarization angle ($\psi = 0$) in Fig.~\ref{fig:AntennaPatterns}. The antenna patterns for non $0$ polarization angle can be gotten by combining these two with a rotation:

\begin{equation}
\label{eq:AntennaPattern_rotation}
    \begin{pmatrix}  F_+(\theta,\phi,\psi) \\ F_\times(\theta,\phi,\psi) \end{pmatrix}
    =
    \begin{pmatrix}
    \cos{2\psi} &  -\sin{2\psi} \\
    \sin{2\psi} &  \cos{2\psi} 
    \end{pmatrix}
    \begin{pmatrix}  F_+(\theta,\phi, 0) \\ F_\times(\theta,\phi, 0) \end{pmatrix} \, .
\end{equation}

In Fig.~\ref{fig:AntennaPatterns} we can see how $F_+(\theta,\phi, 0)$ and $F_\times(\theta,\phi, 0)$ vary depending on the incoming angle of the gravitational wave. At $\theta = \pi$ with $\phi = \frac{\pi}{4}, \frac{3\pi}{4}, \frac{5\pi}{4}, \frac{7\pi}{4}$ we will have that the two antenna factors vanish, and since they vanish for $\psi=0$, from Eq.~\eqref{eq:AntennaPattern_rotation} we have that they vanish for all polarizations. These four points are called the blind spots of the detector, since gravitational waves coming from positions in the sky close to them will be highly suppressed and almost impossible to detect.

\begin{figure*}[t!]
\centering
\includegraphics[width=0.55\textwidth]{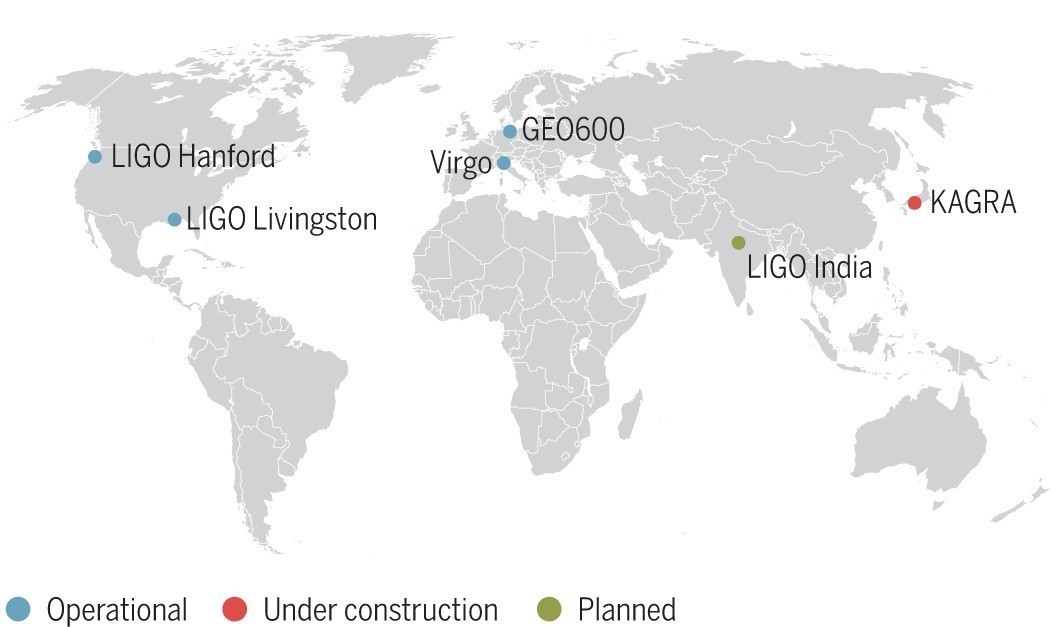}
\label{fig:RealProyect:DetectorPositions}
\includegraphics[width=0.4\textwidth]{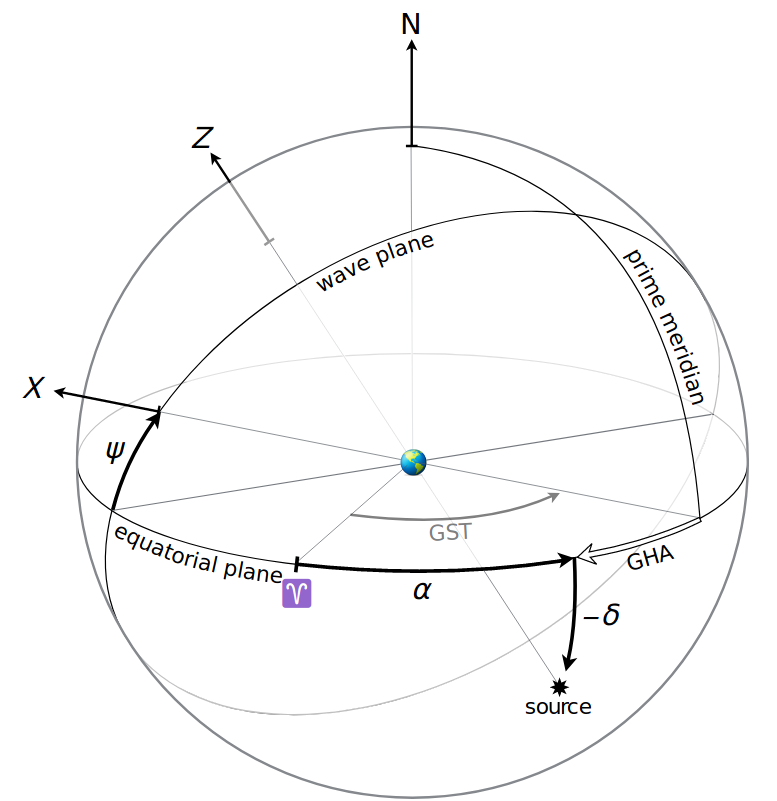}
\label{fig:RealProyect:EarthEquatorialFrame}
\caption{Left: We show the position of the different interferometric gravitational wave detectors. Image from Ref.~\cite{DetectorLocations}. Right: We show the Earth equatorial frame used to locate sources in the sky. (X,Y,Z) is the wave frame, $\delta$ is the declination, $\alpha$ is the right ascension and GHA is the greenwich hour angle. Image from Ref.~\cite{lalsuite}.}
\label{fig:RealProyect}
\end{figure*}

We will be interested in the signal one gravitational wave will imprint in all the different detectors on Earth. These different detectors will be in different points of the globe, shown in Fig.~\ref{fig:RealProyect}, and placed with different orientations that vary over time due to the orbit and rotation of the Earth. We will locate a particular event in the sky using Earth's equatorial frame, shown in Fig.~\ref{fig:RealProyect}, in which the wave will have a right ascension $\alpha$, declination $\delta$ and polarization angle $\psi$. We will have to relate these coordinates with the detector coordinates defined in Fig.~\ref{fig:AntennaCoord}, which will allow us to compute the antenna patterns and project the wave onto the detector. It will also be very important to take into account the fact that the wave will arrive at different times to each detector due to the distance between the detectors. This time difference is known as the light travel time between detectors and depending on the incoming direction of the wave it can be as long as 27ms between LIGO Hanford and Virgo, 26ms between LIGO Livingston and Virgo and 10ms between the two LIGO detectors.

In practice we will do the projection of the gravitational waves into the detector using \texttt{lalsuite} \cite{lalsuite} which is a software developed by the LIGO collaboration that can be used for this purpose. It takes into account all the antenna pattern and light travel time effects using accurate positions and orientations of the detectors as well as small Doppler shift corrections due to the Earth's rotation and orbit. 

\begin{figure*}[t!]
\begin{center}
\includegraphics[width=0.8\textwidth]{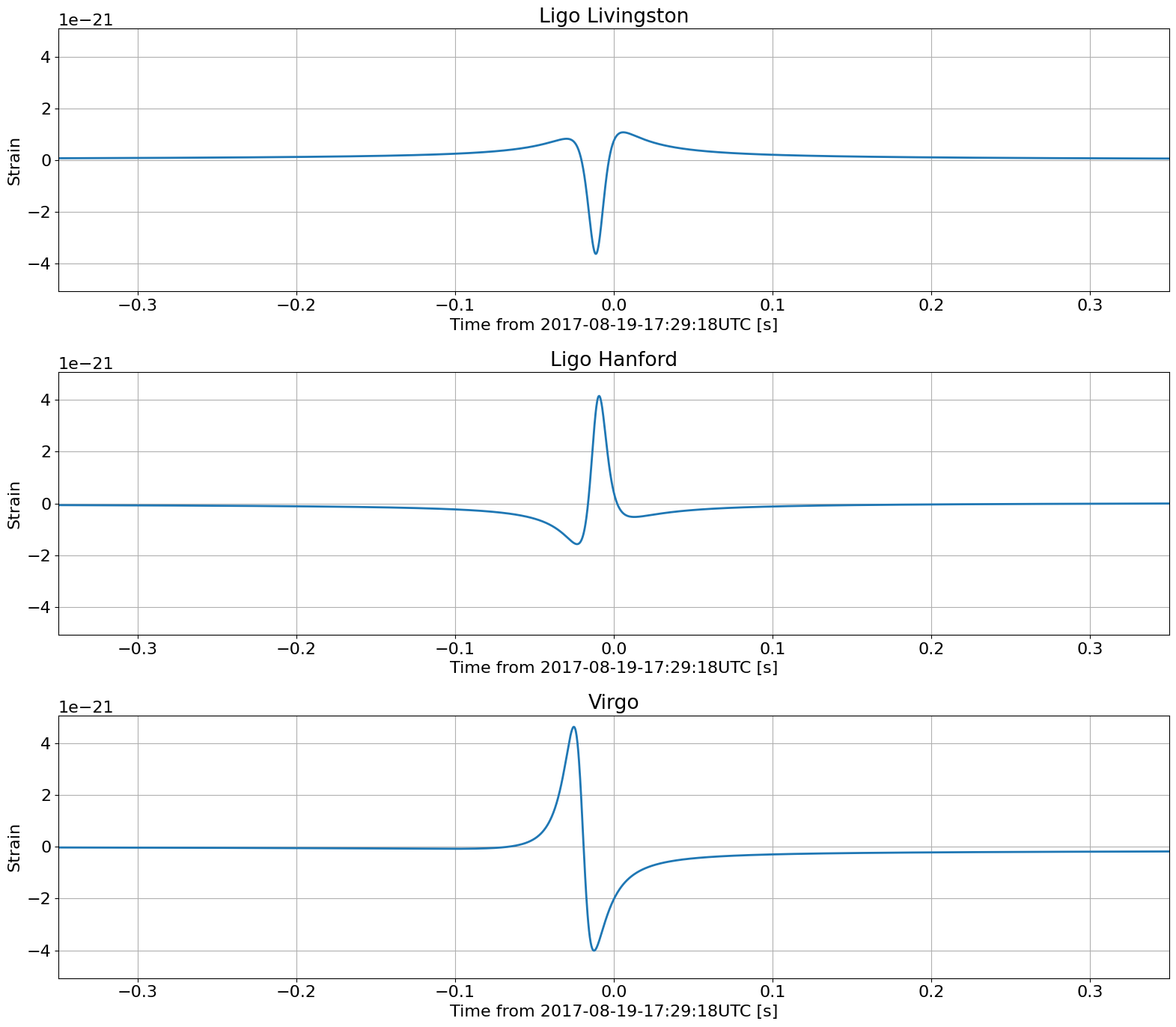}
\end{center} 
\caption{Result of projecting the gravitational waves of Fig.~\ref{fig:PolarizationsTime} into the gravitational wave detectors, assuming that they come from $\delta = 1.0$ rad, $\alpha = 3.7$ rad, with $\psi = 0.2$ rad and with the periastron time ($t=0$ in the simulation) taking place at 17:29:18 UTC of 2017-08-19 at the center of the Earth.}
\label{fig:ProyectedTime}
\end{figure*}

To exemplify how the projected waves look, in Fig.~\ref{fig:ProyectedTime} we show the result of projecting the gravitational waves of Fig.~\ref{fig:PolarizationsTime} into the gravitational wave detectors, assuming that they come from $\delta = 1.0$ rad, $\alpha = 3.7$ rad, with $\psi = 0.2$ rad and with the periastron time ($t=0$ in the simulation) taking place at 17:29:18 UTC of 2017-08-19 at the center of the Earth. We observe that even though it is the same gravitational wave in all interferometers, because of the antenna factors, it will have a different amplitude and shape in each of them. In the specific case shown here, the biggest amplitude is recorded in Virgo.

\subsubsection{Characterizing the stationary noise}
\label{sec:detec:laser:noise}

From the interference pattern one observes in the photodiode of the interferometer, one can in principle reconstruct the value of $\delta L/L$ and thus obtain the value of the strain amplitude $h(t)$. In practice, the output of any real detector will also contain noise, so the reconstructed quantity $s(t)$ will actually be given by:

\begin{equation}
    s(t) = h(t) + n(t) \, ,
    \label{eq:DetectorOutput}
\end{equation}

\noindent where $h(t)$ will be the part of the detector output coming from the gravitational wave signal, while $n(t)$ is the noise and it accounts for all the rest of the output coming in general from a variety of sources.

The noise $n(t)$ can usually be assumed to be stationary, this means that its properties do not vary in time. If the noise is stationary, the different Fourier components are uncorrelated, and thus:

\begin{equation}
    \langle \tilde{n}^{*}(f) \tilde{n}(f') \rangle \propto \delta(f-f') \, \rightarrow \, \langle \tilde{n}^{*}(f) \tilde{n}(f') \rangle \equiv \frac{1}{2} S_{n}(f) \delta(f-f') \, ,
    \label{eq:PSD_def}
\end{equation}

\noindent where $\langle ... \rangle$ denotes the ensemble average, that is, the average over many realizations of the same system. Eq.~\eqref{eq:PSD_def} defines the noise power spectrum distribution $S_n(f)$. Since $n(t)$ is real, $\tilde{n}(-f) = \tilde{n}^{*}(f)$ and therefore $S_n(-f) = S_n(f)$. In addition, $n(t)$ is dimensionless like the strain, and therefore $S_n(f)$ has dimensions of $\text{Hz}^{-1}$.

\begin{figure}[t!]
\begin{center}
\includegraphics[width=0.45\textwidth]{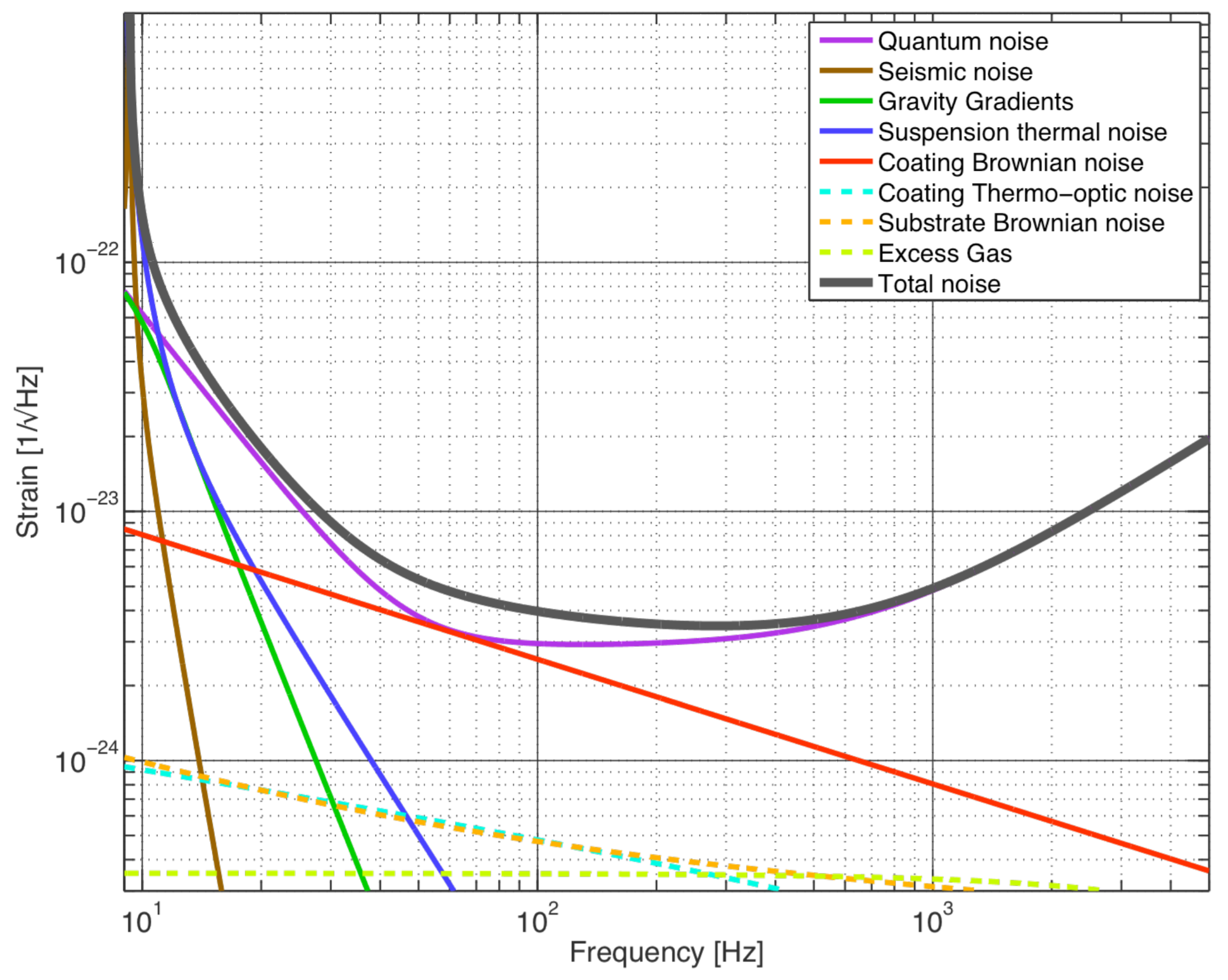}
\end{center} 
\caption{Advanced LIGO design strain sensitivity detailing the different modeled sources of noise. Image taken from Ref.~\cite{AdvLIGO_design}}
\label{fig:aLIGO_ASD}
\end{figure}

Stationary noise will be fully characterized by $S_n(f)$. To get an idea of the sensitivity of the detector, one defines the strain sensitivity as $\sqrt{S_n(f)}$, which gives an idea of the amplitude of the noise at a given frequency. Values of the strain smaller than $\sqrt{S_n(f)}$ at a given frequency will be difficult to measure. In Fig.~\ref{fig:aLIGO_ASD} we have represented the design strain sensitivity for the advanced LIGO detector \cite{AdvLIGO_design}. This is very representative of modern gravitational wave interferometric detectors. Below $20 \text{ Hz}$ the sensitivity is very limited by the seismic noise as well as by the thermal noise in the suspension of the mirrors. Above $10^3 \text{ Hz}$ the sensitivity starts to worsen again due to the quantum noise, which encompasses the effects of statistical fluctuations in detected photon arrival rate (shot noise) and radiation pressure due to photon number fluctuations. 

The noise power spectrum distribution $S_n(f)$ will be extensively used in the data analysis and therefore it will be extremely useful to compute it. In practice we only have one realization of the system and the amount of time we measure is finite and sampled in a discrete grid. Because of this, doing the ensemble average of Eq.~\eqref{eq:PSD_def} will not be a viable way to get $S_n(f)$. We will estimate $S_n(f)$ using Welch's method \cite{welch}.

The basic idea of this method is to divide the signal into successive overlapping segments, computing the periodogram for each segment and averaging. Let $x(n), \; n=0,...,N-1$ be the vector whose power spectra we want to compute. We take windowed segments of this vector $x_k(j)$, possibly overlapping, of length $L$, and with the start of consecutive segments being $D$ units apart. That is:

\begin{equation}
    x_m(n) = w(n)x(n+mD) \quad n=0,..,L-1 \, ,
    \label{eq:xk_welch}
\end{equation}

\noindent where $w(n)$ is the window function. In our analysis we will always use the Hanning window to smooth discontinuities at the beginning and end of the sampled signal. It is given by:

\begin{equation}
    w(n) = \frac{1}{2}\left( 1- \cos\left( \frac{2 \pi n}{L-1} \right) \right) \quad n=0,..,L-1 \, .
    \label{eq:w_hanning}
\end{equation}

We define $K$ as the total number of segments we have ($(K-1)D+L=N$), and thus $m = 0,1,...K-1$. We compute the periodogram for each one of the segments:

\begin{equation}
    P_m(f_k) = \frac{1}{U} \left|\text{FFT}_k(x_m) \right|^2 \equiv \frac{1}{U} \left| \sum_{n=0}^{L-1} x_m(n) e^{-i2 \pi n k/L}\right|^2 \, ,
    \label{eq:periodo_welch}
\end{equation}

\noindent where $\text{FFT}$ denotes the fast Fourier transform and $U$ is a normalization that has to be introduced to take into account the effect of the window and it is given by:

\begin{equation}
    U = \sum_{n=0}^{L-1} w^2(n) \, .
    \label{eq:w_norm}
\end{equation}

The Welch estimate for the power spectral density is then given by the average of the periodograms over all the segments, where we also have to take into account that the spacing between consecutive points of $x(n)$ is $\Delta t$ and the fact that we defined $S(f)$ with a factor $\frac{1}{2}$ in Eq.~\eqref{eq:PSD_def}:

\begin{equation}
    S(f_k) = \frac{2 \Delta t}{K} \sum_{m=0}^{K-1} P_m(f_k) \, ,
    \label{eq:S_welch}
\end{equation}

\noindent which has the correct $\text{Hz}^{-1}$ units.

In general for the computation of the noise power spectrum distribution of a gravitational wave detector we will always use the detector output for the strain $s(t)$, which in principle contains the signal $h(t)$ and the noise $n(t)$. This is justified because the data of current gravitational wave detectors are vastly dominated by noise, except when a gravitational wave event takes place.

When we have a gravitational wave event candidate in our data and we want to analyze it, to compute the power spectrum distribution we will exclude the windows that we suspect have gravitational wave signal, and we will just compute the power spectrum around them.

\begin{figure}[t!]
\begin{center}
\includegraphics[width=0.485\textwidth]{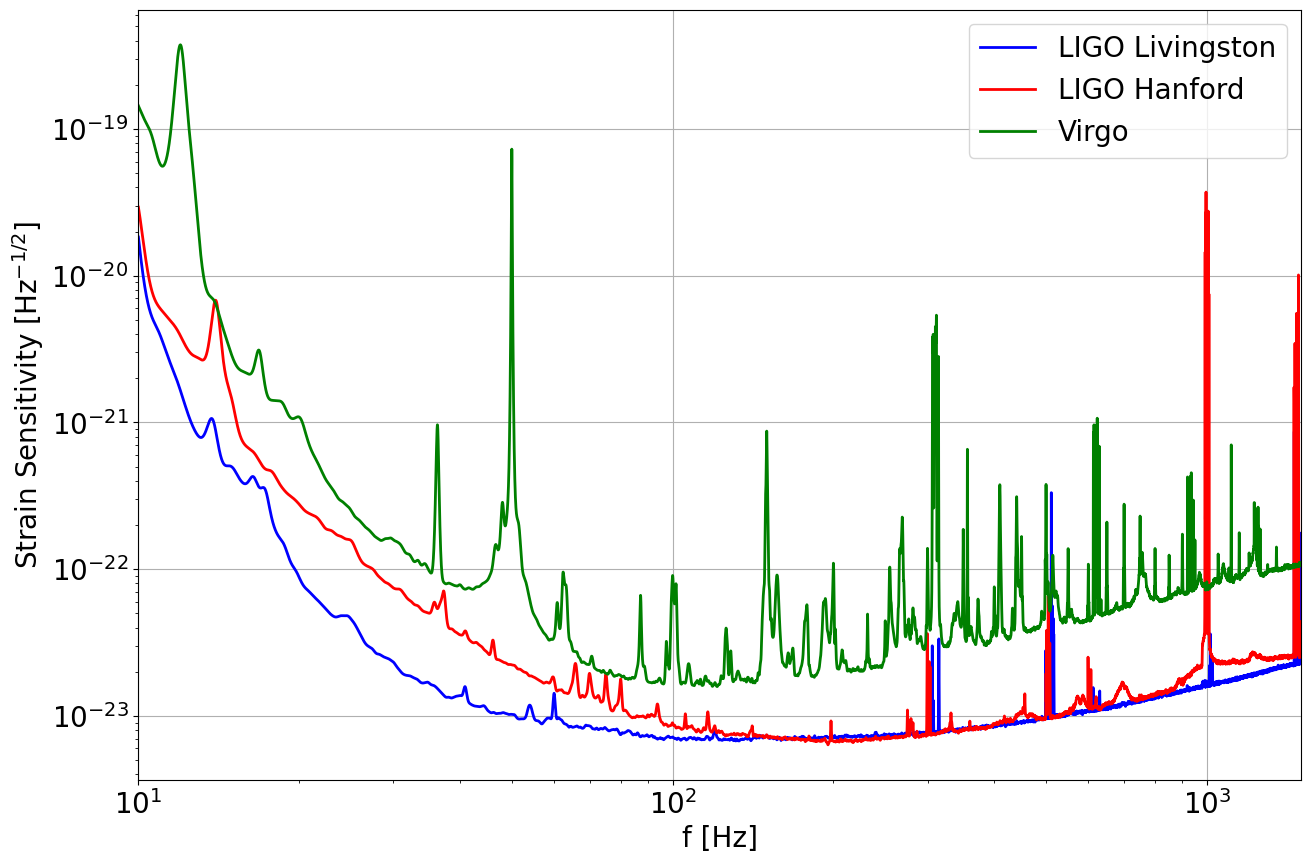}
\end{center} 
\caption{Strain sensitivity for LIGO Livingston, LIGO Hanford and Virgo computed using Welch's method between 16:55:10 UTC and 18:03:26 UTC of 2017-08-19. The data is sampled at a frequency of 4096Hz and in Welch's method we use a segment length of 4s ($L = 16384$) with a segment stride of 2s ($D = 8192$).}
\label{fig:ASDs}
\end{figure}

In Fig.~\ref{fig:ASDs} we show the strain sensitivity $\sqrt{S(f)}$ obtained when applying Welch's method to $4096$s of measured strain of the two LIGO detectors and the Virgo detector. We observe that even though the general shape of the strain sensitivity is very similar to the design one in Fig.~\ref{fig:aLIGO_ASD}, the experimental curve has a lot of spikes on top of the design curve. These spikes come from so called ``technical'' sources of noise, which comprises a large variety of environmental sources such as the noise coming from the pumps that keep the vacuum inside the interferometer arms, the coupling of electrical circuits, etc.

Note in Fig.~\ref{fig:ASDs} that the LIGO detectors are more sensitive than the Virgo detector, having about a factor of 3 better sensitivity along all the frequency spectrum. From the LIGO detectors, Livingston has better sensitivity than Hanford, specially at low frequencies ($f \lesssim 100$ Hz), due its superior suspension system. At high frequencies ($f \gtrsim 100$ Hz) both LIGO interferometers have similar strain sensitivities. 

\subsection{Signal processing}
\label{sec:detec:procesing}

As we have seen in Sec.~\ref{sec:detec:laser:noise}, detector data will not only contain the gravitational wave signals we are looking for, but also noise. For current detectors such as LIGO-Virgo, we expect the amount of noise to be very significant when compared with the amplitude of gravitational waves. In this section we will explain the methods that will be used to reduce the impact of the noise and represent the data in such a way that the gravitational wave signals from close hyperbolic encounters are enhanced.

\begin{figure*}[t!]
\begin{center}
\includegraphics[width=0.8\textwidth]{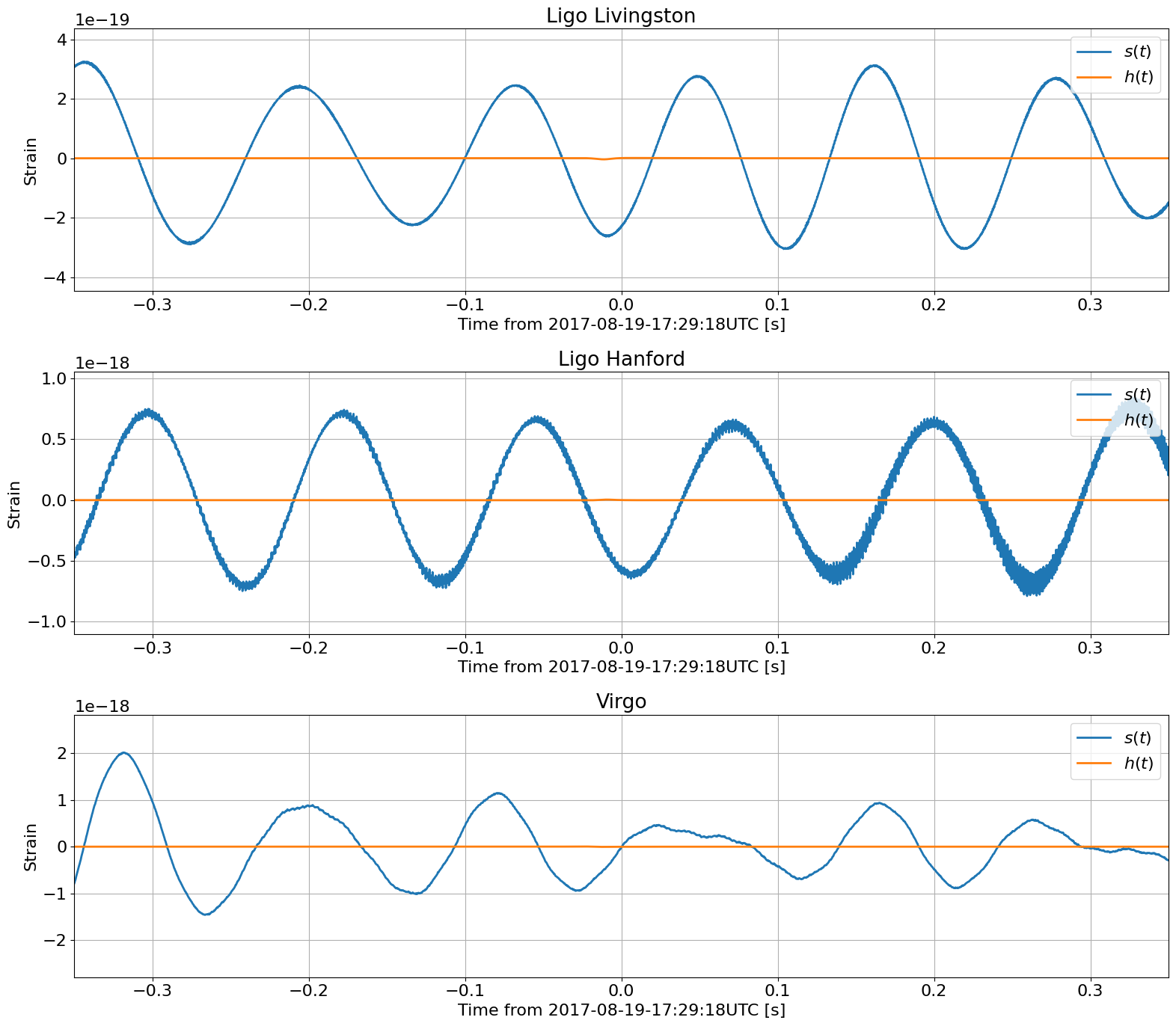}
\end{center} 
\caption{In blue we show the result $s(t)$ of injecting the gravitational waves of Fig.~\ref{fig:ProyectedTime} into the experimental strain of the different detectors. In orange we show the gravitational wave signal $h(t)$. The noise is so dominating that the features of the gravitational wave signal can not be seen. }
\label{fig:InjectedRAW}
\end{figure*}

To exemplify the results of each step of the signal processing we will use the example gravitational wave from a hyperbolic encounter that was shown in Fig.~\ref{fig:ProyectedTime}, which we will inject into the experimental output of the detector. Since we are assuming that the response of the detector is linear, injecting the gravitational wave will equate to adding it to the experimental strain ($s(t) = s_{exp}(t)+h(t)$). The result of doing this is shown in Fig.~\ref{fig:InjectedRAW}, where we have also plotted the gravitational wave signal $h(t)$, whose amplitude is so much smaller than the noise that its features can not be seen.

\subsubsection{Filtering}
\label{sec:detec:procesing:filtering}

The gravitational wave data we are using is sampled at a frequency of 4096Hz. Because of this, we can explore Fourier modes below the Nyquist frequency of 2048Hz. Nonetheless, not all the frequencies will give us useful information, and therefore we will filter the ones that are dominated by noise. 

In the strain sensitivity of Fig.~\ref{fig:ASDs} we observe how at frequencies below 20Hz the noise greatly increases, mainly because of the seismic and thermal noises. Because of this we will apply a 20Hz high-pass filter to the data to remove frequencies below 20Hz, which will always be completly dominated by noise.

Because gravitational waves from hyperbolic encounters only perform one oscillation, the characteristic frequency of the waves will be closely related with the length of this oscillation ($f_c \sim 1/\Delta t$) and since this single oscillation will be well located in time, due to the Fourier uncertainty principle, it will have a high frequency spread. Since we are using the post Newtonian approximation, the characteristic length scale of the problem will have to be bigger than tens of times the Schwarzschild radius and the characteristic speed of the black holes will have to be much smaller than the speed of light. All in all, since $\Delta t  \sim \Delta x/v$, the largest frequency we can probe in the limit of validity of our post Newtonian approximation can be estimated to be around $f_c \sim 1/\Delta t \sim \frac{c}{100 R_s}$, which for typical black holes of $5M_\odot$ is $f_c \sim 200$Hz, but generally it will be lower.

Since high frequencies will not have gravitational wave signal from hyperbolic encounters and will thus be dominated by noise, we will ignore them by applying a 800Hz low-pass filter. Together with the 20Hz high-pass filter, this will equate to a 20-800Hz band-pass filter.

\begin{figure*}[t!]
\begin{center}
\includegraphics[width=0.8\textwidth]{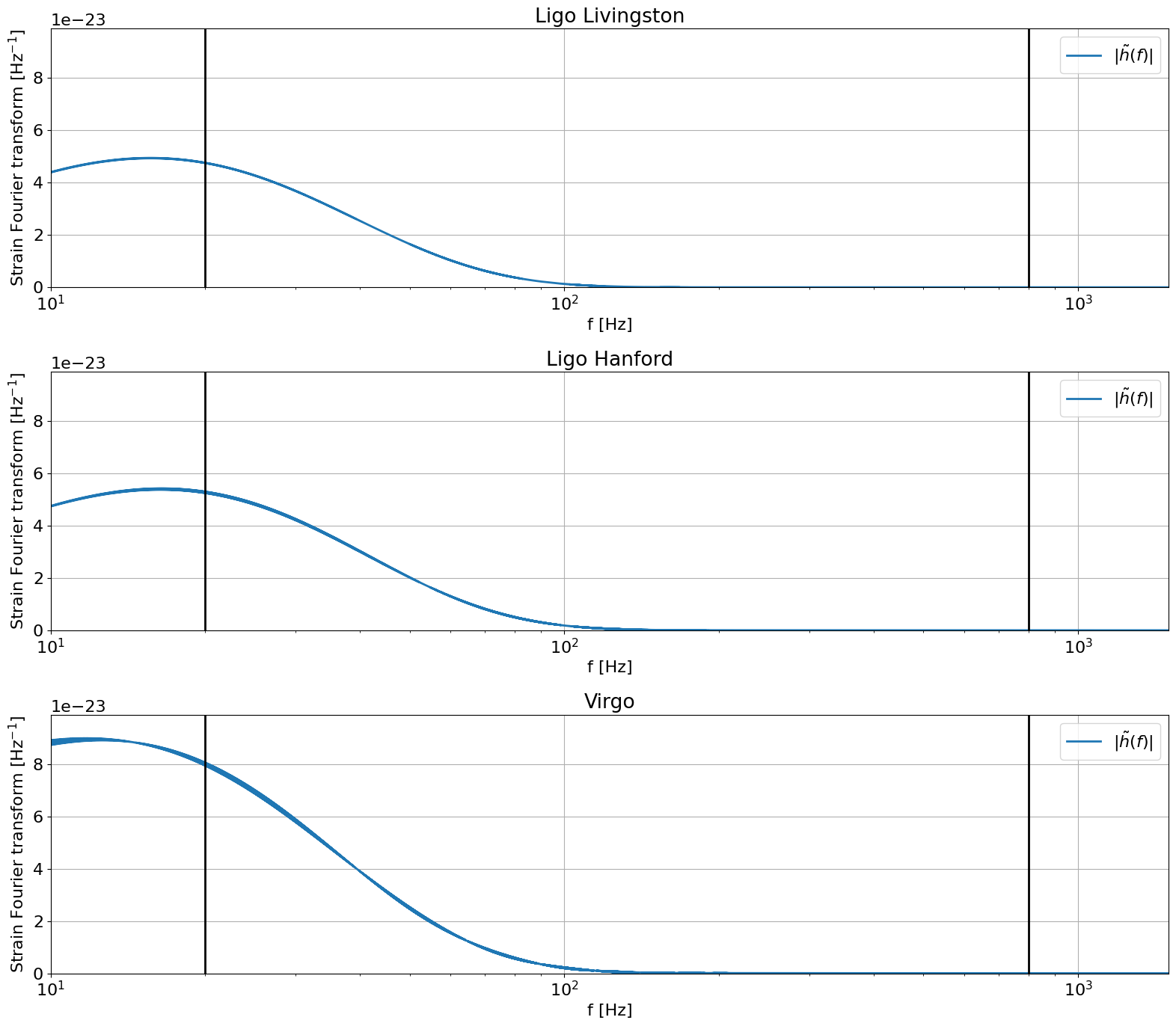}
\end{center} 
\caption{Fourier transform of the gravitational wave shown in Fig.~\ref{fig:ProyectedTime}. We have marked with black lines the 20-800Hz frequency window we are using to band-pass the strain.}
\label{fig:ProyectedFrequency}
\end{figure*}

In Fig.~\ref{fig:ProyectedFrequency} we show the Fourier structure of the gravitational wave of Fig.~\ref{fig:ProyectedTime} that we have been using as an example, together with the 20-800Hz window we are using to band-pass filter the data. This simulated gravitational wave came from black holes of $20M_\odot$ and $15M_\odot$ and we can see how the signal becomes negligible at high values of the strain, well below the 800Hz value.

\begin{figure*}[t!]
\begin{center}
\includegraphics[width=0.8\textwidth]{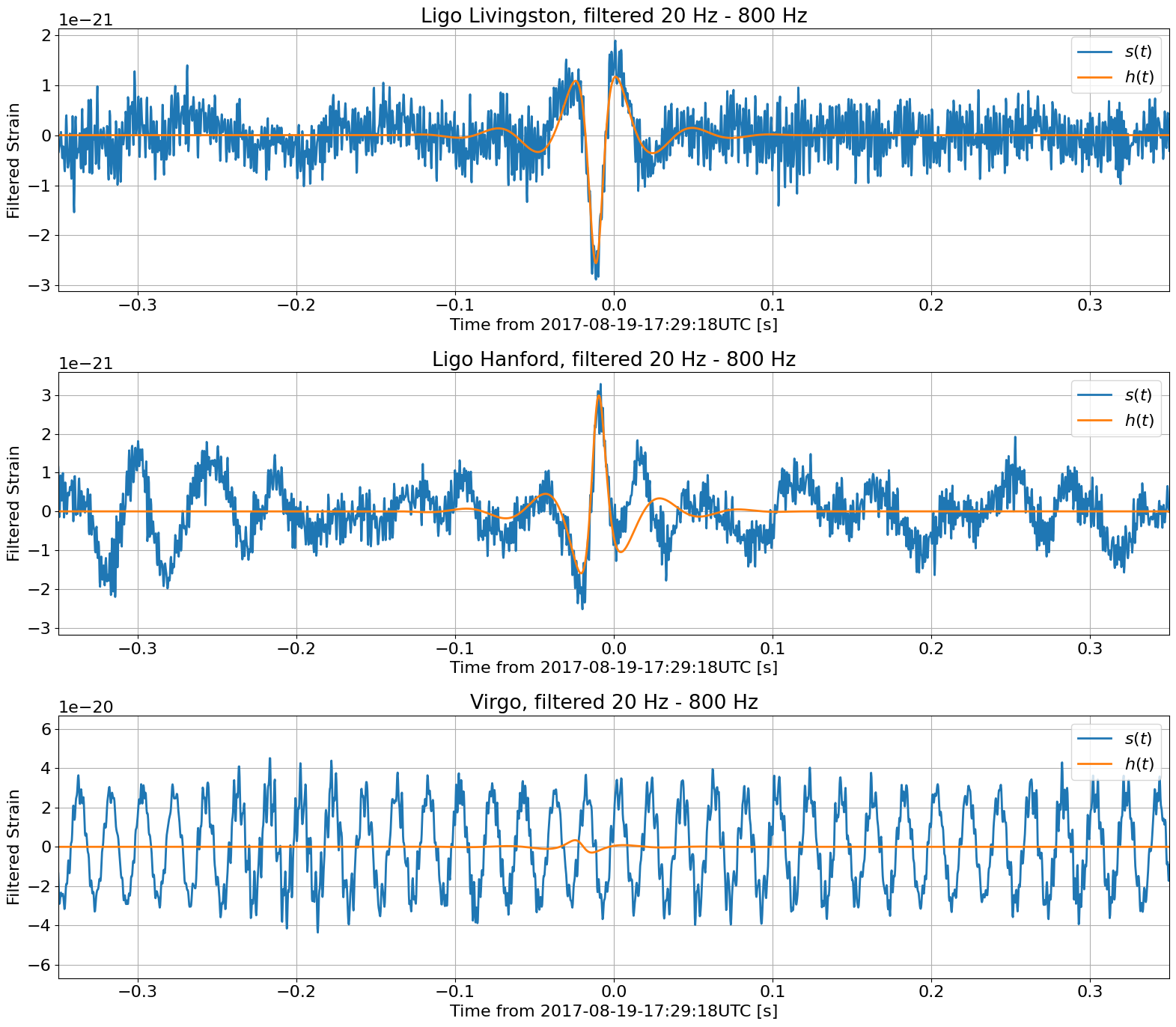}
\end{center} 
\caption{Result of band-pass filtering the signals of Fig.~\ref{fig:InjectedRAW} within a 20-800Hz window.}
\label{fig:InjectedFiltered}
\end{figure*}

In Fig.~\ref{fig:InjectedFiltered} we show the result of applying the 20-800Hz band-pass to $s(t)$ and $h(t)$ of Fig.~\ref{fig:InjectedRAW}. Now the signal can be seen much better in the strain of the LIGO detectors, since we have removed the frequencies that are the source of the most noise. Nonetheless if we did not have the actual injected signal in Hanford to guide the eye, it would not be so easy to see its presence in this detector, since Hanford still has large fluctuations of noise. Lastly, we observe that Virgo is still dominated by noise

\subsubsection{Whitening}
\label{sec:detec:procesing:whitening}

After filtering away the very low and very high frequencies, which are completely dominated by noise, we note that the remaining noise power spectrum at the laser interferometer depends strongly on frequency (see Fig.~\ref{fig:ASDs}), because of this we say that the noise is ``colored''.

At the frequencies at which the noise power spectrum is bigger, large strain values will be less significant. Because of this, we will want to weigh down the strain at the parts of the frequency range with the higher values of the noise power spectrum. We will do this in such a way that the transformed strain would have a flat power spectrum if it contained only noise. This will be the whitened strain. The name comes from the fact that noise with a flat power spectrum is said to be ``white''. 

Since in Eq.~\eqref{eq:PSD_def} the power spectrum is defined proportional to the average of the modulus squared of the Fourier transform, the whitened strain will just be:

\begin{equation}
    \tilde{s}_{\text{whiten}}(f) = \frac{\tilde{s}(f)}{\sqrt{S_n(f)}} \, ,
    \label{eq:whiten_strain}
\end{equation}

\noindent where to get the whitened strain in the time domain we just compute the inverse Fourier transform of Eq.~\eqref{eq:whiten_strain}. 

\begin{figure*}[t!]
\begin{center}
\includegraphics[width=0.8\textwidth]{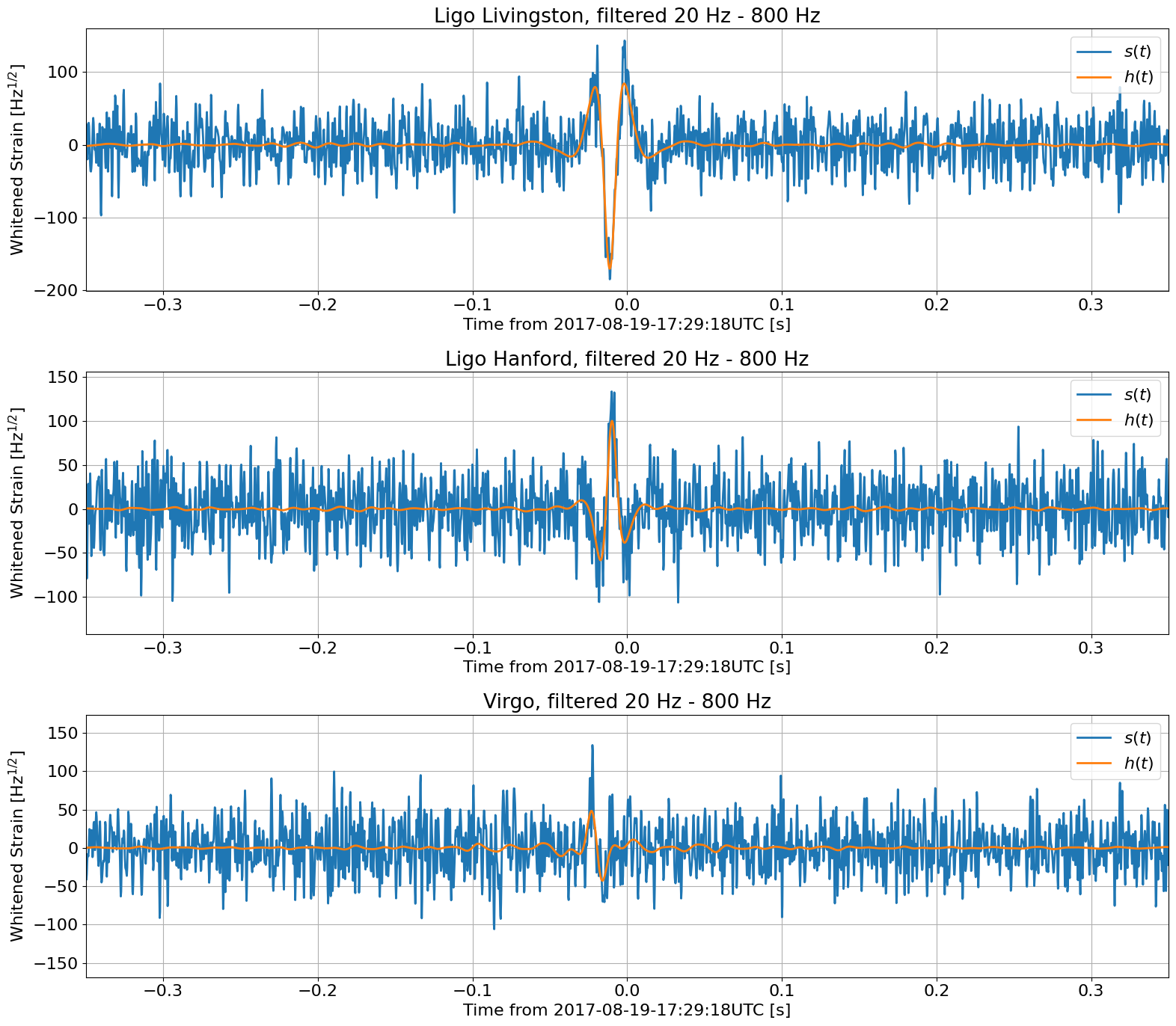}
\end{center} 
\caption{Result of whitening the signals of Fig.~\ref{fig:InjectedFiltered}.}
\label{fig:InjectedWhitened}
\end{figure*}

The result of doing this whitening to the filtered signals of Fig.~\ref{fig:InjectedFiltered} is shown in Fig.~\ref{fig:InjectedWhitened}. We can see that whitening the strain further enhances the signal with respect to the noise. Now the gravitational wave event can be clearly seen in the two LIGO detectors, even if we did not have the injected event to guide the eye. The situation in Virgo is also much more improved, since we have suppressed the frequencies that contained the most noise.

\subsubsection{Q transform}
\label{sec:detec:procesing:Qtransform}

Since gravitational waves from close hyperbolic encounters only perform one oscillation, they will not only have a characteristic spread in time, but they will also have a large characteristic spread in frequencies, as shown in Fig.~\ref{fig:ProyectedFrequency}. Because of this it will be interesting to study the signals in the time-frequency domain.

To convert our whitened data into the time-frequency domain we will use the Q transform \cite{Q_transform_LIGO}. The Q transform can be seen as a series of filters centered at different frequencies $f$, each having a bandwidth $\delta f$, such that the quality factors of the filters has a constant value Q \cite{Q_transform_Judith}:

\begin{equation}
    Q = \frac{f}{\delta f} \, .
    \label{eq:Q_definition}
\end{equation}

In practice, the way it will be computed will be by taking the Fourier transform of the whitened strain $\tilde{s}(\phi)$, applying a window centered at frequency $f$ with characteristic size $\delta f = Q/f$ and then computing the inverse Fourier transform, to obtain $s(t,f)$. That is:

\begin{equation}
    s(t,f) = \int_{-\infty}^{\infty} \tilde{s}(\phi) \tilde{w}(\phi,f) e^{i 2 \pi \phi t} d\phi \, ,
    \label{eq:Q_transform_definition}
\end{equation}

\noindent where one possible convention for the window function \cite{PyCBC}, that will be used during all this analysis is the bi-square window:

\begin{align}
    & \tilde{w}(\phi,f) = \sqrt{\frac{315 Q}{128\sqrt{11} f}} \left[ 1 - \left( \frac{\phi - f}{\sqrt{11} f/Q} \right)^2 \right]^2, \nonumber\\
    &  \text{with} \; \phi \in \left[ f - \frac{\sqrt{11} f}{Q} , f + \frac{\sqrt{11} f}{Q} \right] \, ,
    \label{eq:bi-square_window}
\end{align}

\noindent which satisfies the requirement of having $\delta f = f/Q$ and is normalized. For our purposes we will not be interested in the phase of the Q transformed strain, we will be mainly interested in the energy deposited at each point of the time-frequency domain, which will be defined as the modulus squared of Eq.~\eqref{eq:Q_transform_definition}. 

In addition, we will want to normalize this energy so that the noise floor is set at unity. This is done by dividing by the median value of the energy. The median value is well suited for this normalization because it will correctly estimate the noise floor without being very affected by possible large fluctuations in the Q transform due to large signals being present. We then have that the normalized energy in the time frequency-domain will be:

\begin{equation}
    P(t,f) = \frac{|s(t,f)|^2}{\text{median}(|s(t,f)|^2)} \, .
    \label{eq:Normalized_Energy}
\end{equation}

\begin{figure}[t!]
\begin{center}
\includegraphics[width=0.48\textwidth]{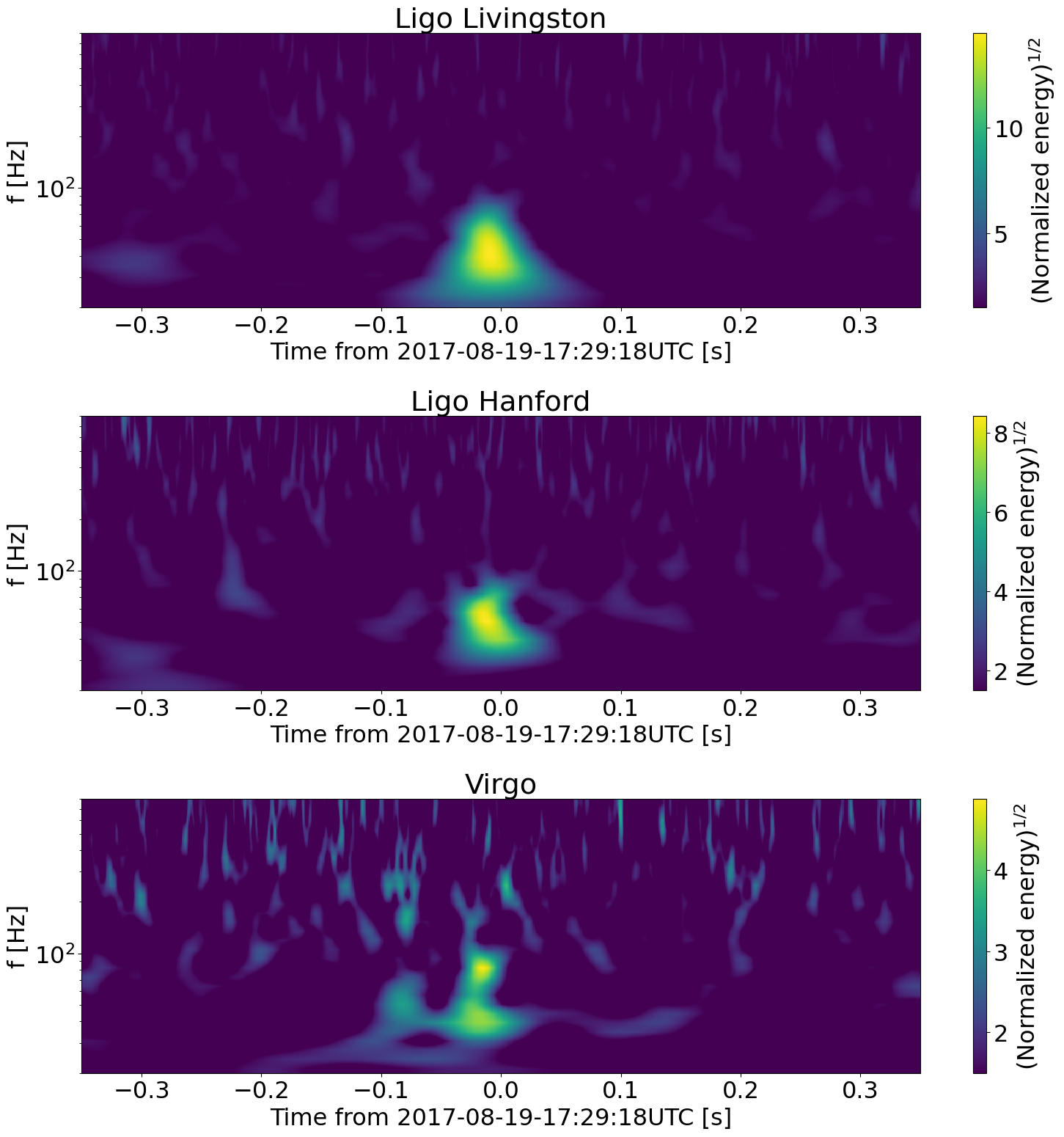}
\end{center} 
\caption{Square root of the normalized energy derived from the Q transforms of the whitened signals of Fig.~\ref{fig:InjectedWhitened}. The Q transform is computed using a quality factor $Q=8$.}
\label{fig:InjectedQtransform}
\end{figure}

In Fig.~\ref{fig:InjectedQtransform} we show the normalized energy computed from the whitened strain of our example shown in Fig.~\ref{fig:InjectedWhitened}. The Q transform of this example, as well as in the rest of the analysis, will be computed using a quality factor of $Q=8$ which was found to be the typical value that maximized the maximum of the normalized energy of Eq.~\eqref{eq:Normalized_Energy} when injecting close hyperbolic encounters within the parameter space studied (see appendix \ref{sec:anex:InjectionParams}) into simulated Gaussian noise from advanced LIGO at design sensitivity \cite{AdvLIGO_design}.

In this representation, the gravitational wave event clearly stands out above the noise in all three interferometers, and we can see its structure in the time-frequency domain. In general, gravitational waves from close hyperbolic encounters will look similar to the ones shown in Fig.~\ref{fig:InjectedQtransform}, the normalized energy is symmetric in time and at low frequencies is suppressed because of the large value of the power spectrum distribution by which we have divided when we whitened it. At large frequencies the normalized energy decays because as we saw in Sec.~\ref{sec:detec:procesing:filtering}, hyperbolic events are heavily skewed towards small frequencies. The final resulting shape of hyperbolic events is teardrop like, with the actual shape of the teardrop depending on the parameters of the close hyperbolic encounter as well as the detector sensitivity at that moment.

\subsubsection{Signal to noise ratio}
\label{sec:detec:procesing:SNR}

To characterize how loud a signal is with respect to the noise, the most commonly used metric in gravitational wave astronomy is the signal to noise ratio \cite{Maggiore_Vol1}. It is computed by looking at how much signal there is at each frequency and weighting it by the noise power spectrum. That is:

\begin{equation}
    \left( \frac{S}{N} \right)^2 = 4 \int_{f_{\text{min}}}^{f_{\text{max}}} \frac{|\tilde{h}(f)|^2}{S_n(f)}df \, ,
    \label{eq:SNR_def}
\end{equation}

\noindent where $\tilde{h}(f)$ is the Fourier transform of the gravitational wave signal projected into the detector and $f_{\text{min}}$ and $f_{\text{max}}$ are the lower and upper limits of the frequency range under study (in our case they would be 20Hz and 800Hz respectively).

With this definition, the signal to noise ratio of the example event that we have been analyzing throughout this section, shown in Fig.~\ref{fig:InjectedQtransform}, would be of 20.1 in Livingston, 11.1 in Hanford and 6.7 in Virgo. To get an idea of the significance of an event in a detector network, the signal to noise ratios in the different detectors are summed in quadrature \cite{Maggiore_Vol1}. That is:

\begin{equation}
    \left( \frac{S}{N} \right)^2_{\text{tot}} = \sum_{i} \left( \frac{S}{N} \right)^2_i \, .
    \label{eq:SNR_network}
\end{equation}

Using this formula, the total network signal to noise ratio of the example event, shown in Fig.~\ref{fig:InjectedQtransform}, would be 23.9.

\section{Search for close hyperbolic encounters in LIGO-Virgo data}
\label{sec:Data}

Now that in Sec.~\ref{sec:detec} we have determined the kind of signals that gravitational waves from close hyperbolic encounters leave in laser interferometers and how to process the data to make these signals stand out over the noise, we will want to look for them in available data of current detectors. 

For this purpose we will have to select the data to be analyzed, which will be done based on data quality considerations. We will look for gravitational waves from close hyperbolic encounters in the selected data using a two step trigger. The first step of the trigger will take an approach similar to standard burst searches \cite{cWB}, being a theory independent loose preselection of possible candidates based in correlations between detectors in the time-frequency domain. In the second step of the trigger we will use the templates developed in Sec.~\ref{sec:waveforms} to train a neural network to look at which of the preselected events look like gravitational waves from close hyperbolic encounters.

\subsection{Data selection and quality}
\label{sec:Data:Quality}

Gravitational wave astronomy benefits a lot from having multiple detectors online at the same time, since the more detectors a signal is seen in, the more confidence can be had that it was of astrophysical origin, as opposed of coming from a terrestrial noise source. Because of this we will limit our search to data taken with the three detectors (LIGO Livingston, LIGO Hanford and Virgo) operational and that is publicly available \cite{OpenData_O1_O2}. At the time the analysis was done, this limits our search to less than 4 weeks of data at the end of the second observing run O2, between 2017-08-01 and 2017-08-26.

\begin{figure*}[t!]
\begin{center}
\includegraphics[width=0.8\textwidth]{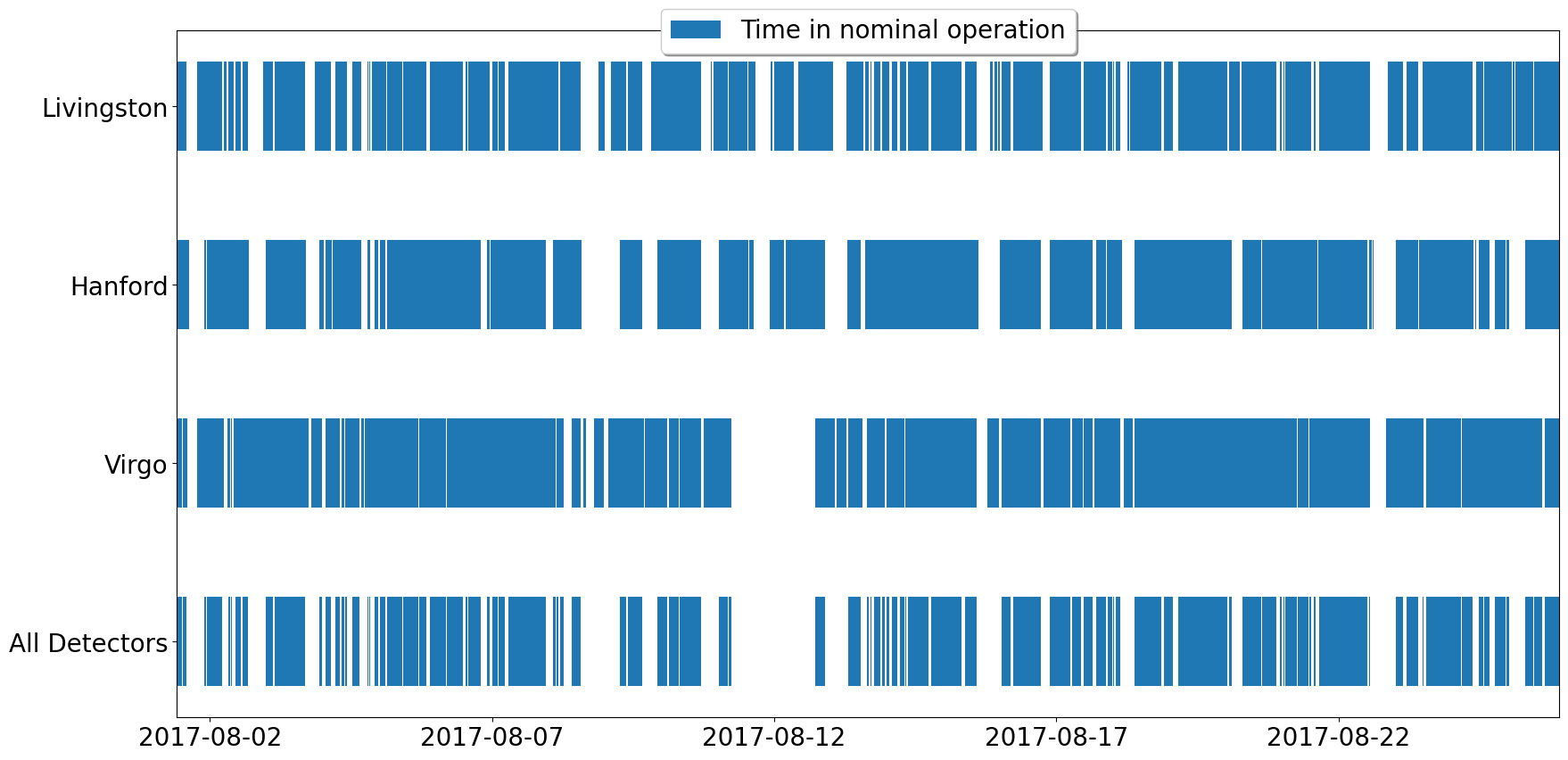}
\end{center} 
\caption{Times between 2017-08-01 and 2017-08-26 at which each detector (LIGO Livingston, LIGO Hanford and Virgo) was in nominal operation. At the bottom we show the times at which the three detectors were in nominal operation at the same time. These are the times that will be analyzed.}
\label{fig:DetectorInNominalOperation}
\end{figure*}

Nonetheless, laser interferometers are very delicate and sensitive machines, and because of it, the LIGO and Virgo detectors are not always in observing mode, and when they are, sometimes the data quality is not good enough for analysis. For our search we will only use the data in which the three detectors were operating under nominal conditions, that is, with no known hardware issues. The times that satisfy this condition are shown in the last row of Fig.~\ref{fig:DetectorInNominalOperation}, where we can see that the amount of data rejected for our analysis is quite large, fracturing the timeline into many segments of very different lengths. From the 24.5 days that the three detectors were measuring at the same time, only 15.3 days (62\% of data) will satisfy the data quality requirements and will be used for the analysis.

\subsection{Non-stationary noise: glitches }
\label{sec:Data:glitches}

Real laser interferometer data will not only contain stationary sources of noise as the ones discussed in Sec.~\ref{sec:detec:laser:noise}, but it will also contain some transient sources of noise. These are short duration perturbations in the strain measured by the detector that come from a terrestrial origin, usually because of environmental or instrumental factors.

\begin{figure*}[t!]
\begin{center}
\includegraphics[width=.95\textwidth]{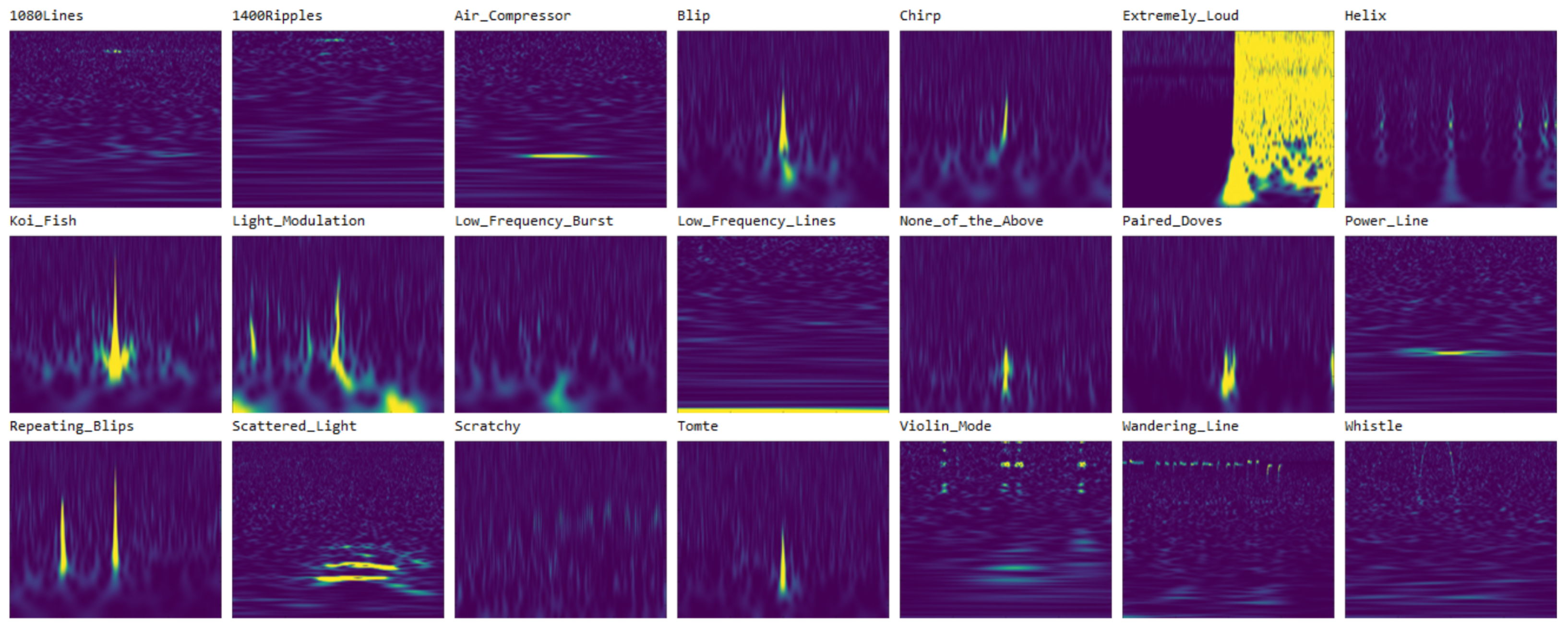}
\end{center} 
\caption{Different classes of glitches as classified by Ref.~\cite{Glitches}. The glitches are represented in the time-frequency domain following a signal processing very similar to the one used to make Fig.~\ref{fig:InjectedQtransform}. The x axis corresponds to time, the y axis to frequency and the color corresponds to the square root of normalized energy as defined in Eq.~\eqref{eq:Normalized_Energy}. Image taken from Ref.~\cite{Glitches}.}
\label{fig:Glitches}
\end{figure*}

Transient sources of noise are much harder to clean using the signal processing methods discussed in Sec.~\ref{sec:detec:procesing} because, since they are strain perturbations that have a small duration in time and are only observed once, in many ways they will behave like transient gravitational wave signals. The large transient noise events will be called ``glitches''. There are many different types of glitches, classified by their morphology. In Fig.~\ref{fig:Glitches} we show different classes of glitches as classified by Ref.~\cite{Glitches}. The glitches are represented in the time-frequency domain following a signal processing very similar to the one explained in Sec.~\ref{sec:detec:procesing} to make Fig.~\ref{fig:InjectedQtransform}. 

Some of the glitches, such as blip glitches, low frequency burst glitches or chirp glitches can look similar to the gravitational waves from close hyperbolic encounters we are looking for, and they will constitute the main source of background in our search.

\subsection{Image preselection: correlation trigger}
\label{sec:Data:presel}

As we have seen in previous sections, gravitational wave detectors will have a lot of noise coming from many different terrestrial sources. Nonetheless, because the detectors are hundreds of kilometers apart and do not interact with each other, we would expect that their noise will be in general uncorrelated. 

In opposition to this, whenever a gravitational wave passes through the Earth, it will leave a simultaneous imprint in all the detectors (within the milliseconds of light travel time between detectors). Because of this, gravitational waves will induce correlations between the measurements of the different detectors. We will exploit this feature to generate a theory independent trigger to look for gravitational wave candidates.

Even though the antenna factors will change the amplitude and phase of the gravitational wave strain observed in different detectors, the duration and frequency of the signal will not change, and we expect to see an excess of normalized energy in the same region of the time-frequency domain for the three detectors. 

We will thus look for positive correlations between the normalized energies of the detectors, that is, excess normalized energy in the same time-frequency region (within the light travel time between detectors) in the three detectors. As we will see, this method works well to look for gravitational waves from close hyperbolic encounters since in this case the strain only performs one oscillation and because of this, the normalized energy has a large area in the time-frequency domain. 

The correlation between the normalized energy of two different detectors will be quantified using Pearson correlation coefficient $r$ \cite{PearsonCorrCoef}. If $x_n $ and $y_n$ ($n = 0,..., N-1$) are the two vectors between which we want to compute the correlation, their Pearson correlation coefficient will be defined as:

\begin{equation}
    r[x,y] = \frac{\text{cov}(x,y)}{\sigma_x \sigma_y} = \frac{\sum_{n=0}^{N-1}(x_n - \overline{x})(y_n - \overline{y})}{\sqrt{\sum_{n=0}^{N-1}(x_n - \overline{x})^2}\sqrt{\sum_{n=0}^{N-1}(y_n - \overline{y})^2}} \, .
    \label{eq:Pearson_def}
\end{equation}

This coefficient satisfies that if $x_n = |\alpha|y_n$, then $r = 1$ and if $x_n = -|\alpha|y_n$, then $ r = -1$. If there is no correlation between $x$ and $y$, then $r = 0$. In general the more positive correlation there is, the closer to 1 that $r$ will be.

To create a trigger implementing the correlation between detectors, we will first apply the signal processing discussed in Sec.~\ref{sec:detec:procesing}. That is, we will band-pass filter the signal in a 20-800Hz range, whiten it, and compute the normalized energy in the time-frequency domain using the Q transform.

Since Livingston is the detector with the best sensitivity, specially at low frequencies where hyperbolic encounters deposit most of their energy, we will use it as the reference interferometer, and we will compute the correlation of Hanford and Virgo with it.

We will divide Livingston's normalized energy in images such as the one shown in Fig.~\ref{fig:InjectedQtransform} that will be 0.3 seconds in length and creating an image every 0.15 seconds (50\% overlap). The images will be 200x200 pixels in size. Once we have the image in Livingston defined, we generate corresponding images at the same time at Hanford and Virgo, allowing for small time shifts due to the light travel time ($\pm 0.010$s with Hanford and $\pm 0.026 s$ with Virgo).

We then compute the correlation between the normalized energy in the pixels of the images using Pearson's correlation coefficient of Eq.~\eqref{eq:Pearson_def} and we pick the Hanford and Virgo time shifts that yield the highest value of the coefficient, since we are looking for positive correlations between the normalized energies. This yields a value for the correlation between Livingston and Hanford $r_{L1-H1}$ and between Livingston and Virgo $r_{L1-V1}$ \footnote{It is customary to denote the Livingston detector by L1, the Hanford detector by H1 and the Virgo detector by V1. Nowadays there is only one detector in each site, so the number is redundant. But there used to be more detectors in the midpoints of the detector arms, leading to L2, H2...}. The trigger will accept as candidate events the images that satisfy:

\begin{equation}
    D = \frac{2}{3} r_{L1-H1} + \frac{1}{3} r_{L1-V1} > 0.3 \, ,
    \label{eq:discriminant_formula}
\end{equation}

\noindent where $D$ is the discriminant of the trigger. The factors of $\frac{2}{3}$ and $\frac{1}{3}$ were chosen to give a higher weight to the correlation with Hanford, since this detector has better sensitivity than Virgo and because of it, the correlation of Livingston with Hanford is more likely to have a gravitational wave origin. With this correlation trigger we wanted to have a preselection of gravitational wave candidates to then feed to the neural network that will do the final clasification. We put the minimum value of the discriminant $D$ to be accepted at 0.3 to reject a large amount of the noise that is randomly correlated between detectors while accepting most of the loud enough gravitational waves.

This choice of parameters in Eq.~\eqref{eq:discriminant_formula} is discussed in greater detail in appendix \ref{sec:anex:PearsonCoef}.

\subsubsection{Trigger validation with injections}
\label{sec:Data:presel:Validation}

To test how the trigger works at detecting gravitational waves from close hyperbolic encounters, we will inject signals with random parameters (as specified in Table \ref{table:InjectParam} of appendix \ref{sec:anex:InjectionParams}) at random times in the experimental data. We will then run the trigger on the data containing these injections to study how many of them we are able to detect.

\begin{figure*}[t!]
\centering
\includegraphics[width=0.48\textwidth]{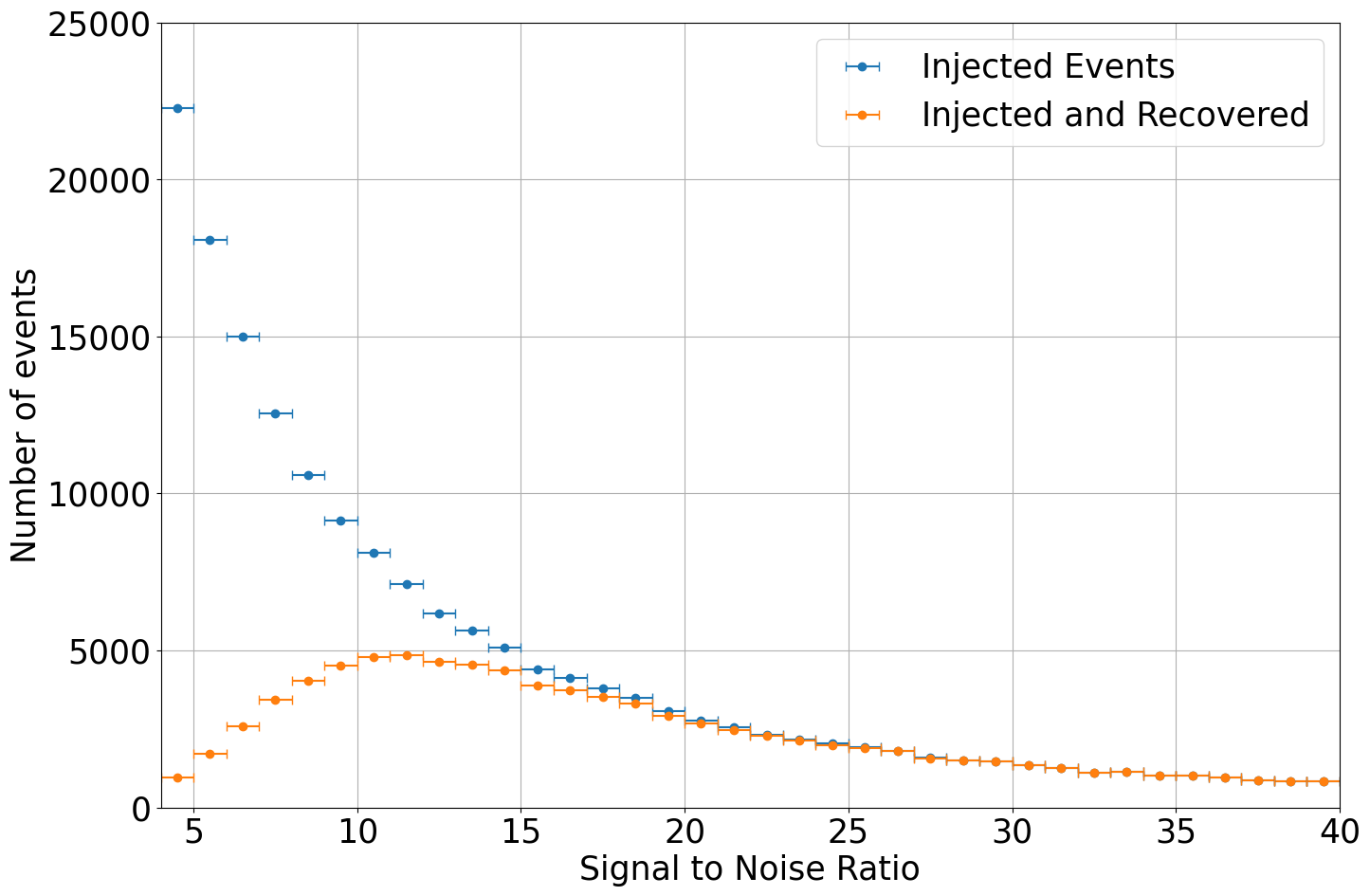}
\includegraphics[width=0.48\textwidth]{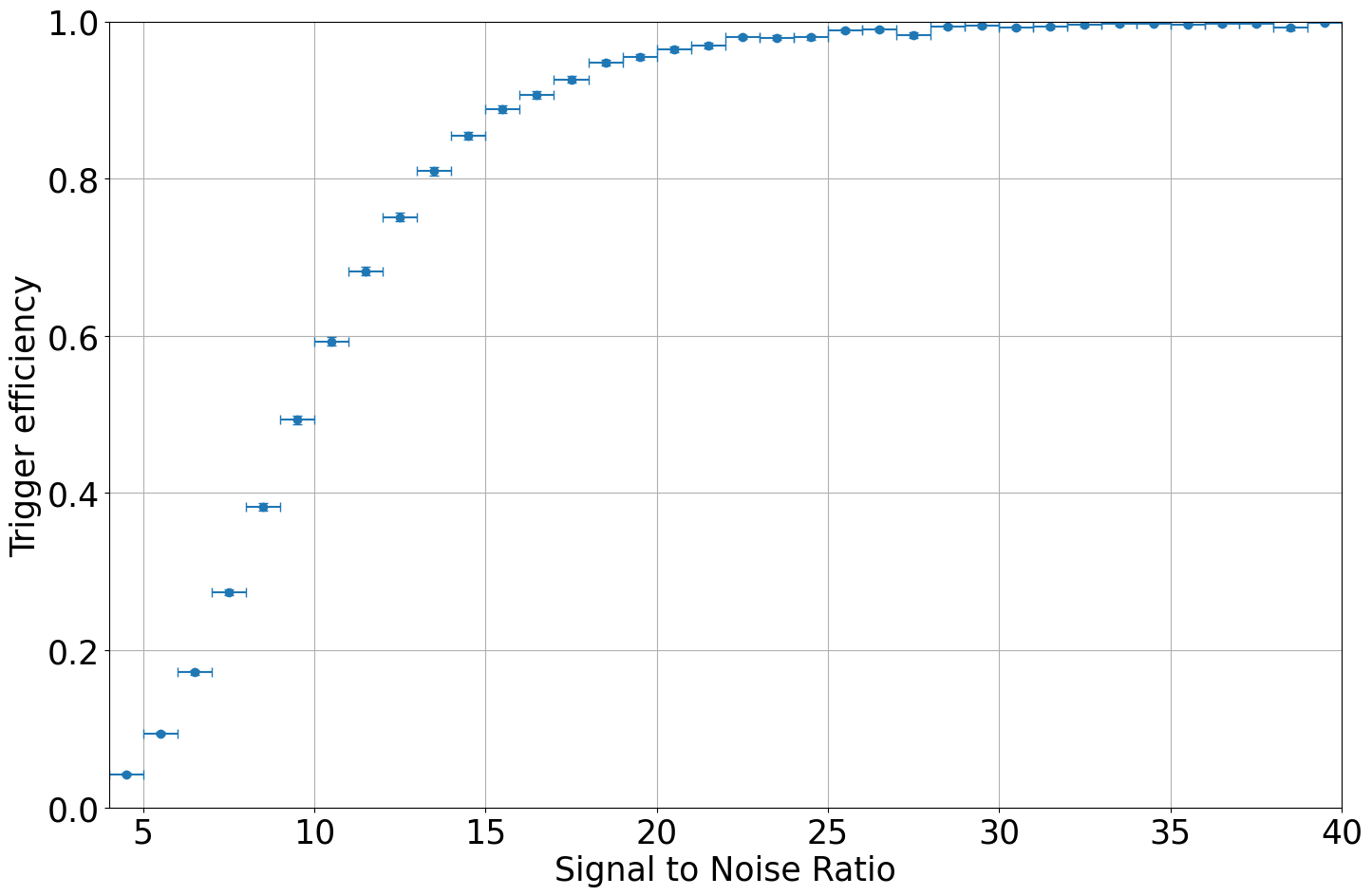}	
\caption{Results of the correlation trigger running on data containing injected events with random parameters as specified in Table \ref{table:InjectParam} of appendix \ref{sec:anex:InjectionParams}. Left: Number of events injected and number of events injected and recovered as a function of the signal to noise ratio. Right: Trigger efficiency as a function of the signal to noise ratio of the injected events.}
\label{fig:TrigerInjected}
\end{figure*}

We will inject a total of 169108 signals with total signal to noise ratio between 4 and 40. The distribution of the injections with the signal to noise ratio is shown in Fig.~\ref{fig:TrigerInjected}, where we see that it follows a power law, in accordance with what is predicted in the literature \cite{SNR_PowerLaw} for any gravitational wave source with randomly generated parameters. We have also plotted in Fig.~\ref{fig:TrigerInjected} the events that are recovered from the injections. To see how good the trigger is at recovering events as a function of the signal to noise ratio, in Fig.~\ref{fig:TrigerInjected} we have plotted the efficiency of the trigger, defined as the fraction of injections that are recovered at a given signal to noise ratio. 

\begin{figure}[t!]
\centering
\includegraphics[width=0.48\textwidth]{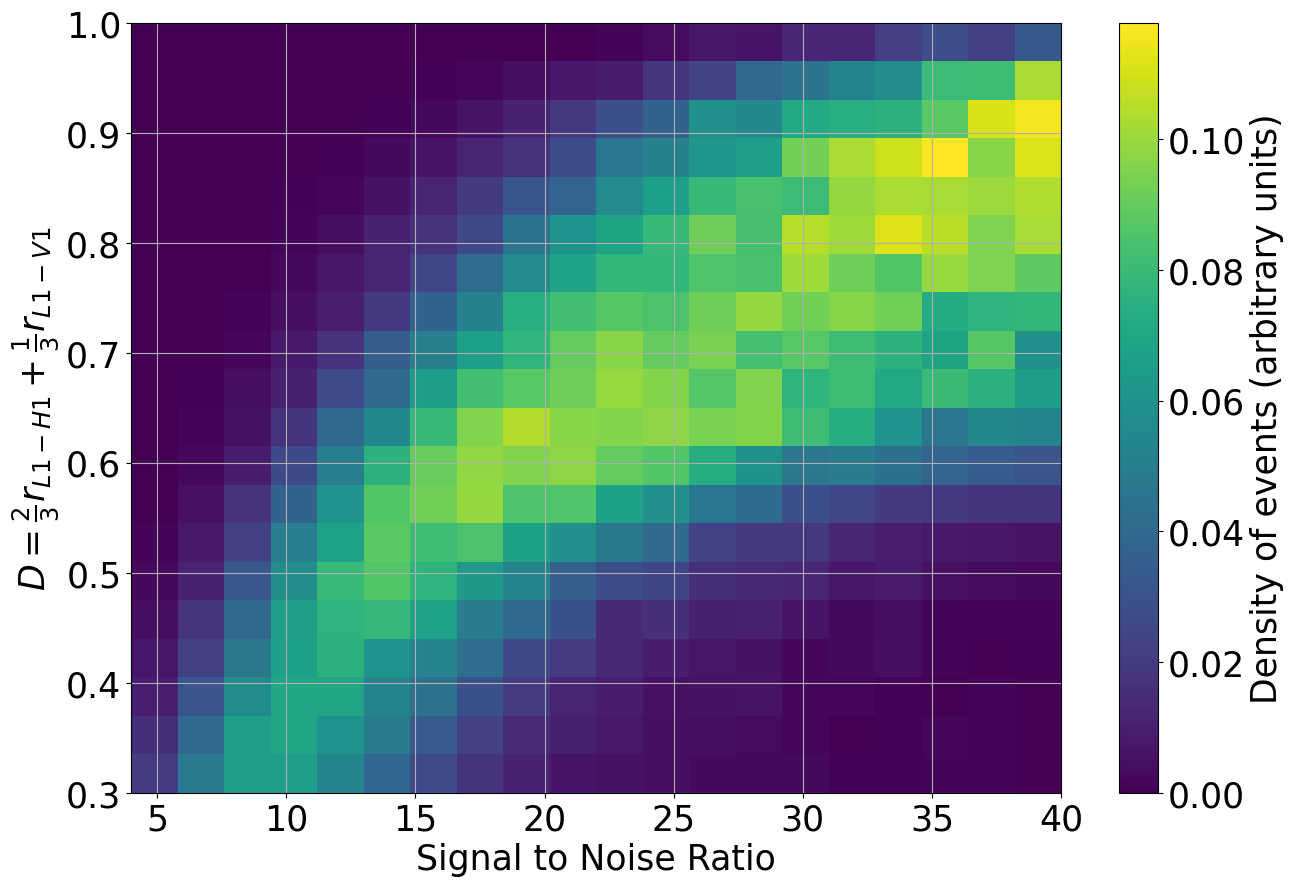}
\caption{Two dimensional histogram of the trigger discriminant and the signal to noise ratio for injected events. To obtain the trigger discriminant as a function of the signal to noise ratio and mitigate the effects of injecting different number of events at different signal to noise ratios, we have weighted each event by the inverse of the number of events injected at its signal to noise ratio.}
\label{fig:TrigerInjected2}
\end{figure}

At low values of the signal to noise ratio we recover only a small fraction of the injected events, because they are not loud enough to induce strong correlations between the detectors. As the signal to noise ratio of the events increases, the efficiency increases because the louder an event is, the more correlations it will induce and the bigger the discriminant will be. Above a signal to noise ratio of 10, we recover more than $50\%$ of the injected events.

The discriminant as a function of the signal to noise ratio is shown in Fig.~\ref{fig:TrigerInjected2}. We can see that the discriminant monotonically increases as the signal to noise ratio increases. Nonetheless, it has quite a large spread, this is partly due to the effects that the different antenna factors of the interferometers have. For example if the wave comes from the blind spot of one interferometer, the correlation between this interferometer and the others will be suppressed. In addition, if the gravitational wave signal happens close to a large fluctuation of noise in a detector, since this large fluctuation of noise will not appear in the other detectors, it will diminish the correlation. This trigger will also have different sensitivities towards events with different parameters. Events at low frequencies will be easier to detect, since these low frequencies also correspond to longer times and the correlated area in the time-frequency domain will be larger.

Nonetheless, after testing the trigger with injections we can confidently say that the basic idea works for gravitational waves from close hyperbolic encounters, since the discriminant monotonically increases as the loudness of the event increases and it is able to recover most of the events above a signal to noise ratio of 10.

As an additional validation of the trigger, we note that in the 15.3 days of data we are analyzing, there were 4 gravitational wave signals claimed by the LIGO-Virgo collaboration \cite{GWTC-1}, all of them coming from the coalescence of compact binary objects. Since the correlation trigger we have developed is theory independent, it should also recover these events.

The result of checking the output of the trigger at the time of these events is shown in Table \ref{table:CBC_correlation}. Our trigger recovers all events except GW170818. This event is not recovered because it was mostly seen in Livingston, since it was close to the blind spots of Virgo and Hanford, suppressing the correlations with these interferometers.

We also note that even though GW170817 has a very large signal to noise ratio of 33, the trigger discriminant is not as high as we would expect from Fig.~\ref{fig:TrigerInjected}, this is because this event is a binary neutron star merger \cite{GW170817} which happens at very high frequencies and with a very large quality factor. Because of this, the area of its normalized energy will be small, making the discriminant of our trigger smaller.

\begin{table}[t!]
\centering
\begin{tabular}{c|c|c|c}
Event Name  & S/N & $D = \frac{2}{3} r_{L1-H1} + \frac{1}{3} r_{L1-V1}$ & Triggered \\  
\hline
GW170809 & 11.0 & 0.337 & Yes \\
\hline
GW170814 & 15.9 & 0.469 & Yes \\
\hline
GW170817 & 33.0 & 0.403 & Yes \\
\hline
GW170818 & 11.3 & 0.158  & No \\
\hline
\end{tabular}
\caption{Results of the correlation trigger on the confirmed gravitational wave signals present in the analyzed data.}
\label{table:CBC_correlation}
\end{table}

\subsubsection{Trigger false alarm rate}
\label{sec:Data:presel:FAR}

The correlation trigger that we have designed will not only accept gravitational wave events, but it will also accept noise that is randomly correlated between detectors. To estimate how many noise events will trigger on average, we will run our trigger in the LIGO-Virgo data, but we will shift the timeline of the Hanford and Virgo detectors by several seconds in such a way that if there are correlations between the data, they can not possibly come from gravitational waves (gravitational waves must be correlated within milliseconds). 

By shifting the timeline of the detectors by different amounts we can get an almost arbitrarily large amount of ``noise'' data. To study how the trigger reacts to the randomly correlated noise in this data, the main quantity that we will look at is the false alarm rate (FAR). The false alarm rate quantifies how many events with trigger discriminant $D$ higher than some value we can expect per unit of observing time. That is:

\begin{equation}
    \text{FAR}(x) = \frac{N_{\text{events}}(D>x)}{T_{\text{noise}}} \, ,
    \label{eq:FAR_def}
\end{equation}

\noindent where $N_{\text{events}}(D>x)$ is the number of events with trigger discriminant higher than $x$ we observe in the noise data, and $T_{\text{noise}}$ is the duration of the noise data.

\begin{figure}[t!]
\begin{center}
\includegraphics[width=0.45\textwidth]{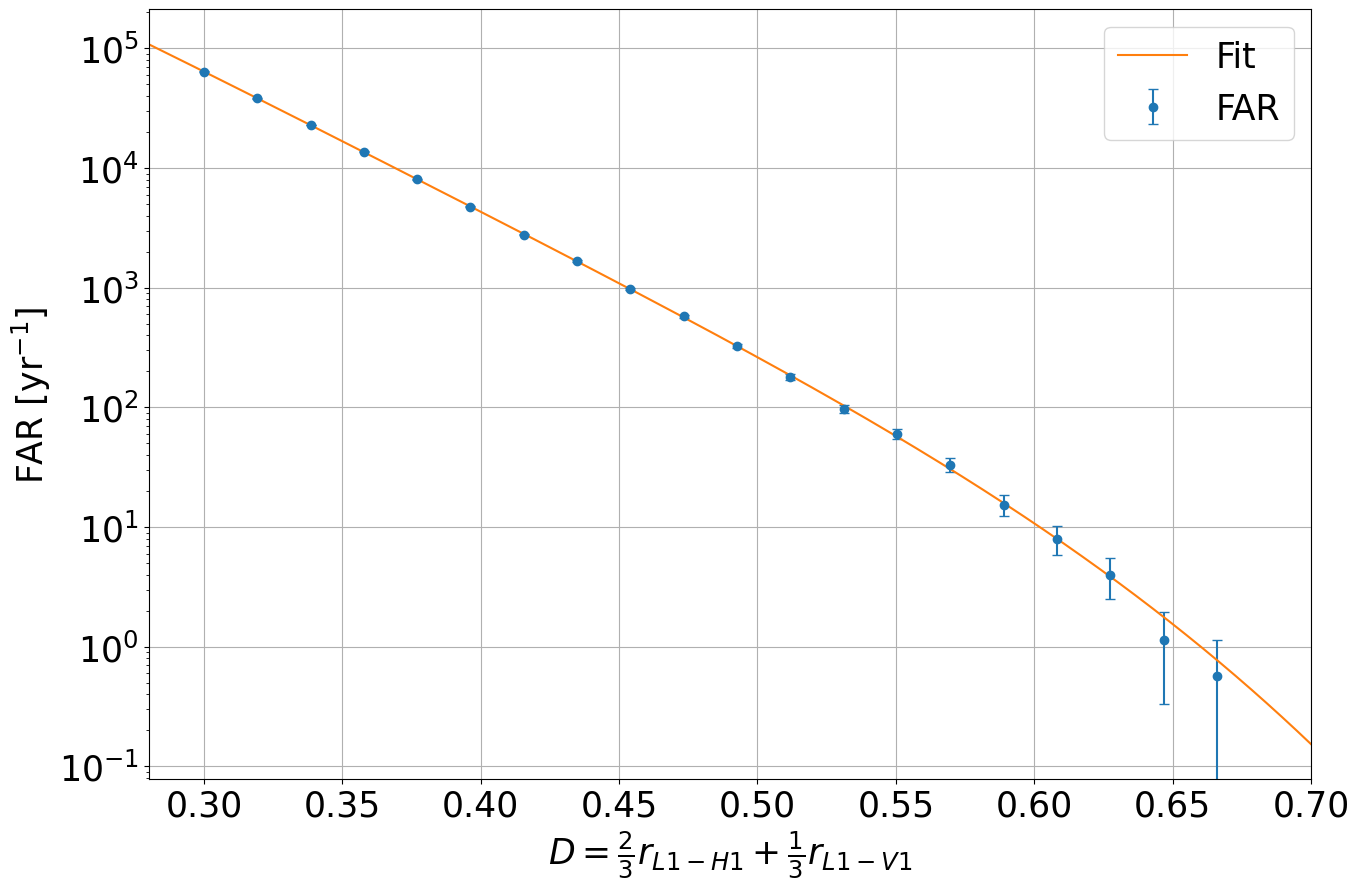}
\end{center} 
\caption{False alarm rate as a function of the discriminant for 1.75yr of noise data. The error bars are set assuming Poisson statistics, ${\Delta N_{\text{events}}(D > x) =  \sqrt{N_{\text{events}}(D > x)}}$. We have fitted a smooth curve to guide the eye.}
\label{fig:FAR}
\end{figure}

To study the false alarm rate with enough statistics, in total we have used the 15.3 days of true data to obtain 642 days ($T_{\text{noise}} = 1.75$ yr) of noise data by shifting the timelines of Hanford and Virgo by different amounts. The result for the false alarm rate obtained is shown in Fig.~\ref{fig:FAR}, where we have fitted it to a smooth curve to guide the eye. We can observe that the value of the false alarm rate exponentially decreases as the discriminant increases. This is what we would expect of a well behaved trigger, events with high values of the discriminant are increasingly more rare because randomly having very strong correlations between detector noise is unlikely to happen.

Having the false alarm rate, we can estimate how many events we could expect to see above a certain discriminant as a function of the observing time $T_{\text{observed}}$:

\begin{equation}
    N_{\text{expected}}(D>x) =\text{FAR}(x) T_{\text{observed}} \, .
    \label{eq:N_expected}
\end{equation}

\begin{figure}[t!]
\begin{center}
\includegraphics[width=0.45\textwidth]{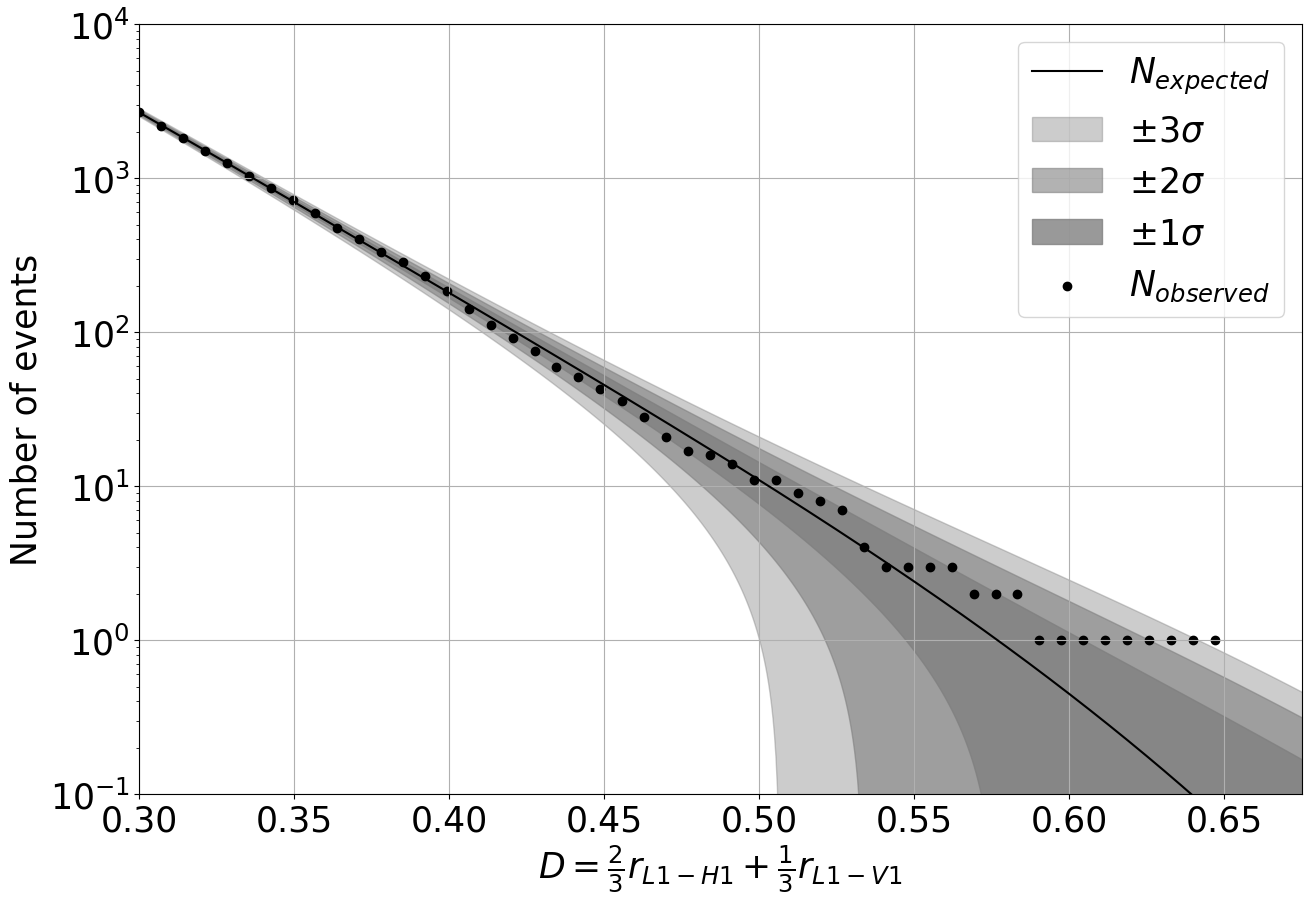}
\end{center} 
\caption{Number of events above a certain discriminant observed and expected. Since we are assuming Poisson statistics for the number of expected events, $\sigma = \sqrt{N_{expected}}$.}
\label{fig:Nobs_vs_Nexp}
\end{figure}

This expected number of events can be compared with the number of events one observes when running the trigger in the data without applying any time shifts. When doing this we got 2704 observed events having the distribution with the trigger discriminant shown in Fig.~\ref{fig:Nobs_vs_Nexp}.

We can see that there is a more than 3$\sigma$ excess in the observed events at large correlation values. This is due to one event with a discriminant value of $D = 0.647$, which from Fig.~\ref{fig:FAR} we can see it corresponds to a false alarm rate of 1.5 yr$^{-1}$, that is, we expect one such event every 240 days, but we got one in 15.3 days. 
 
\begin{figure*}[t!]
\begin{center}
\includegraphics[width=0.95\textwidth]{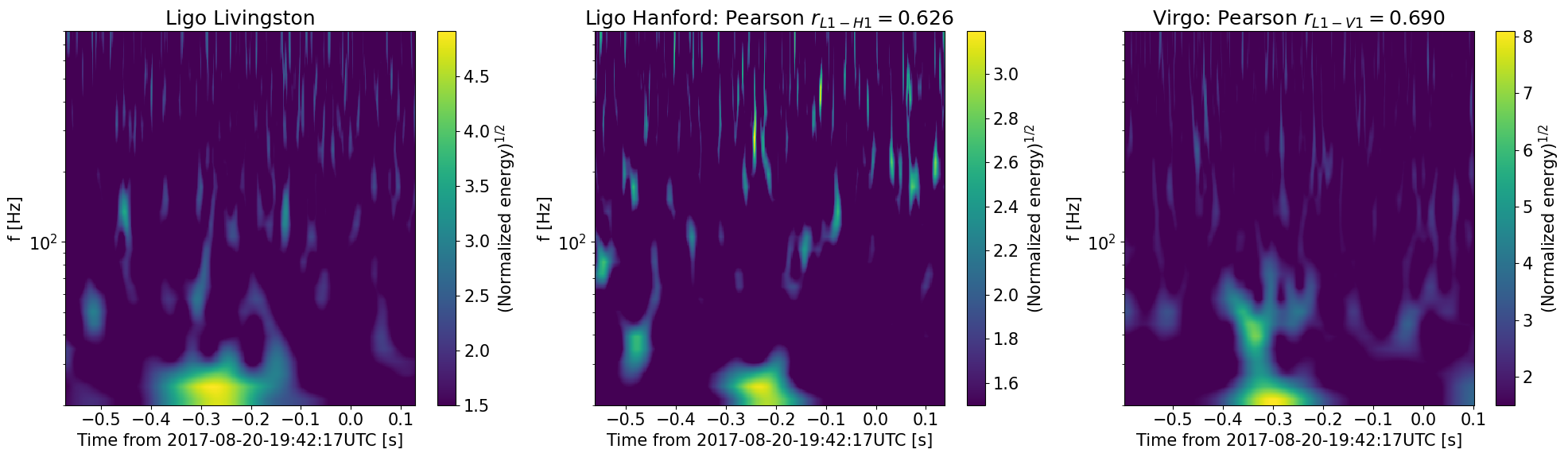}
\end{center} 
\caption{Event with the highest value of the trigger discriminant found in the analyzed data, with a value  $D  = 0.647$. From our false alarm rate analysis we expect an event like this every 240 days but we got one in 15.3 days of data.}
\label{fig:3sigma_event}
\end{figure*}

This event is shown in Fig.~\ref{fig:3sigma_event}. Even though it has a very large value of the correlation, this does not necessarily mean it will be a gravitational wave signal. As was discussed in Sec.~\ref{sec:detec:procesing:Qtransform}, we would expect Virgo to have in general the smallest normalized energy, since it has by far the largest noise power spectrum distribution and we are dividing by this quantity when whitening the signal. Nonetheless, in this case it has the largest normalized energy. This could be due to the fact that the gravitational wave is coming from close to a blind spot of Hanford and Livingston but right in the most sensitive part of the sky for Virgo. To confidently determine whether or not this triggered event, as well as all the others, can come from gravitational waves of close hyperbolic encounters, we will need to further analyze the triggered events with a method that takes into account the properties of the signal. This will be done in next section using a neural network.

\subsection{Neural network}
\label{sec:Data:CNN}

To determine which of the 2704 events that were selected by the correlation trigger could have been produced by gravitational waves coming from close hyperbolic encounters, we will need to compare them with the theoretical predictions developed in Secs.~\ref{sec:waveforms} and \ref{sec:detec}. This will be done by training a neural network to classify the images of the normalized energy selected by the correlation trigger into two classes, ``noise'' images and ``CHE'' (close hyperbolic encounter) images. 

To do this we will first have to specify the characteristics of the images we want to classify, and based on this we will design the architecture of our neural network. We will then have to train this neural network and test and validate its performance. Finally, when we are confident on the validity of the approach, we will classify the data images and find the most promising close hyperbolic encounter candidates.

\subsubsection{Image samples}
\label{sec:Data:CNN:Samples}

We will need to generate image samples to train and validate our neural network as well as to make predictions on the data. Careful considerations need to be taken to generate all the images in the same way so as to not introduce biases in our samples. This is done by running the same image generation code on different sets of data.

\begin{figure}[t!]
\centering
\includegraphics[width=0.15\textwidth]{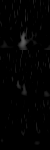}
\label{fig:ExampleSamples:Injections}
\includegraphics[width=0.15\textwidth]{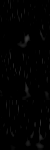}
\label{fig:ExampleSamples:Noise}
\includegraphics[width=0.15\textwidth]{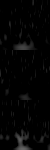}
\label{fig:ExampleSamples:Data}
\caption{Examples of images used to train and validate our neural network as well as to make predictions on the data. Left: We show a gravitational wave with a signal to noise ratio of 10.9 that was injected in data where the Hanford timeline had been shifted by +2s and the Virgo one by -1s. Center: We show an event that triggered when running the program on the data where the Hanford timeline had been shifted by +5s and the Virgo one by -5s. Right: We show the same candidate of Fig.~\ref{fig:3sigma_event} that was the most correlated event we found when running the trigger on the untouched data.}
\label{fig:ExampleSamples}
\end{figure}

The code in question will run the correlation trigger described in Sec.~\ref{sec:Data:presel} and for each event that triggers it will generate an image centered around that event, such as the ones shown in Fig.~\ref{fig:ExampleSamples}. 

The images are $50 \times 150$ pixels in size because we vertically stack a $50 \times 50$ pixel image of each of the three detectors into the same image (Livingston in the first row, Hanford in the second row and Virgo in the bottom). This is done to feed the information about the three detectors at the same time to the neural network, and in this way be able to classify the event with all the available information. For each detector, the x axis corresponds to 0.7 seconds around the trigger time, while the y axis corresponds to logarithmically spaced frequencies between 20 and 800 Hz. 

With the pixel color we represent the natural logarithm of the normalized energy on a fixed scale between 1 and 9. This fixed scale is desirable for a neural network, since it gives an absolute meaning to the color. We put the minimum value of the scale at 1 (which corresponds to normalized energy 2.7), above the noise floor to avoid confusing the neural network with the many small fluctuations around this noise floor. The natural logarithm is chosen to increase the dynamic range we can fit in the image. Finally, the image will be represented in grayscale because we are only considering one quantity, the normalized energy, which can be represented in grayscale using only one channel. This grayscale will reduce the memory requirements by a factor of 3 with respect to using a colored ``RGB'' representation.

The noise samples are generated by running the trigger and image generator on the same 1.75yr of data of Sec.~\ref{sec:Data:presel:FAR} with the Hanford and Virgo detectors shifted in time. We obtain 112153 noise images from this shifted data, from which 64028 (corresponding to 1yr of data) will be used to train the neural network and 48125 (corresponding to 0.75yr of data) will be used for validation.

The close hyperbolic encounter (CHE) samples are generated by running the trigger and image generator on data containing the same 169108 injections of Sec.~\ref{sec:Data:presel:Validation}, with signal to noise ratios between 4 and 40 and whose parameters are randomly generated as specified in Table \ref{table:InjectParam} of appendix \ref{sec:anex:InjectionParams}. Before injecting the signals, the data of the Hanford and Virgo detectors is shifted in time, to avoid having the same background as for the true data samples. We will use different time shifts from the ones used to generate the noise images. In this way we obtain 89597 CHE images from the shifted data with injections, from which 45356 will be used to train the neural network and 44241 will be used for validation.

The data samples are generated by running the trigger and image generator on the synchronized LIGO-Virgo data. This will yield 2704 images corresponding to the accepted events of the first level trigger that we want to classify.

\subsubsection{Neural network design}
\label{sec:Data:CNN:design}

Neural networks are mathematical models used to tackle complex data analysis problems with Machine Learning. The design of the neural networks has been motivated by the functioning of the brain and just like in the brain, the basic building block of a neural network is the neuron \cite{NNreview}. In this context, a neuron is a function that takes a series of inputs $\vec{x}$, weighs them with  a series of weights $\vec{w}$, adds a bias $b$ and evaluates the result on an activation function $\phi$. That is:

\begin{equation}
    y = \phi(\vec{w} \cdot \vec{x} + b) \, ,
    \label{eq:Neuron_def}
\end{equation}

\noindent where $\vec{x}$ and $\vec{w}$ are vectors of $\mathbb{R}^n$, with $n$ being the number of inputs of the neuron. For the activation function different choices can be made. Some frequently used activation functions are shown in Table \ref{table:NeuronActivation}.

\begin{table}[t!]
\centering
\begin{tabular}{c|c}
Activation function & $\phi(z)$ \\
\hline
& \\[-.67em]
Linear & L(z) = z \\[.5em]
\hline
& \\[-.67em]
Sigmoid & $S(z) = \frac{1}{1+e^{-z}}$ \\[.5em]
\hline
& \\[-.67em]
ReLu & $R(z) = 
\begin{cases}
       0 &\; z < 0\\
       z &\; z \geq 0\\
\end{cases} $\\[1.5em]
\hline
\end{tabular}
\caption{Common neuron activation functions.}
\label{table:NeuronActivation}
\end{table}

\begin{figure}[t!]
\begin{center}
\includegraphics[width=0.42\textwidth]{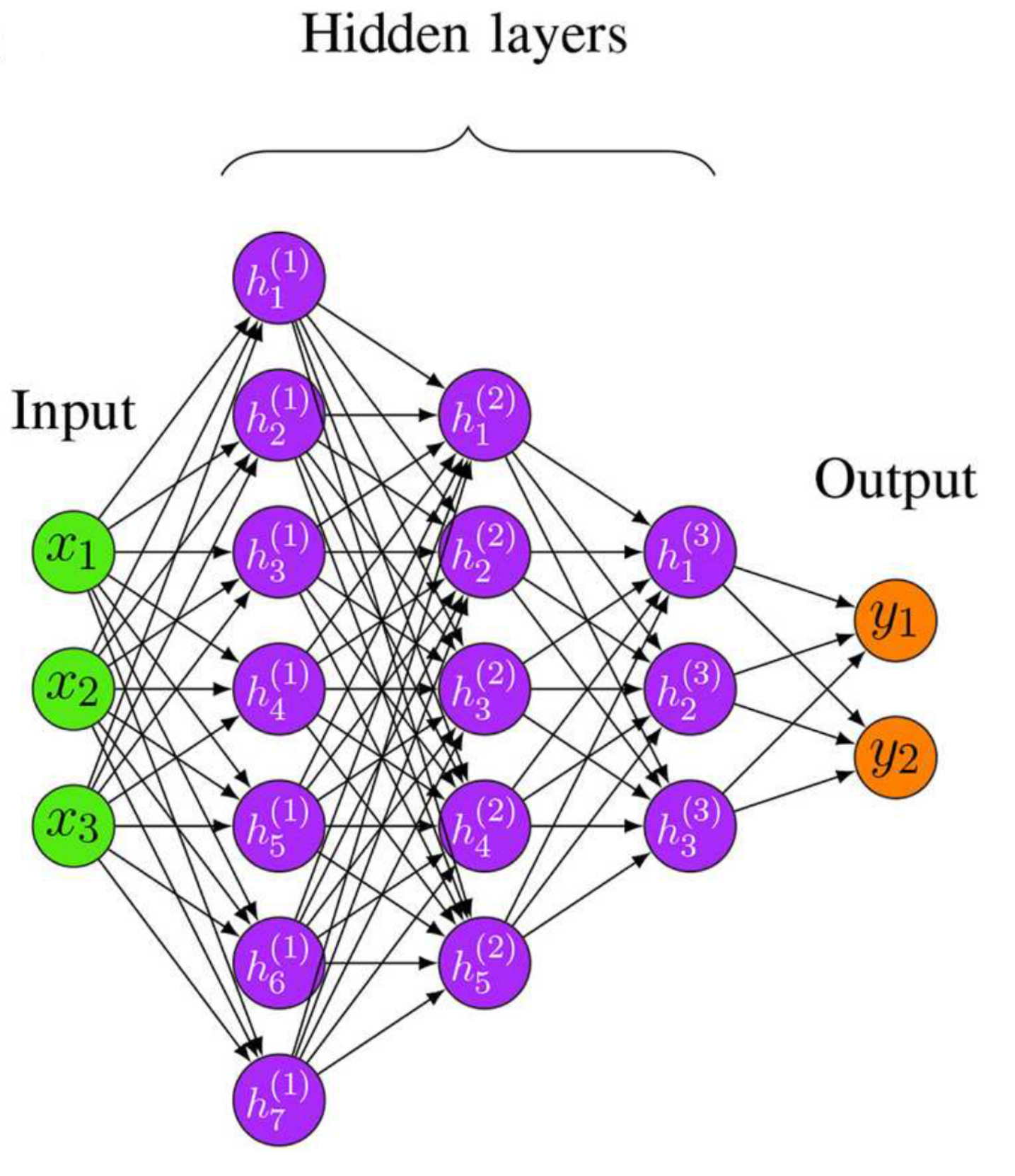}
\end{center} 
\caption{Example of a Feedforward Neural Network with 3 hidden layers. Image taken from Ref.~\cite{NNreview}.}
\label{fig:FeedforwardNeuralNetwork}
\end{figure}

To form a neural network, neurons have to be connected with each other. This can be done in many ways, but for our analysis we will focus on Feedforward Neural Networks as the one shown in Fig.~\ref{fig:FeedforwardNeuralNetwork}. In this type of neural network, neurons are grouped into layers. The output of the neurons of each layer is fed as an input for the neurons of the next layer.

The first layer of the network is called the input layer, since it represents the data we want to feed to the neural network to analyze. In the case of image recognition, this input corresponds to the values of the image pixels. 

The last layer of the network is the output layer, which will give the result of the neural network. In a problem of binary classification we will only need one output, between 0 and 1, that will represent the probability for the input image to come from a close hyperbolic encounter.

The neurons in the layers between the input and the output are used only for the internal calculations of the network. This is why these layers are called hidden layers.

At initialization, we will have to set the connections and the activation functions of the neurons. Depending on the design of the neural network, some of the weights and biases can also be set at initialization, but usually a large amount of weights and biases are left as free internal parameters. These internal parameters will be tuned during the training process to optimize the network at fitting the training data. 

The best suited type of network for image classification is the Convolutional Neural Network, which will be the one that will be used in our analysis to classify the data images. Convolutional Neural Networks are a type of Feedforward Neural Network characterized by the fact that neurons of one layer are not connected with all the neurons of the next layer, but they are only connected with the nearby neurons. This is desired because in images, pixels are usually locally very correlated, that is, their value strongly depends on nearby pixel values. Instead of analyzing all the pixels at the same time without considering their position in the image, with Convolutional Neural Networks we will extract local patterns in progressively large scales of the image and relate them to each other. This local connectivity will greatly reduce the number of free parameters and make the training more simple and generalizable. To construct a Convolutional Neural Network we will make use of three different types of layers \cite{NNreview}. 

The first type of layer we will need will be the Convolutional layer. This layer will be defined by a number $K$ of ``kernels'', each one of which is a matrix of size $N \times N$ whose parameters will be determined during training. The neurons of the convolutional layer will perform the convolution operation between one of the $K$ kernels and a $N \times N$ region of the previous layer matching the kernels size. We will also need to specify the stride $S$, which is the number of pixels in the previous layer by which the kernel center is moved to perform the next convolution. If the stride is greater than 1, it can be used to make the output of the convolutional layer smaller than its input and thus compress the information. To avoid miss-match effects we can define some zero-padding $P$, to pad zeros around the border of the input. Additionaly, a non-linear activation function such as ReLu or Sigmoid (Table \ref{table:NeuronActivation}) can be applied to the result of the convolution.

The fact that neurons in a convolutional layer share the same weights through the kernels greatly reduces the number of free parameters. The sharing of weights means that the convolutional layer will extract local pixel patterns in a location independent way. This is useful because usually local pixel values are highly correlated (in our case they vary smoothly) and their individual values do not matter as much as the local pattern. Additionally, the relevant features we are looking for are usually location invariant in the image. When the neural network is optimized, it finds the kernels that highlight the most important features for the next layers.

Convolutional Neural Networks will also make use of Pooling layers, which are usually placed after convolutional layers to reduce the dimension of their output conserving as much information as possible. Pooling layers work in a similar way to convolutional layers, having a stride $S$, a zero-padding $P$ and a pooling window size $N \times N$ that need to be specified. Nonetheless, the pooling operation is specified, rather than learned. The most common pooling methods are averaging over the pooling window or computing the maximum on each pooling window. To reduce the dimension of the input conserving as much information as possible, usually the stride has a value larger than 1, and the window size is larger than the stride. Pooling can introduce spatial invariance to the neural network, since pooling operations such as averaging or finding the maximum do not depend on the positions of the inputs inside the pooling window.

Lastly, Convolutional Neural Networks will also have fully connected layers, which are layers like the ones shown in Fig.~\ref{fig:FeedforwardNeuralNetwork}, where the neurons of the fully connected layer are connected with all neurons of the previous layer. After specifying an activation function for the neurons, all weights and biases are free to vary independently. These layers offer great learning flexibility because of the many degrees of freedom they have from the maximal amount of connections. Having so many degrees of freedom can come at the cost of leading to overfitting (when the network fits the training data too specifically and can not generalize to different data). The increase in the number of parameters can also be a problem, not only because of the much greater amount of time it will take to train the network, but also because the training can be degraded.

This degraded training is a general problem of neural networks with many layers and free parameters and was noted in Ref.~\cite{ResNet}. The basic problem is that if a network is sufficiently complex to find and relate all the relevant features of an image, adding more complexity will not benefit the result in any way, but it will just make the learning process more complicated and will result in general in worse performance.

\begin{figure}[t!]
\begin{center}
\includegraphics[width=0.45\textwidth]{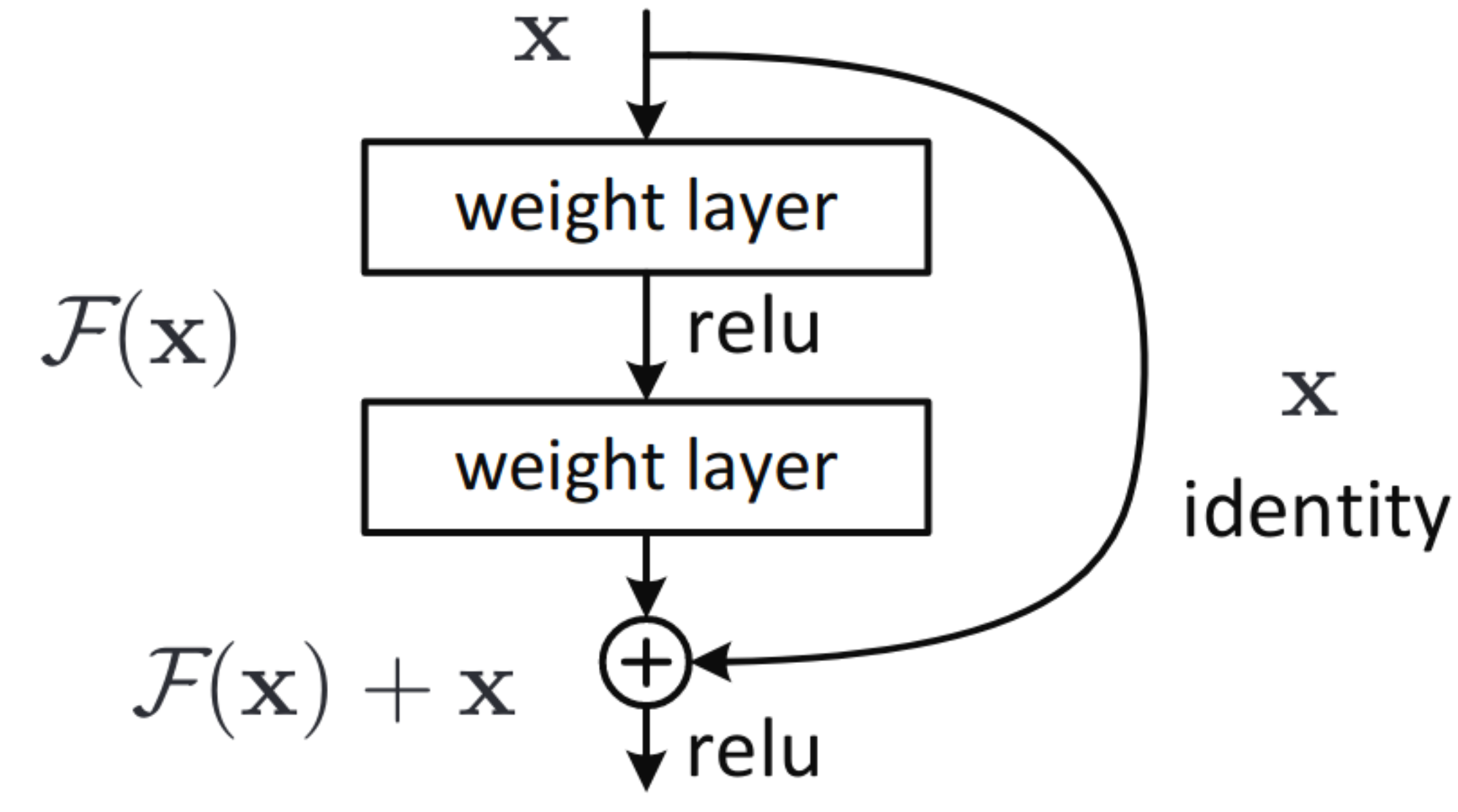}
\end{center} 
\caption{Schematic representation of a residual block. Image taken from Ref.~\cite{ResNet}.}
\label{fig:ResidualBlock}
\end{figure}

To tackle this problem, the authors of Ref.~\cite{ResNet} developed a type of Convolutional Neural Network called Residual Neural Network (ResNet). The building block of these networks is the Residual block, schematically shown in Fig.~\ref{fig:ResidualBlock}. The input of the block $x$ will go through two paths, one path passes through different layers that transform it into $\mathcal{F}(x)$, which is called the residual, while the other path does not get transformed. At the end of the block, the two signals are summed. If the network already has computed all relevant features of the image, it will be easy for the network to do the identity mapping without degrading the signal by optimizing the residual $\mathcal{F}(x)$ towards 0. Because of this, residual neural networks usually do not get worse with increased number of layers, they plateau.

To classify the input images into noise or gravitational waves from close hyperbolic encounters, we will use a neural network based on one of the most refined architectures of Residual Neural Networks there is, called \emph{ResNet-50} (because it has 50 hidden layers). The actual network architecture we will use is detailed in Table \ref{table:ResNet50}, where the first layer of the network will take as its input the pixels from the $50 \times 150$ pixel image we want to classify. Finally, note that the last layer will return one single number, and since the activation of this last neuron is a Sigmoid function, this number will be between 0 and 1 (see Table \ref{table:NeuronActivation}). It will be interpreted as the probability for the image to be generated by gravitational waves from close hyperbolic encounters.

\begin{table*}[t!]
\centering
\begin{tabular}{c|c|c}
Structure & output size & Specifications \\
\hline
Convolutional layer & $25 \times 75$ &$7 \times 7$, 64, stride 2 \\
\hline
Pooling layer & $13 \times 38$ & $3 \times 3$ max pool, stride 2 \\
\hline
& \\[-.75em]
Residual blocks 1-3 & $13 \times 38$ & $\begin{bmatrix}
1 \times 1\text{, 64}\\
3 \times 3\text{, 64} \\
1 \times 1\text{, 256} \\
\end{bmatrix} \times 3$ \\[1.5em]
\hline
& \\[-.75em]
Residual blocks 4-7 & $7 \times 19$ & $\begin{bmatrix}
1 \times 1\text{, 128}\\
3 \times 3\text{, 128} \\
1 \times 1\text{, 512} \\
\end{bmatrix} \times 4$ \\[1.5em]
\hline
& \\[-.75em]
Residual blocks 8-13 & $4 \times 10$ & $\begin{bmatrix}
1 \times 1\text{, 256}\\
3 \times 3\text{, 256} \\
1 \times 1\text{, 1024} \\
\end{bmatrix} \times 6$ \\[1.5em]
\hline
& \\[-.75em]
Residual blocks 14-16& $2 \times 5$ & $\begin{bmatrix}
1 \times 1\text{, 512}\\
3 \times 3\text{, 512} \\
1 \times 1\text{, 2048} \\
\end{bmatrix} \times 3$ \\[1.5em]
\hline
Pooling layer & $1 \times 1$ & global average pool, Sigmoid activation function \\
\hline
\end{tabular}
\caption{Architecture of the neural network used for our analysis based on the \emph{ResNet-50} architecture. The nomenclature of the convolutional layers is ``window size ($N \times N$), number of different kernels ($K$), stride ($S$)'' and when not specified we are using a ReLu activation function. We have put the convolutional layers of each residual block between brackets and the multiplicative number outside the brackets represents how many times this block is repeated. The first convolutional layer of blocks 4, 8 and 14 will have a stride $S =2$ to halve the width and height of the output, while the other convolutional layers in the residual blocks have a stride $S=1$.}
\label{table:ResNet50}
\end{table*}

The actual implementation of the neural network will be done using the \texttt{tensorflow} \cite{tensorflow} library of \texttt{Python}.

\subsubsection{Training and validation}
\label{sec:Data:CNN:TrainAndVal}

Even though we have defined in the previous section the architecture of the neural network that will be used to classify the images, this network design will contain a lot of free parameters with undetermined values. We will have to determine the optimal values of these parameters such that whenever we input an image with the format described in Sec.~\ref{sec:Data:CNN:Samples} that contains gravitational waves from a close hyperbolic encounter, the neural network returns a 1 and whenever we input images that contain only noise, the neural network returns a 0.

These optimal values of the free parameters are determined by ``fitting'' the neural network to the training images, described in Sec.~\ref{sec:Data:CNN:Samples}, whose class (``noise'' or ``CHE'') is known. How well the neural network fits the data will be quantified via a loss function that we will want to minimize. As our loss function we will use the binary cross-entropy between true label values and the neural network predictions. This is the most commonly used loss function for binary classifiers \cite{CrossEntropy} and it is computed in the following way:

\begin{equation}
    H(p,q) = -\frac{1}{N} \sum_{i = 1}^{N}  q_i \log p_i + (1 - q_i) \log(1-p_i) \, ,
    \label{eq:BinaryCrossEntropy}
\end{equation}

\noindent where N is the total number of images we are evaluating, $q_i$ are the true labels (0 for noise and 1 for CHE) and $p_i$ is the output of the neural network (we want it to be 0 for noise and 1 for CHE). Note that the closer the predictions $p_i$ are to the real values $q_i$, the smaller the binary cross-entropy will be, taking a value of 0 whenever $p_i = q_i$. Minimizing this loss function will thus produce predictions closer to the real values.

\begin{figure*}[t!]
\centering
\includegraphics[width=0.45\textwidth]{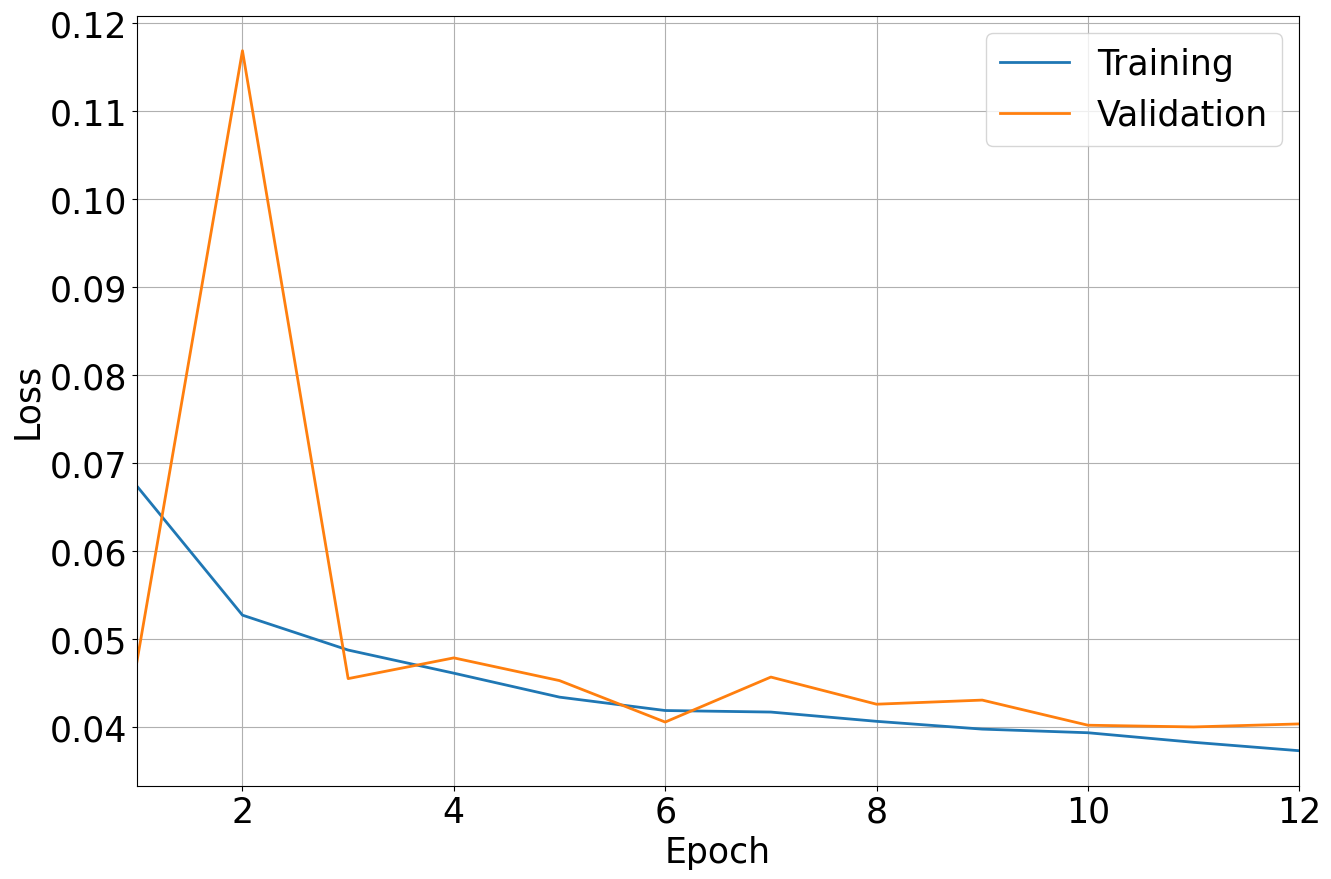}
\label{fig:Training:Loss}
\includegraphics[width=0.45\textwidth]{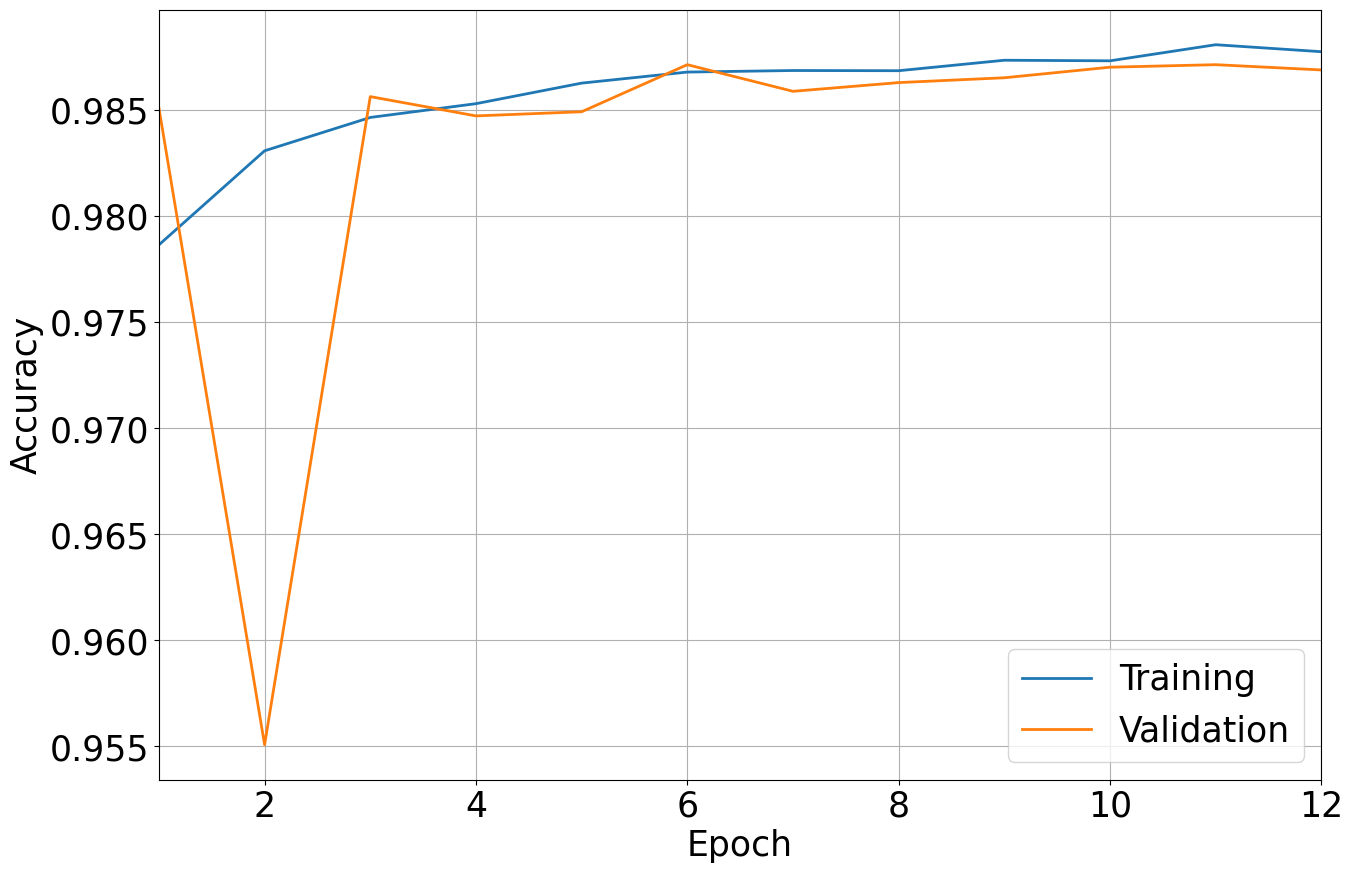}
\label{fig:Training:Accuracy}
\caption{Training results of the neural network validated on the test samples. Left: Neural Network loss for the validation as well as the training samples as a function of the training epoch. Right: Neural Network accuracy for the validation as well as the training samples as a function of the training epoch.}
\label{fig:Training}
\end{figure*}

For the optimization of the parameters of the neural network to minimize the loss function on the training data we will use the \emph{Adam} method for stochastic optimization \cite{Adam}. \emph{Adam} is an algorithm for fist-order gradient-based optimization of stochastic objective functions, based on adaptive estimates of lower order moments. We will use the \texttt{Python} implementation of this algorithm in \texttt{tensorflow} \cite{tensorflow}, setting a learning rate of $\alpha = 0.001$ a batch size of 32 and applying 12 training epochs. The result of training the neural network on the 64028 noise training images and 45356 CHE training images is shown in Fig.~\ref{fig:Training}. 

On Fig.~\ref{fig:Training} we have plotted the loss as a function of the training epoch, computed both for the training as well as for the test images used for validation. We can observe how the training loss monotonically decreases, as we would expect from the fact that we are optimizing the network to minimize this quantity. More interesting is to observe the behavior of the loss when testing the network on the validation samples, which are not being looked at to optimize the network. We observe that even though this test loss fluctuates somewhat at small training epochs, at larger epochs it converges towards small values of the loss, around 0.04, close to the training loss. This means that the neural network is being able to correctly generalize from the training data to data it has not previously looked at and that it is not suffering from overfitting.

On Fig.~\ref{fig:Training} we also represent the evolution of the accuracy of the neural network as a function of the number of training epochs, where the accuracy is the number of images correctly classified (assuming that $p \leq 0.5$ corresponds to noise and $p>0.5$ corresponds to CHE) divided by the total number of images. The accuracy of the neural network on the training images generally increases with the training epoch, reaching a final value of 98.8\%, which means that the optimization of the loss is translating into a very accurate classification of the images. Looking at the accuracy in the test images, we observe that after the 12 epochs it reaches 98.7\%, close to the training value. This further validates the fact that the neural network is learning the correct patterns to classify the images during training and it is able to apply them to data it has not previously looked at.

\begin{figure}[t!]
\begin{center}
\includegraphics[width=0.45\textwidth]{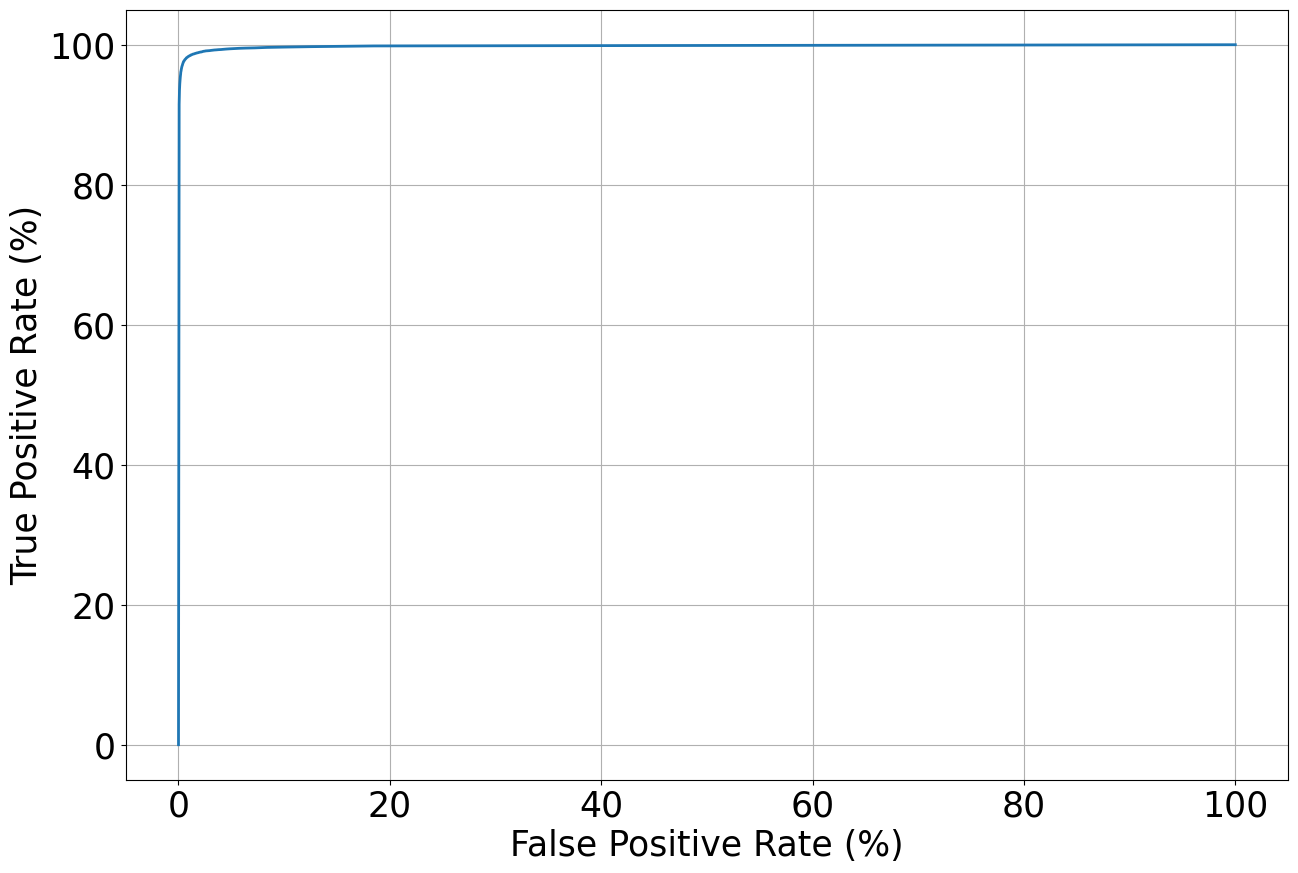}
\end{center} 
\caption{Receiver operating characteristic curve of the neural network on the test images.}
\label{fig:ROC}
\end{figure}

A very useful representation to visualize the performance of a binary classifier is through the ``receiver operating characteristic'' (ROC) curve \cite{ROCcurve}. This is a parametric curve in which we show the evolution of the rate of true positives \footnote{The rate of true positives is the number of test CHE images correctly identified as CHE images divided by total number of CHE images tested.} and of the rate of false positives \footnote{The rate of false positives is the number of test noise images incorrectly identified as CHE images divided by total number of noise images tested.} when varying between 0 and 1 the minimum probability we require to consider an event to be from a close hyperbolic encounter. The ROC curve of our neural network is shown in Fig.~\ref{fig:ROC}. If we look at the upper left corner of Fig.~\ref{fig:ROC}, we can see that given the right probability threshold, we can obtain very large values of the true positive rate at very low values of the false positive rate. This is the desired operation point of any classifier and it is indicative of the fact that our neural network is a very good binary classifier. 

For our purposes we will choose a high value of the probability threshold to consider that an event is a CHE, requiring that $p > 0.9$. This will be done to reduce a lot the number of false positives that we will obtain and thus select only the most promising events. The truth table one obtains when running the neural network in the validation samples with this 0.9 threshold is shown in Table \ref{table:TruthTable}, where we can observe how the Neural Network with the 0.9 threshold greatly suppresses the false positives without having too many false negatives. Since we obtained 135 false positives in the 48125 test noise images, corresponding to 275 days of noise data, the false alarm rate of this neural network trigger will be 0.49 days$^{-1}$, that is, if the data contained only noise, we would expect a false trigger every 2 days. 

\begin{table}[t!]
\centering
\begin{tabular}{c|c|c|}
  & True & False \\ 
\hline
Positive & 42733 & 135 \\
\hline
Negative & 47990 & 1508 \\
\hline
\end{tabular}
\caption{Truth table of the neural network when considering the requirement $p > 0.9$ to classify an event as CHE. Positive/Negative means that the event has been classified as CHE/noise. True/False means that the event has been correctly/incorrectly classified. The false positive rate derived from this table will be 0.28\%, while the true positive rate will be 96.59\%.}
\label{table:TruthTable}
\end{table}

\begin{figure*}[t!]
\centering
\includegraphics[width=0.45\textwidth]{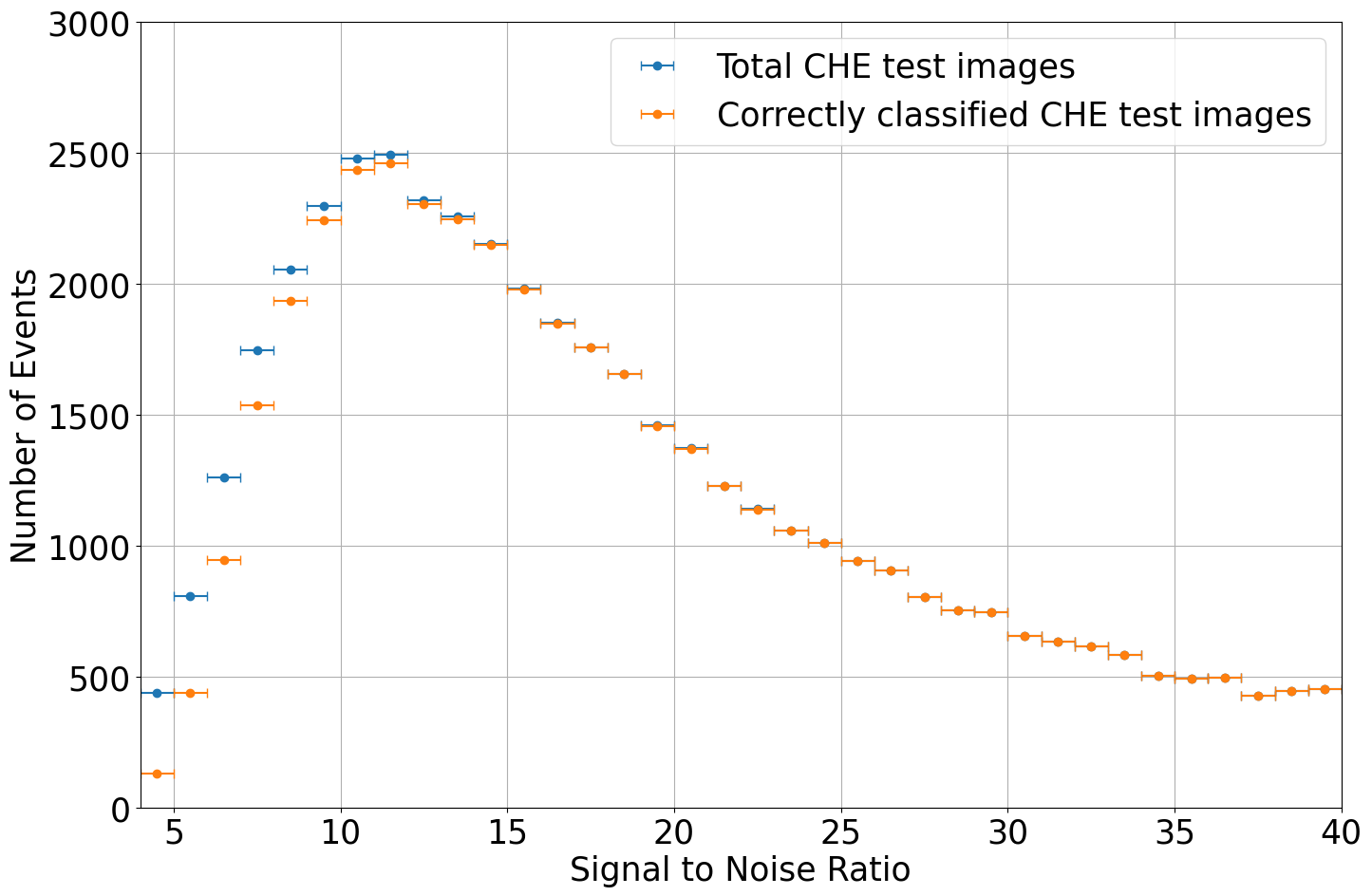}
\label{fig:MLTrigerInjected:ML_SNR_distribution}
\includegraphics[width=0.45\textwidth]{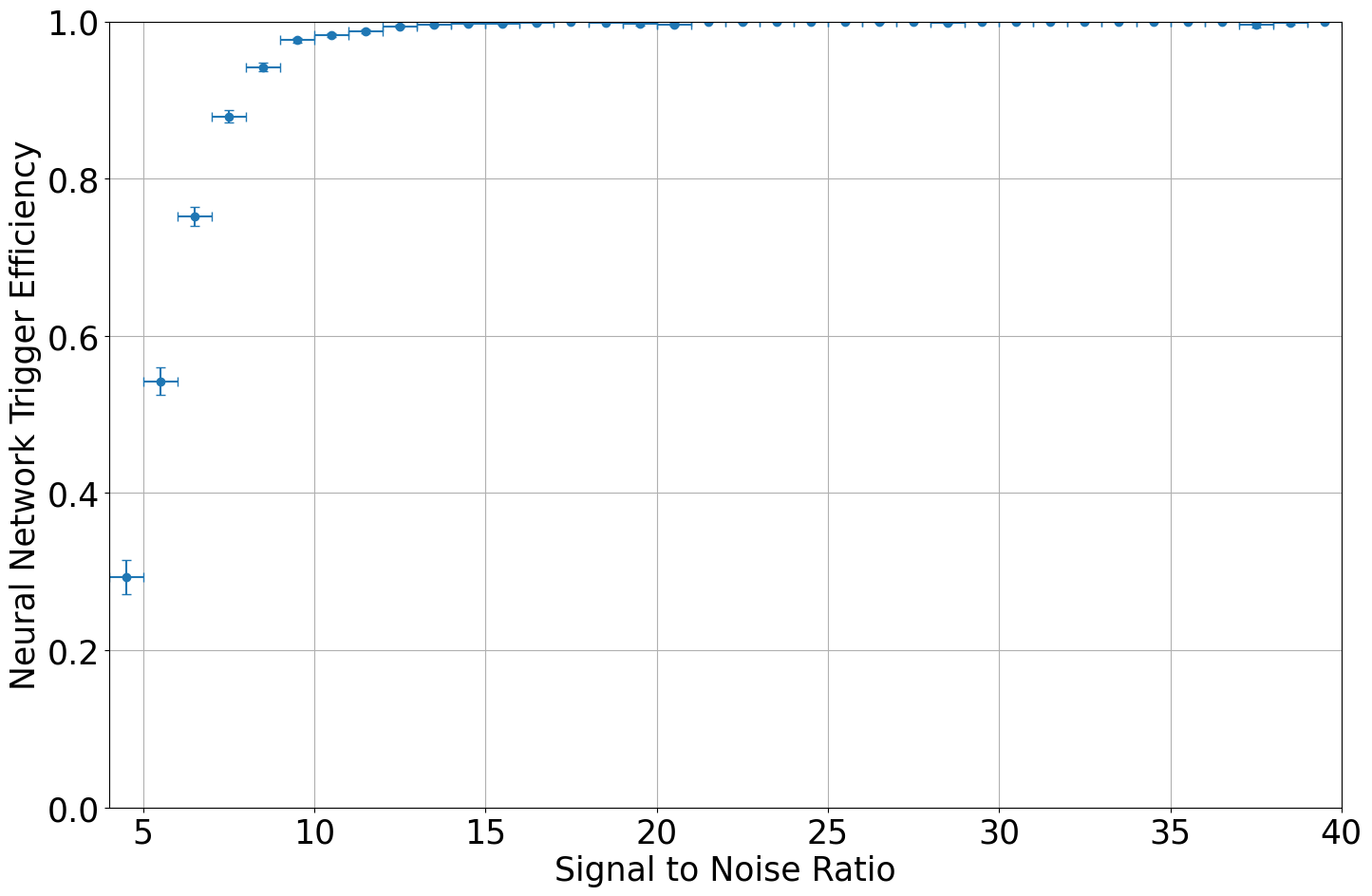}
\label{fig:MLTrigerInjected:ML_Efficiency}
\caption{Results of the neural network as a function of the signal to noise ratio for the CHE validation samples. Left: Number of test CHE images and number of test CHE images correctly identified as a function of the signal to noise ratio. Right: Neural Network efficiency as a function of the signal to noise ratio of the CHE test events.}
\label{fig:MLTrigerInjected}
\end{figure*}

Nonetheless, we do not expect all the signals to be equally easy to detect. Just as in the case of the correlation trigger, we would expect that the close hyperbolic encounters with high signal to noise ratio will be easier to identify. To test this, in Fig.~\ref{fig:MLTrigerInjected} we show the distribution in signal to noise ratio of the CHE images used for testing the neural network as well as the distribution of the subset of these images that is correctly identified by the neural network. Dividing the number of events detected at each signal to noise ratio by the total number of events tested at that signal to noise ratio, we can get the ``trigger efficiency'' of our neural network. This is shown in Fig.~\ref{fig:MLTrigerInjected}, where we see that the efficiency at very low values of the signal to noise ratio is quite small, meaning that we detect only a small fraction of faint CHE events. Nonetheless, it rises quickly with the increasing signal to noise ratio and we are able to detect more than 50\% of the events above a signal to noise ratio of 5. 

The relatively bad performance at very low signal to noise ratio is expected because of the fact that such faint events leave a very small signal in the detector that is hard to distinguish from the correlated noise fluctuations.

\subsubsection{Results on the data}
\label{sec:Data:CNN:Data}

Now that we have validated the neural network, checking that it recovers most of the close hyperbolic encounters (CHE) with signal to noise ratio above 5 while having a false alarm rate of 0.49 days$^{-1}$, we will want to run the neural network on the 2704 data events that passed the correlation trigger requirements, and check how many of them are determined by the neural network to have a probability $p > 0.9$ to be close hyperbolic encounters.

\begin{figure}[t!]
\begin{center}
\includegraphics[width=0.45\textwidth]{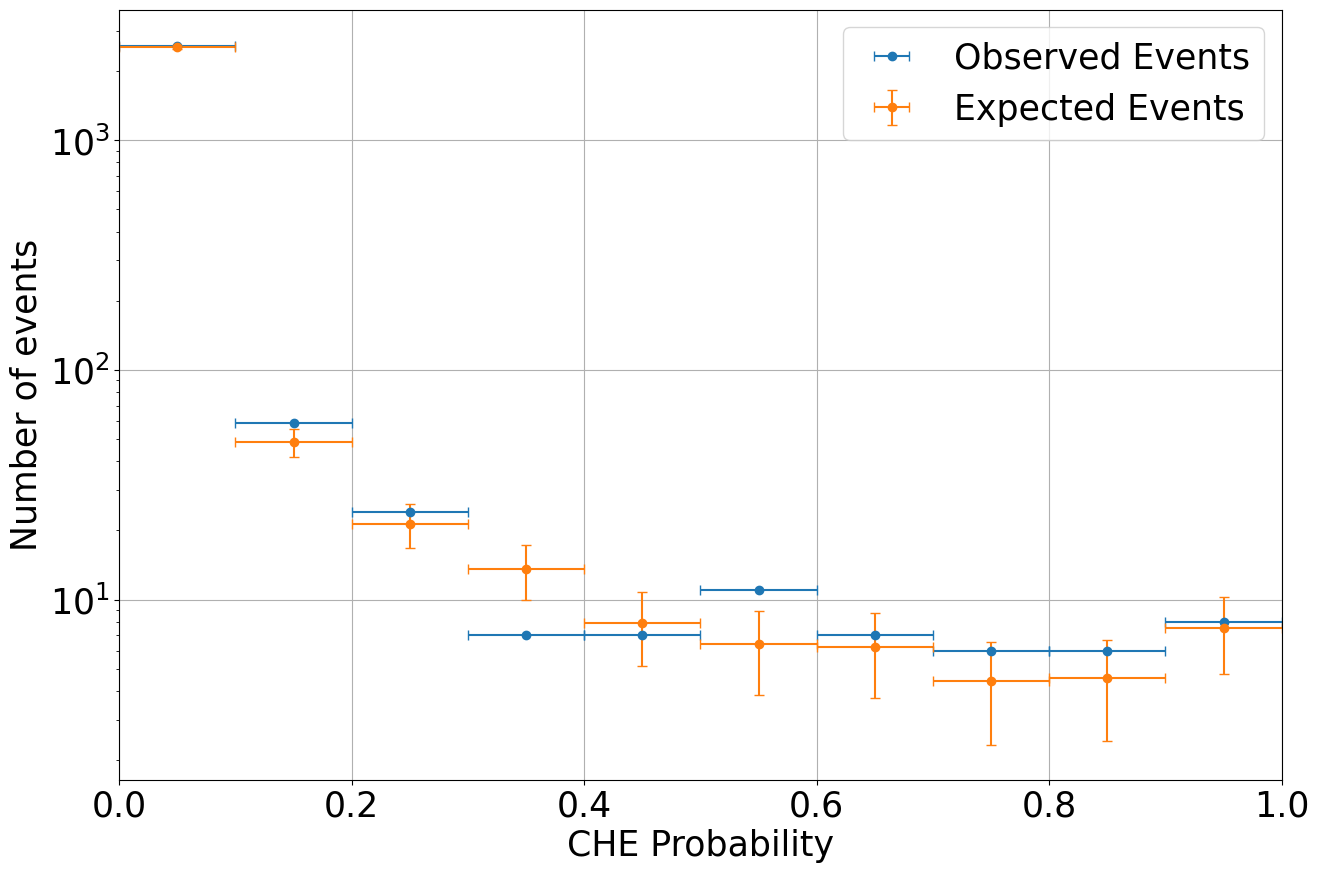}
\end{center} 
\caption{Number of observed events and expected noise events as a function of the probability given by the neural network to be a close hyperbolic encounter. The number of expected noise events refers to the number of events we would expect to see if the data only contained noise and it has been obtained with the results of the neural network on the 48125 test noise images, corresponding to 275 days of noise data. The uncertainty on the number of expected noise events is computed assuming Poisson statistics, $\sigma = \sqrt{N_{\text{expected}}}$. }
\label{fig:hprob_GW}
\end{figure}

In Fig.~\ref{fig:hprob_GW} we show a histogram of the number of events we obtain at each value of the CHE probability when running the neural network on the data. If we compare the number of observed events with the number of events we would expect if only noise was present on the data, we can observe that at high values of the CHE probability, we consistently observe more events than expected, but almost all excesses are within the one standard deviation region and are thus not statistically significant.

\begin{figure*}[t!]
\begin{center}
\includegraphics[width=0.8\textwidth]{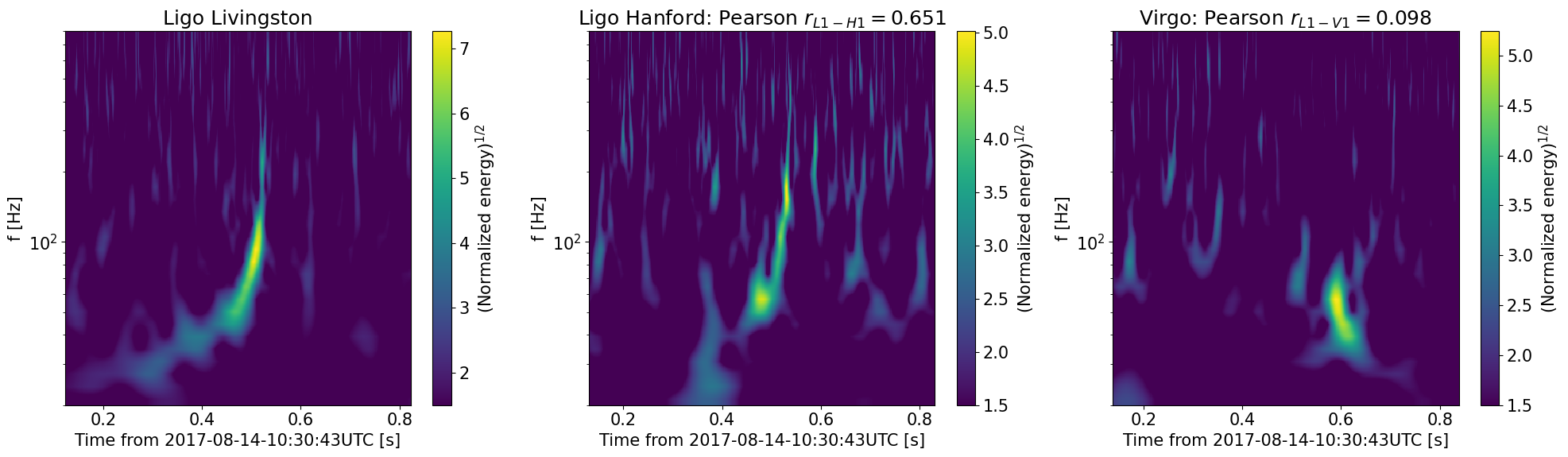}
\end{center} 
\caption{Event with neural network CHE probability $p=0.997$ and correlation trigger discriminant $D = 0.469$. This event corresponds to GW170814 \cite{GWTC-1}, which is claimed by LIGO-Virgo to be the coalescence of two black holes of masses $31M_{\odot}$ and $25M_{\odot}$.}
\label{fig:hiper_0.997}
\end{figure*}

\begin{figure*}[t!]
\begin{center}
\includegraphics[width=0.8\textwidth]{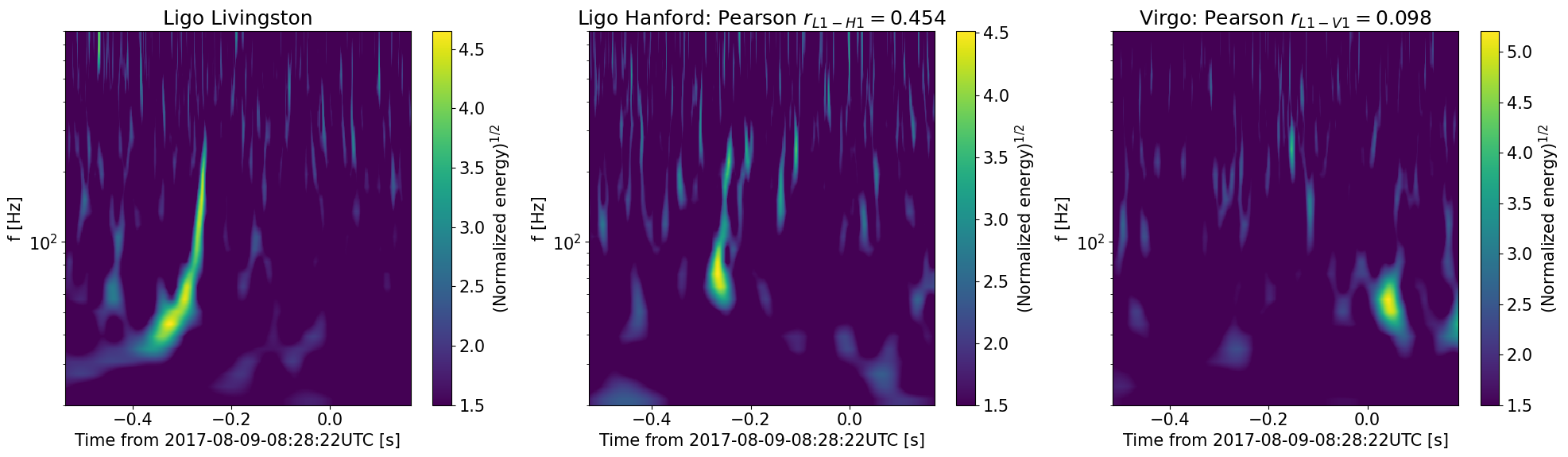}
\end{center} 
\caption{Event with neural network CHE probability $p=0.991$ and correlation trigger discriminant $D = 0.337$. This event corresponds to GW170809 \cite{GWTC-1}, which is claimed by LIGO-Virgo to be the coalescence of two black holes of masses $35M_{\odot}$ and $24M_{\odot}$.}
\label{fig:hiper_0.991}
\end{figure*}

\begin{figure*}[t!]
\begin{center}
\includegraphics[width=0.8\textwidth]{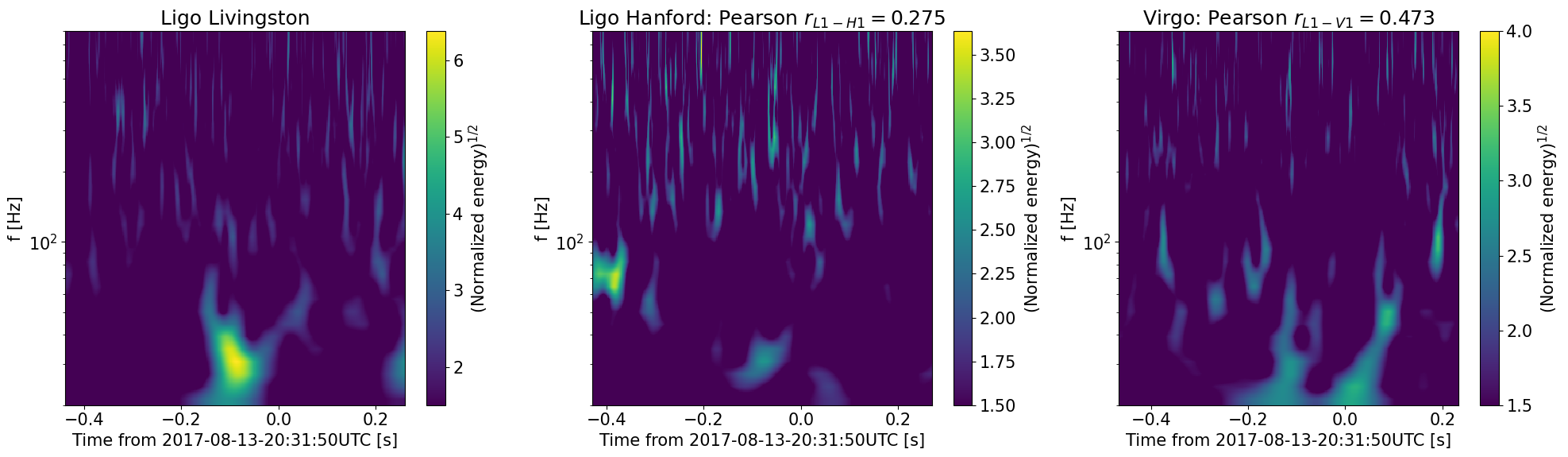}
\end{center} 
\caption{Event with neural network CHE probability $p=0.980$ and correlation trigger discriminant $D = 0.341$.}
\label{fig:hiper_0.980}
\end{figure*}

\begin{figure*}[t!]
\begin{center}
\includegraphics[width=0.8\textwidth]{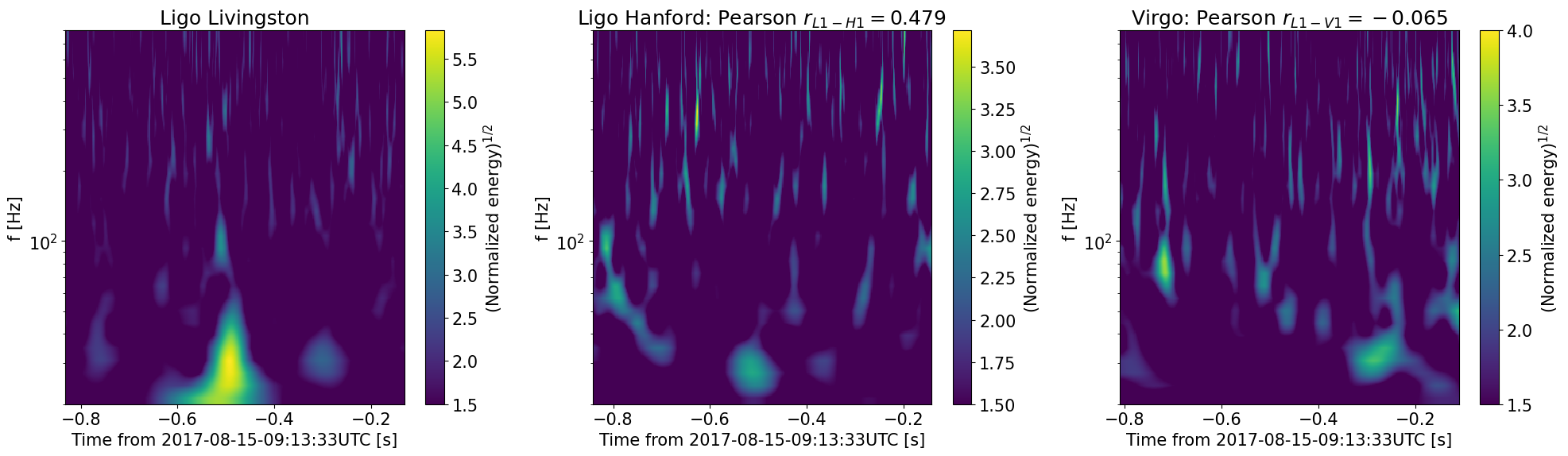}
\end{center} 
\caption{Event with neural network CHE probability $p=0.976$ and correlation trigger discriminant $D = 0.321$.}
\label{fig:hiper_0.976}
\end{figure*}

\begin{figure*}[t!]
\begin{center}
\includegraphics[width=0.8\textwidth]{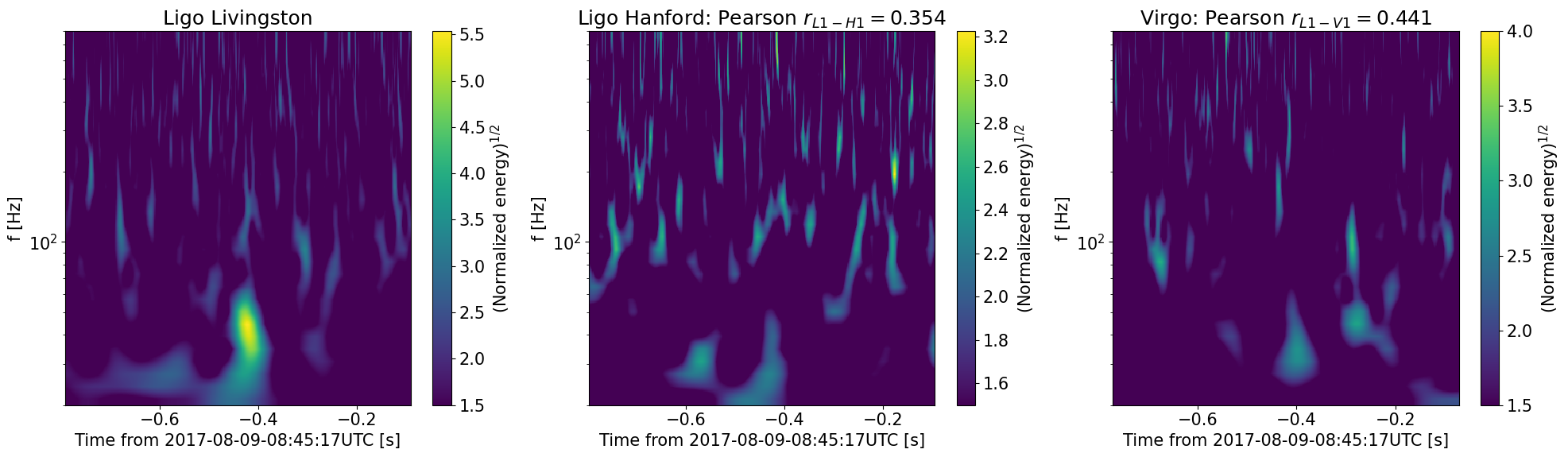}
\end{center} 
\caption{Event with neural network CHE probability $p=0.974$ and correlation trigger discriminant $D = 0.382$.}
\label{fig:hiper_0.974}
\end{figure*}

\begin{figure*}[t!]
\begin{center}
\includegraphics[width=0.8\textwidth]{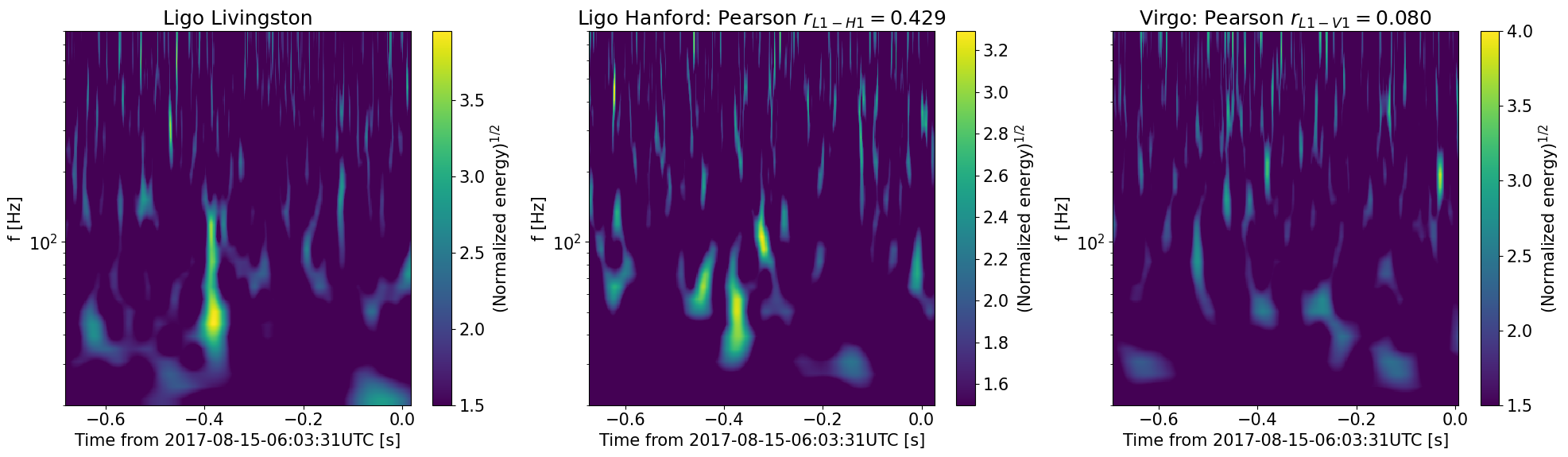}
\end{center} 
\caption{Event with neural network CHE probability $p=0.942$ and correlation trigger discriminant $D = 0.314$.}
\label{fig:hiper_0.942}
\end{figure*}

\begin{figure*}[t!]
\begin{center}
\includegraphics[width=0.8\textwidth]{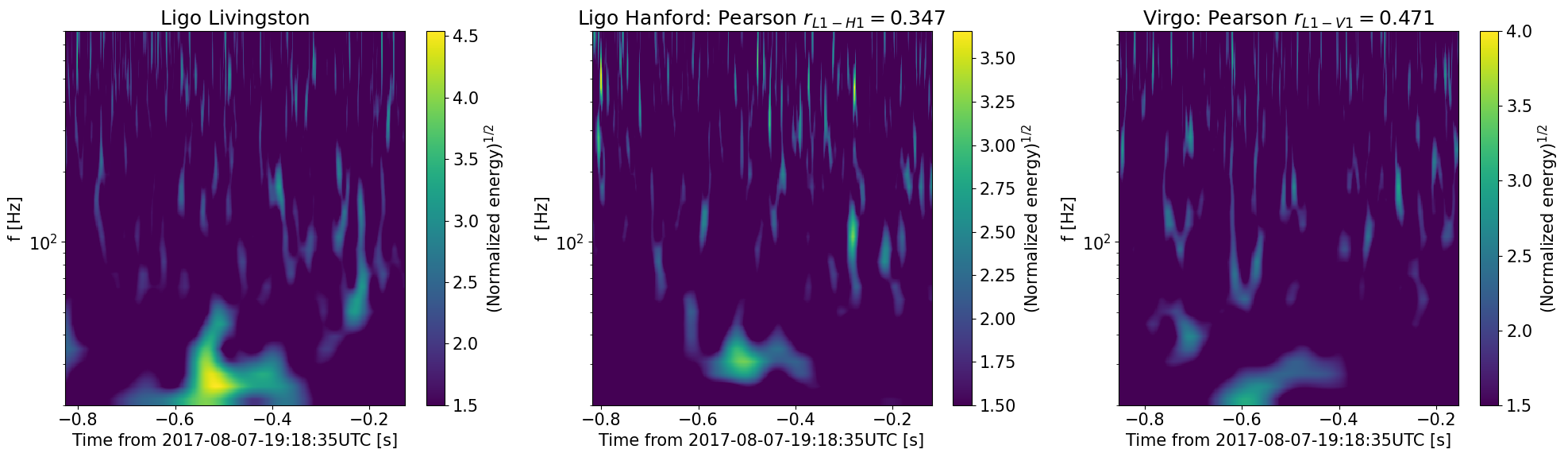}
\end{center} 
\caption{Event with neural network CHE probability $p=0.927$ and correlation trigger discriminant $D = 0.388$.}
\label{fig:hiper_0.927}
\end{figure*}

\begin{figure*}[t!]
\begin{center}
\includegraphics[width=0.95\textwidth]{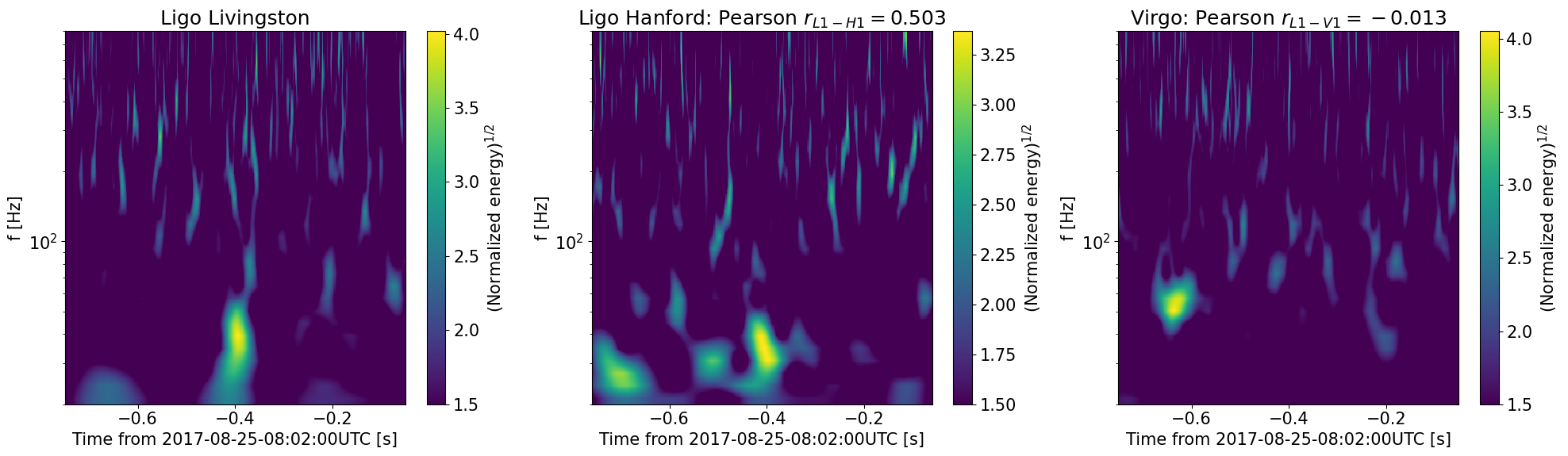}
\end{center} 
\caption{Event with neural network CHE probability $p=0.914$ and correlation trigger discriminant $D = 0.337$.}
\label{fig:hiper_0.914}
\end{figure*}

The last bin of Fig.~\ref{fig:hprob_GW} corresponds to the events with $p>0.9$ that are determined by the neural network to be the most likely to come from close hyperbolic encounters. From the validation of the neural network on the noise test images we would expect to have $7.5 \pm 2.7$ events above this 0.9 probability if the 15.3 days of analyzed data contained only noise. When we actually run the neural network on the data we obtain 8 such events, whose properties are detailed in Table~\ref{table:Triggers} and are shown in Figs.~\ref{fig:hiper_0.997}-\ref{fig:hiper_0.914}. This number of observed events is well within the expected range if there was only noise in the data. Nonetheless, this will not mean that individual events within the selected candidates will not be able to come from hyperbolic encounters.

\begin{table*}[t!]
\centering
\begin{tabular}{c c c c}
\hline \hline
    GPS trigger time (s)  & CHE probability & Correlation trigger discriminant & Fig. \\  
    \hline
    1186741861.47 & 0.997 & 0.469 & \ref{fig:hiper_0.997} \\
    1186302519.81 & 0.991 & 0.337 & \ref{fig:hiper_0.991} \\
    1186691527.91 & 0.980 & 0.341 & \ref{fig:hiper_0.980} \\
    1186823630.52 & 0.976 & 0.321 & \ref{fig:hiper_0.976} \\
    1186303534.56 & 0.974 & 0.382 & \ref{fig:hiper_0.974} \\
    1186812228.67 & 0.942 & 0.314 & \ref{fig:hiper_0.942} \\
    1186168732.52 & 0.927 & 0.388 & \ref{fig:hiper_0.927} \\
    1187683337.60 & 0.914 & 0.337 & \ref{fig:hiper_0.914} \\
\hline \hline
\end{tabular}
\caption{Close hyperbolic encounter candidates accepted by the correlation trigger and determined by the neural network to have a CHE probability greater than 0.9. In the last column we give the figure where the normalized energy of each event is shown.}
\label{table:Triggers}
\end{table*}

The two most promising events according to the neural network are in Fig.~\ref{fig:hiper_0.997} and Fig.~\ref{fig:hiper_0.991}. However, if we look at them, these events have a chirp-like shape, with the frequency increasing with time. This is because they correspond to GW170814 and GW170809 \cite{GWTC-1} respectively, which are claimed by LIGO-Virgo collaboration to come from the coalescence of black hole binaries. Nonetheless, since they are both very massive binaries, the last oscillations before merger can look like close hyperbolic encounters, which is tricking the neural network. Additionally, these two signals come from gravitational waves, and therefore have the correct time and intensity relationships between interferometers that the neural network will be looking for. The fact that we recover these events so strongly is a further validation of the neural network. If we wanted to reject events coming from black hole binary coalescence, we could train the neural network with examples of these events labeled as ``noise''.

The event GW170817 that did pass the correlation trigger (Table \ref{table:CBC_correlation}) is not accepted by the neural network because it comes from the coalescence of neutron stars and therefore the merger happens at very large frequencies (well above 800Hz). This means that the last oscillations happen outside the image, which makes this type of event more difficult to mistake with a close hyperbolic encounter.

The rest of the events found by the neural network, shown in Figs.~\ref{fig:hiper_0.980}-\ref{fig:hiper_0.914} are not claimed by the LIGO-Virgo collaboration. These events look much more like the close hyperbolic encounters we are looking for, described in Sec.~\ref{sec:detec:procesing}. Nonetheless, the number of events we find is consistent with them coming from random correlated glitches. To decide whether or not these events can come from close hyperbolic encounters, further analysis would be required. This further analysis usually consists on estimating the parameters of the encounter using Bayesian inference \cite{BayesianInference} and checking that the results obtained for the parameters make physical sense and that the Bayes factor against the noise hypothesis is large. 

\section{Summary and conclusions}

The objective of this paper was to search for black hole hyperbolic encounters in current gravitational wave detectors. These encounters are expected to happen in dense black hole clusters when two black holes gravitationally scatter off each other. If the black holes get sufficiently close to each other during the scattering they will emit bremsstrahlung gravitational waves that can be observed at current Earth-based laser interferometers.

To determine the gravitational radiation we could expect from these close hyperbolic encounters, in Sec.~\ref{sec:waveforms} we developed 1.5 post newtonian accurate templates for the gravitational waves emitted by spinning compact binaries in hyperbolic orbits, taking into account up to leading order spin effects, and we showed how to numerically implement these templates in an efficient way using \texttt{Python}. 

In Sec.~\ref{sec:detec} we studied how these gravitational waves interact with the network of gravitational wave detectors currently present on Earth, deriving the antenna factors for laser interferometers. Since real detectors are dominated by noise, we discussed how to use data processing techniques to make the signal of the hyperbolic encounters stand out over the noise. Namely we band-passed the data between 20Hz and 800Hz, whitened it with the noise power spectrum distribution and Q transformed it to show the characteristic dependence in the time-frequency domain.

Having determined the signal we were looking for, in Sec.~\ref{sec:Data} we developed a two level trigger to extract the possible hyperbolic candidates from the 15.3 days of publicly available data in which the three detectors were under nominal operation. The first level of the trigger consisted on a loose selection based on correlations between detectors and tuned to accept possible close hyperbolic encounter event candidates while doing a large reduction of the data. This part of the trigger was tested and validated using injections, which showed that we are able to recover most of the events with signal to noise ratio above 10. The second level of the trigger consisted of a convolutional neural network to classify the 2704 events accepted by the first level trigger into either noise or close hyperbolic encounters. This neural network was trained using images of simulated close hyperbolic encounter signals as well as correlation triggers between unsynchronized data. The neural network was validated using test images, which showed that it is able to recover most of the close hyperbolic events that passed the first trigger and that have a signal to noise ratio above 5, while having a false alarm rate of 0.49 days$^{-1}$.

When running the full trigger on the 15.3 days of data we obtained 8 hyperbolic event candidates, consistent with the number of false positives expected if only noise was present on the data, of $7.5 \pm 2.7$. The two most promising events according to the neural network correspond to GW170814 and GW170809 respectively, which come from the coalescence of black hole binaries, whose last oscillation can look like a close hyperbolic encounter, specially if we do not train our network to reject this coalescence type of events. The six remaining candidates did look more like the close hyperbolic encounters we were looking for. Nonetheless, to decide whether or not these events can come from close hyperbolic encounters, further analysis would be required. This further analysis could consist on estimating the parameters of the encounter using Bayesian inference and checking that the results obtained for the parameters make physical sense and that the Bayes factor against the noise hypothesis is large. 

Now that we have developed and validated a method to find close hyperbolic event candidates using a correlation trigger and a neural network, this method can be used to analyze the gravitational wave data of future observing runs. If there are gravitational waves from close hyperbolic encounters at a rate sufficiently large to explain the small excesses seen in Fig.~\ref{fig:hprob_GW}, then analyzing more data we would observe a larger excess of events above the number of expected events, that at some point would be statistically significant. In addition, these future observing runs are also expected to have much better detector sensitivities, which would reduce the background of our search.

To translate the constraints on the rate of close hyperbolic encounters observed into information about cluster dynamics, more work is required in the theoretical understanding of the nature and structure of these clusters, to determine the expected rate and parameters of the hyperbolic encounters that will take place within the clusters.

\section*{Acknowledgements}

The authors thank Sergei Klimenko for his comments as reviewer of this paper in LVK. The authors acknowledge use of the publicly available codes: \texttt{lalsuite} \cite{lalsuite}, \texttt{PyCBC} \cite{PyCBC} and \texttt{tensorflow} \cite{tensorflow} as well as the use of the publicly available LIGO-Virgo open data \cite{OpenData_O1_O2}. They acknowledge support from the research project  PGC2018-094773-B-C32, and the Centro de Excelencia Severo Ochoa Program SEV-2016-059, while  SN also acknowledges support from the Ram\'{o}n y Cajal program through Grant No. RYC-2014-15843. The authors acknowledge use of the Hydra cluster at the Instituto de F\'isica Te\'orica (IFT), on which some of the numerical computations for this paper took place. 

The python codes used by the authors in the analysis of the paper can be found at \href{https://github.com/gmorras/CHE_search_paper}{https://github.com/gmorras/CHE\_search\_paper}.

\appendix

\section{Coefficients in correlation trigger}
\label{sec:anex:PearsonCoef}

In Sec.~\ref{sec:Data:presel} we introduced a way of looking for event candidates using the correlation between detectors. Using Eq.~\eqref{eq:Pearson_def} as prescribed in Sec.~\ref{sec:Data:presel}, we can compute the correlation between Livingston and Hanford $r_{L1-H1}$ and between Livingston and Virgo $r_{L1-V1}$, which will be combined to construct a trigger discriminant $D$. For the sake of simplicity, the discriminant can be chosen as a linear combination of $r_{L1-H1}$ and $r_{L1-V1}$, normalized in such a way that $-1 \leq D \leq 1$:

\begin{equation}
    D = a r_{L1-H1} + (1-a) r_{L1-V1}  \, ,
    \label{eq:opt_discriminant_formula}
\end{equation}

\noindent where $a$ is a free parameter that we will want to optimize. 

Since the trigger discriminant is supposed to determine the significance of an event, it will be natural for it to be closely related with the signal to noise ratio of the event. Because of this, it will be desired to find the value of $a$ that minimizes the variance of the function $D(S/N)$. The computation of this variance is done by doing injections of events at different signal to noise ratios between 4 and 40 with random parameters as specified in Table \ref{table:InjectParam} of appendix \ref{sec:anex:InjectionParams}. With the correlations induced by these injections, the variance can be computed as:

\begin{equation}
    \sigma^2_{\text{tot}} = \sum_{S/N = 4}^{S/N = 40} \sigma^2_{S/N} \, ,
    \label{eq:var_tot}
\end{equation}
where $\sigma^2(S/N)$ is the variance for the events whose floored signal to noise ratio is $S/N$. If $D_{S/N}$ is their discriminant, computed from Eq.~\ref{eq:opt_discriminant_formula}, the variance can be computed as:

\begin{equation}
    \sigma^2_{S/N} = \left\langle (D_{S/N} - \langle D_{S/N} \rangle)^2 \right\rangle \, ,
    \label{eq:var_SNR}
\end{equation}
where here $\langle ... \rangle$ denotes the arithmetic mean.

\begin{figure*}[t!]
\centering
\includegraphics[width=0.45\textwidth]{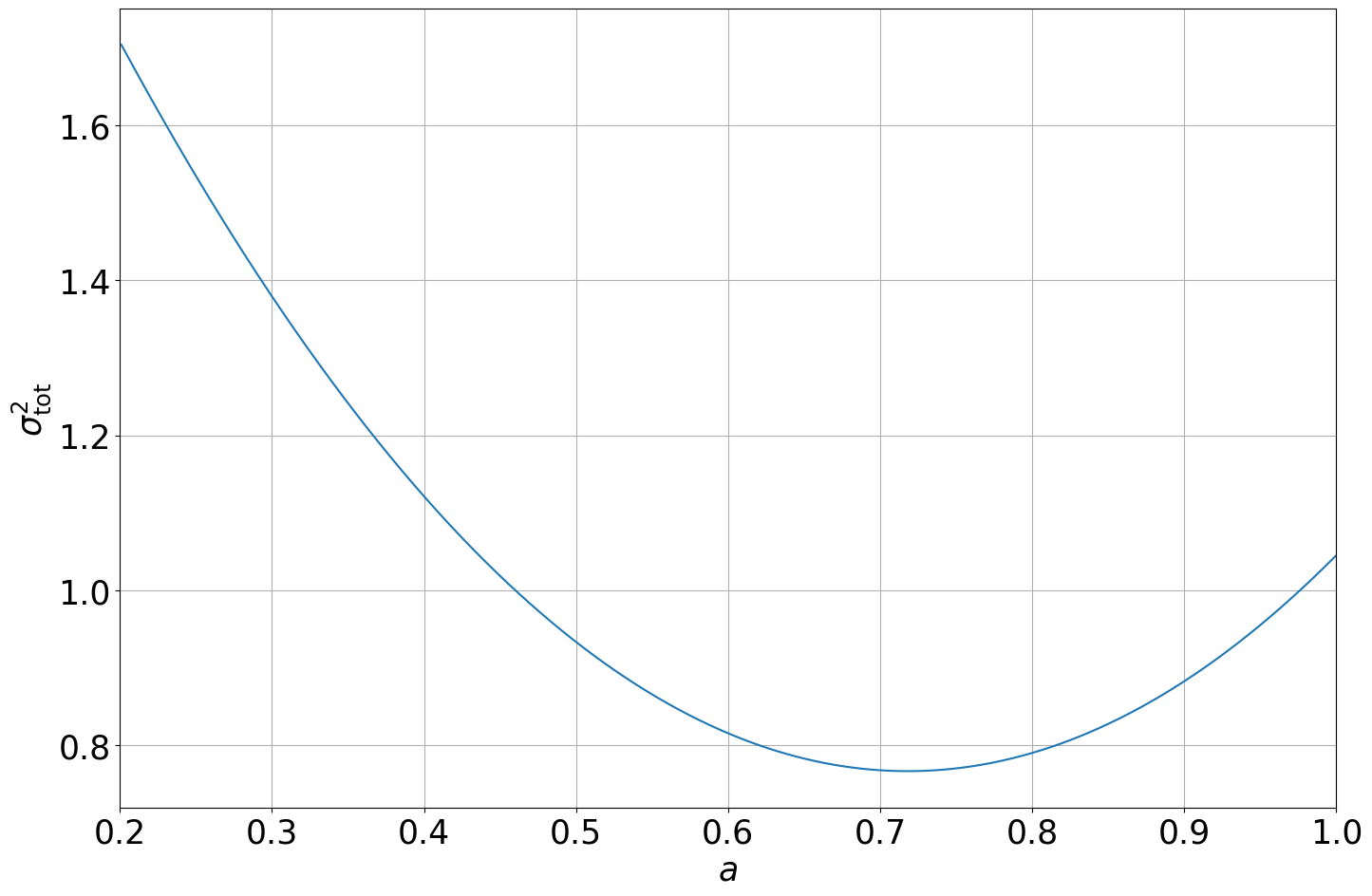}
\label{fig:OptPearson:var}
\includegraphics[width=0.45\textwidth]{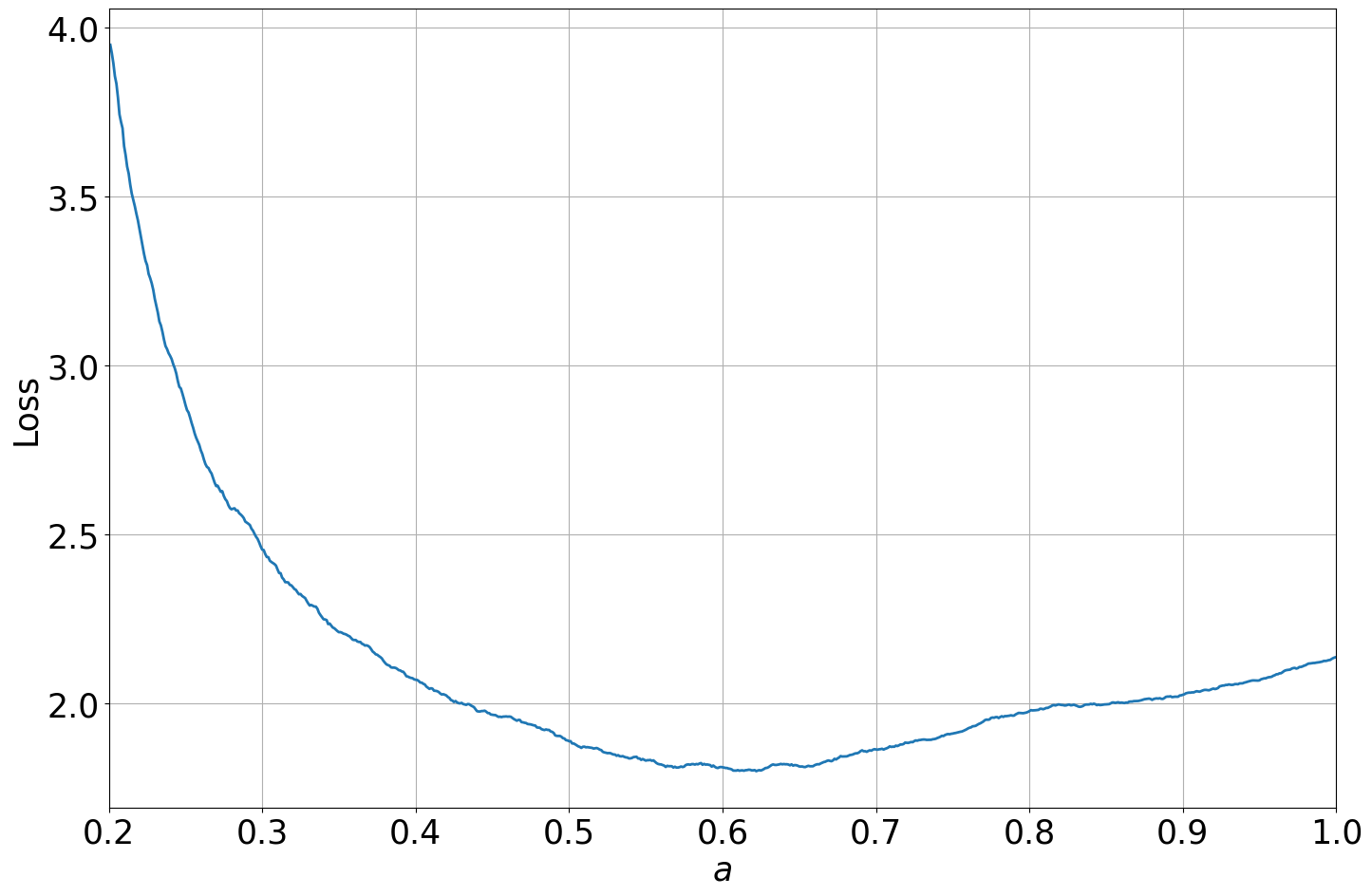}
\label{fig:OptPearson:trig}
\caption{Quantities minimized for the optimization of the correlation trigger. These quantities are shown as a function of the weight of the correlation between Livingston and Hanford $a$. Left: Variance as defined in Eq.~\eqref{eq:var_tot}. Right: Loss as defined in Eq.~\eqref{eq:loss_OptPearson}.}
\label{fig:OptPearson}
\end{figure*}

In Fig.~\ref{fig:OptPearson} we show the variance as a function of the weight of the correlation between Livingston and Hanford $a$, where we observe that the variance has a minimum of $\sigma^2_{\text{tot}} = 0.767$ for $a=0.718$.

Even though minimizing the variance makes sense to more closely relate the signal to noise ratio with the trigger discriminant, in this way we are not optimizing the trigger for its design purpose. The correlation trigger is designed to preselect promising candidates for the neural network by accepting the events satisfying $D > D_0$. Because of this, one other possible approach for optimizing the trigger is to set a reference signal to noise ratio $(S/N)_0$ at which to accept 50\% of the events, and minimize the number of events that are rejected with signal to noise ratio larger than $(S/N)_0$. We chose to set $(S/N)_0 = 9$, because testing smaller values of $(S/N)_0$ the acceptance of the trigger became too large and the quality of the candidates worsened enough to hinder the training of the Neural Network, which reduces the overall performance of the search. For each value of $a$ (Eq.~\eqref{eq:opt_discriminant_formula}) we then find the value of the threshold discriminant $D_0$ such that $N\!\left(D_{(S/N)_0}>D_0\right)/N_{(S/N)_0} = 0.5$. Finally, we define the loss we will want to minimize as:

\begin{equation}
    \text{loss} = \sum_{S/N = (S/N)_0+1}^{40} \frac{N\!\left(D_{S/N}<D_0\right)}{N_{S/N}},
    \label{eq:loss_OptPearson}
\end{equation}

\noindent which is computed for the same injections as the variance.

In Fig.~\ref{fig:OptPearson} we also show the loss as a function of the weight of the correlation between Livingston and Hanford, where we observe that the loss has a minimum of $1.798$ for $a = 0.622$, which has $D_0 = 0.296$.

Looking at Fig.~\ref{fig:OptPearson}, we can observe that both the variance and the loss have their minima near each other. In particular, making the choice of $a = \frac{2}{3}$ that was used in the section Sec.~\ref{sec:Data:presel}, the two quantities take values very close to their minimum, which is why this choice was made. We then choose $D_0 = 0.3$ to keep at around 9 the signal to noise ratio at which 50\% of events are detected.

\section{Injection Parameters}
\label{sec:anex:InjectionParams}

In Table~\ref{table:InjectParam} we show the ranges of parameters that we have used for the injections all through this paper.

\begin{table*}[t!]
\centering
\begin{tabular}{c|c}

    Parameters  & Range  \\  
\hline
    Black hole component masses: $m_1$, $m_2$ & (0.3$M_\odot$, 50$M_\odot$) \\
\hline
    Slope of the hyperbola asymptotes: $j_0 = \sqrt{e_{t0}^2-1}$ & (0.25, 5) \\
\hline
    Maximum velocity: $\text{v}_{\text{max}} = c \overline{\xi}^{1/3} \sqrt{\frac{e_{t0} + 1}{e_{t0} - 1} }$ & (0.1c, 0.4c) \\
\hline
    Distance to the event: $R$ & (0.5Mpc, 50Mpc) \\
\hline
    Magnitude of the component spins: $\chi_1$, $\chi_2$ & (0.0, 0.7) \\
\hline
    Component spins polar angles: $\theta_1^i$, $\theta_2^i$ & (0, $\pi$) \\
\hline
    Component spins azimuthal angles: $\phi_1^i$, $\phi_2^i$ & (0, $2\pi$) \\
\hline
    Orbit inclination: $\Theta$ & (0, $\pi$) \\
\hline
    Initial orbital azimuthal angle: $\Phi_0$ & (0, $2 \pi$) \\
\hline
    Right ascension of the source: $\alpha$ & (0, $2 \pi$) \\
\hline
    Declination of the source: $\delta$ & ($-\pi/2$, $\pi/2$) \\
\hline
    Polarization of incoming gravitational waves: $\psi$ & (0, $2\pi$) \\
\hline
\end{tabular}
\caption{Parameters of the close hyperbolic encounters that have been randomly varied to make the injections, together with the ranges within which we have varied them.}
\label{table:InjectParam}
\end{table*}

\bibliography{Refs}

\end{document}